# The Same and Not the Same: A Machine-Auditable Genealogy for Chemical Adjudication, Explanation, and Hypothesis Generation[1]

Junyi Gong[*a], Zijie Qiu[b,c], and Ben Zhong Tang[*b,c]

[a] Faculty of Chemistry, Shenzhen MSU-BIT University, No. 1 International University Park Road, Shenzhen 518172, Guangdong Province, China.

[b] School of Science and Engineering, Shenzhen Institute of Aggregate Science and Technology, The Chinese University of Hong Kong, Shenzhen (CUHK-SZ), 2001 Longxiang Road, Shenzhen 518172, Guangdong Province, China.

[c] Department of Chemistry, Hong Kong Branch of Chinese National Engineering Research Center for Tissue Restoration and Reconstruction, Institute of Molecular Functional Materials, Division of Life Science and State Key Laboratory of Molecular Neuroscience, Hong Kong University of Science and Technology, Clear Water Bay, Kowloon, Hong Kong, China.

[*] E-mail: gongjunyi@smbu.edu.cn (J.G.), tangbenz@cuhk.edu.cn (B.Z.T.)

**Abstract.** Formalizing individual chemical theories does not establish how their propositions relate, nor how certified relations can guide experiments. We introduce a machine-auditable formal genealogy: a synchronic, proposition-level structure recording translations, premises, observation channels, provenance and typed evidence. For seventeen fixed theory versions, all 136 unordered theory pairs enter a bounded screening comprising 152 named query records. Sixty-seven queries are certified across 53 pairs; unresolved, scoped-not-established and out-of-scope dispositions remain explicit rather than being converted into relations or refutations. Three adjudications delimit the symmetry-factor one-half boundary, separate Kasha's rule from quantum-yield independence while certifying a conditional spectral-shape implication, and distinguish intensity-only from joint intensity–lifetime quenching observations. A fourth case explains aggregation-induced emission as an outcome governed by the nonradiative-to-radiative competition ratio, with restricted conical-intersection access only one conditionally sufficient mechanism. The genealogy also generates model-scoped chemical insights: a Marcus-window condition for positive upper-state emission, a quantitative emission-gap–nonradiative-barrier bridge, spin–orbit and gap-law control of IC/ISC competition, and channel-selective quenching invariants. We translate these certificates into experimental hypothesis cards specifying observables, independent parameters, controls and failure interpretations. Machine checks establish validity within the declared models, not experimental adequacy; experiments test the physical identifications and model applicability rather than the Lean proofs.

## 1 Introduction

Photochemical theory is often transmitted through terse, empirically inflected maxims: the symmetry factor is "usually taken as one-half";[1–3] Kasha's rule "implies that emission is independent of excitation wavelength";[4–6] and static and dynamic quenching "can be distinguished by Stern–Volmer behaviour."[6,7] These statements compress distinct propositions, mechanisms and observation channels into portable rules of thumb. Their scientific use therefore depends on questions that are often left implicit: under which model and premises are they equivalent, when can an experiment distinguish them, and what should be measured when the simple statement fails?

Analogous difficulties arise among theories. Marcus, Hammond and Evans–Polanyi theories can be accommodated within shared free-energy surfaces; Kasha's rule and models of internal conversion can be connected only through explicit rate identifications; and static and dynamic quenching can agree in one observation channel while separating in another. A shared mathematical form does not establish a shared physical proposition. Conversely, a conditional relation can have chemical value without asserting unrestricted identity: it can identify a parameter regime, a measurable consequence or an additional observation channel. Our title echoes Roald Hoffmann's *The Same and Not the Same.*[8]

[1]The Supporting Information is included in this file after the reference list: theory versions and statement provenance, relation semantics and audit statuses, the complete 152-query registry, A1–A4 evidence, theorem cards for the generative case studies, experimental hypothesis protocols, provenance reconciliation, certificate and statement catalogs, and reproduction commands (SI §S1–S13 and Appendices A–D). The complete machine-checked development and machine-readable audit package are available as described under Data availability.

Existing formalization work has largely focused on the correctness of individual theories or theorems.[9–11] Here we address the relational layer and its experimental interpretation. A *formal genealogy* is a synchronic, non-historical, proposition-level structure that records versioned endpoints, translations, shared contexts, premises, observation channels, scopes, provenance and typed evidence. It is not a genealogy of theoretical descent and does not presume that all phenomenological theories reduce to a common physical law.

The infrastructure supports three distinct scientific functions. First, it *adjudicates* inherited claims by supplying exact boundaries, counterexamples and channel-specific evidence. Second, it *explains* mechanistic hypotheses without collapsing an observable phenomenon, its kinetic criterion and a conditionally sufficient mechanism. Third, it *generates* model-scoped hypotheses from certified cross-theory relations. This third function is not unconstrained automated discovery: researchers choose the chemical questions and models, while the machine verifies the resulting statements and preserves their applicability boundaries.

We apply this framework to seventeen fixed theory versions. All 136 unordered theory pairs enter a bounded screening with 152 named query records, of which 67 are certified across 53 pairs. We use A1–A3 to demonstrate adjudication and A4 to demonstrate mechanistic explanation. We then develop four generative case studies: Marcus-window-controlled anti-Kasha emission, an emission-gap–nonradiative-barrier relation, spin–orbit and energy-gap control of phosphorescence, and channel-selective dynamic quenching. Each case is translated into an experimental hypothesis card with observables, independent parameters, controls and an explicit interpretation of failure.

Machine checks establish model-internal validity, not experimental adequacy. Registry coverage is bounded rather than logically exhaustive, and an unresolved, not-established or out-of-scope record is not a nonexistence theorem. Experiments test whether the declared models and physical identifications apply to real systems; they do not refute a Lean certificate by failing to instantiate its premises.

## 2 Results and Discussion

### 2.1 Formal Genealogy as an Inference Infrastructure

The basic unit of a formal genealogy is not a theory name but a versioned proposition with declared provenance. Two statements enter a relational judgement only after their endpoints have been translated into a shared context $\Delta$, with an explicit premise predicate $\Pi : \Delta \to \mathrm{Prop}$, comparison scope and observation convention. A shared context records the variables required for a comparison; it does not assert that the source theories are physically identical.[12,13]

Seven evidence kinds distinguish definitional identity, implication, equivalence, conditional composition, observational equivalence, distinguishability and refutation. Definitional identity is established by definitional unfolding after translation. Implication and equivalence preserve their logical direction. Conditional composition must consume typed parent certificates and align their intermediate proposition. Observational equivalence is indexed by a declared channel, whereas distinguishability requires a separating observation or an identifiability decoder. Refutation requires an explicit counterexample; a failed proof attempt is not a refutation.

The same pair of theories may therefore support several compatible judgements. Static and dynamic Stern–Volmer models, for example, can agree throughout a matched intensity channel while separating in a joint intensity–lifetime channel. Relation kind is a property of a query, context and scope, not a permanent label attached to a pair of theory names.

Every record carries source and target statement versions, translation, premises, scope, evidence and snapshot. Chemical judgement decides what to compare and whether a model is physically reasonable. The machine checks certificate type, premise projections, source integrity and snapshot consistency. Audit outcome is independent of relation kind: the active vocabulary is `certified`, `refuted`, `unresolved`, `notEstablished` and `outOfScope`. The last three do not assert chemical nonexistence, and an unresolved search is never promoted to a refutation.

### 2.2 The Certificate-to-Hypothesis Protocol

A certificate becomes experimentally useful only through a second, explicitly model-dependent translation. For each generative case study we record: (i) the formal statement and certificate identifier; (ii) every load-bearing premise and shared-parameter identification; (iii) the observable implied by the declared channel; (iv) parameters that must be measured independently; (v) controls that test the channel assignment; and (vi) the conclusion that follows if the predicted observation fails.

This protocol distinguishes two gates that are often conflated. The *formal gate* asks whether the theorem follows within the declared model. The *physical-mapping gate* asks whether a real system instantiates the model, parameter identifications and observation channel. An experiment can reject the second gate without affecting the first. In particular, a positive model yield need not exceed an experimental detection limit, and an exact mathematical zero is not interchangeable with a below-detection measurement.

The same protocol also clarifies what an experiment cannot show. Observational equivalence in one channel does not prove mechanism identity; a one-way implication does not predict its converse; and equality of barrier-rate expressions does not by itself identify a conical-intersection mechanism. Figure 1 summarizes the resulting theory–certificate–experiment loop.

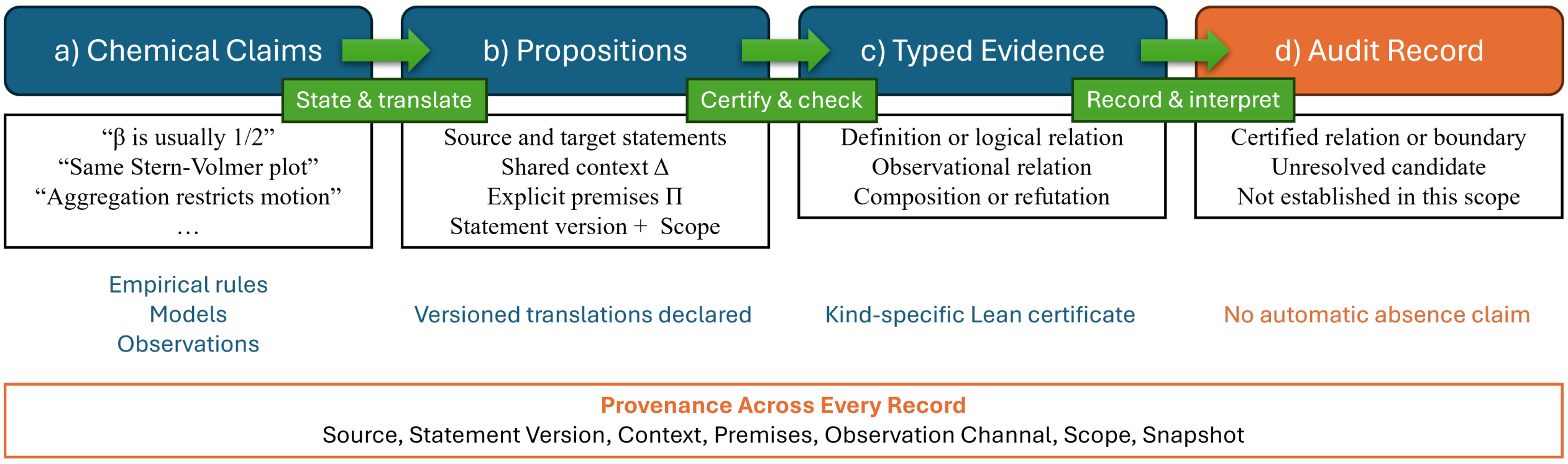

Figure 1: **(a),** Versioned chemical statements are translated into shared contexts with explicit premises, scopes and observation channels. **(b),** Bounded proposition-level screening produces typed certificates or one of the other declared audit outcomes. **(c),** Certified relations support three scientific functions: adjudicating inherited rules, explaining mechanisms as conditional chains, and generating model-scoped hypotheses. **(d),** The certificate-to-hypothesis protocol assigns observables, independent parameters, controls and failure interpretations. Experiments test the physical mapping and can motivate revised queries; they do not replace or directly falsify a Lean proof. Human chemical modelling, machine verification and experimental adequacy remain distinct responsibilities.

### 2.3 The Certified Landscape of Seventeen Theories

We apply the framework to seventeen fixed phenomenological theory versions, primarily from photochemistry and photophysics with comparative examples from catalysis, electrochemistry and crystal chemistry. The collection includes shared free-energy surfaces, excited-state rate ladders, quantum-yield competition, quenching, radiation laws, energy transfer and conical-intersection kinetics. This heterogeneity is intentional: some statements can be compared through explicit translations, others only through shared observables, and some should not be forced into a common context.

All $\binom{17}{2} = 136$ unordered pairs have a named statement-level query or a statement-level `outOfScope` disposition. The active registry contains 152 queries and 152 records: 67 `certified`, zero `refuted`, 39 `unresolved`, two `notEstablished` and 44 `outOfScope`. The 67 certified queries occur over 53 theory pairs. Ninety-four pairs have at least one non-`outOfScope` query, whereas 42 have only a statement-level `outOfScope` record. These are query outcomes and pair-coverage facts, not a claim that all propositions or all chemical relations have been exhausted.

The certified queries comprise 36 observational equivalences, 13 equivalences, 12 definitional identities, four implications and two conditional compositions. The certificate library contains 93 objects: 38 delivered-module certificates and 55 bounded-probe certificates. Active certified evidence uses 55 probe certificates and 12 reused delivered-module certificates. The remaining 26 delivered certificates are retained support or library evidence and do not add active pair or query coverage. Three XD1 candidate declarations remain unresolved and are not certificate-library members.

A separate 151-record provenance registry contains 26 named queries and 125 pair anchors. Its additive disposition index classifies four declaration aliases, three certified-view links and 44 scope relationships. The retained Kasha–Vavilov refutation and Stern–Volmer distinguishability certificates remain valid within their declared scopes, separately from the active bounded outcome distribution with zero `refuted` records and no active distinguishability certificate. Provenance support, active queries and theory-level adjudications are never pooled into a single edge count.

Figure 2 presents the active landscape as a theory-pair matrix rather than as a sparse graph of selected edges, together with the provenance chain that produces it. Panel (a) reports certified-query multiplicity for all 136 pairs without treating multiple queries in one cell as additional theory pairs. Panel (b) traces the frozen 102-file corpus and 1,154 statement rows through 92 active statement endpoints to 152 active query records and 67 certified evidence items, while retaining the 93-object certificate library and the separate 151-record provenance registry as distinct evidence layers. Unresolved and out-of-scope regions identify future translation and observation-channel questions; they do not denote chemical nonexistence. Per-row snapshot provenance is given in SI Appendix C.

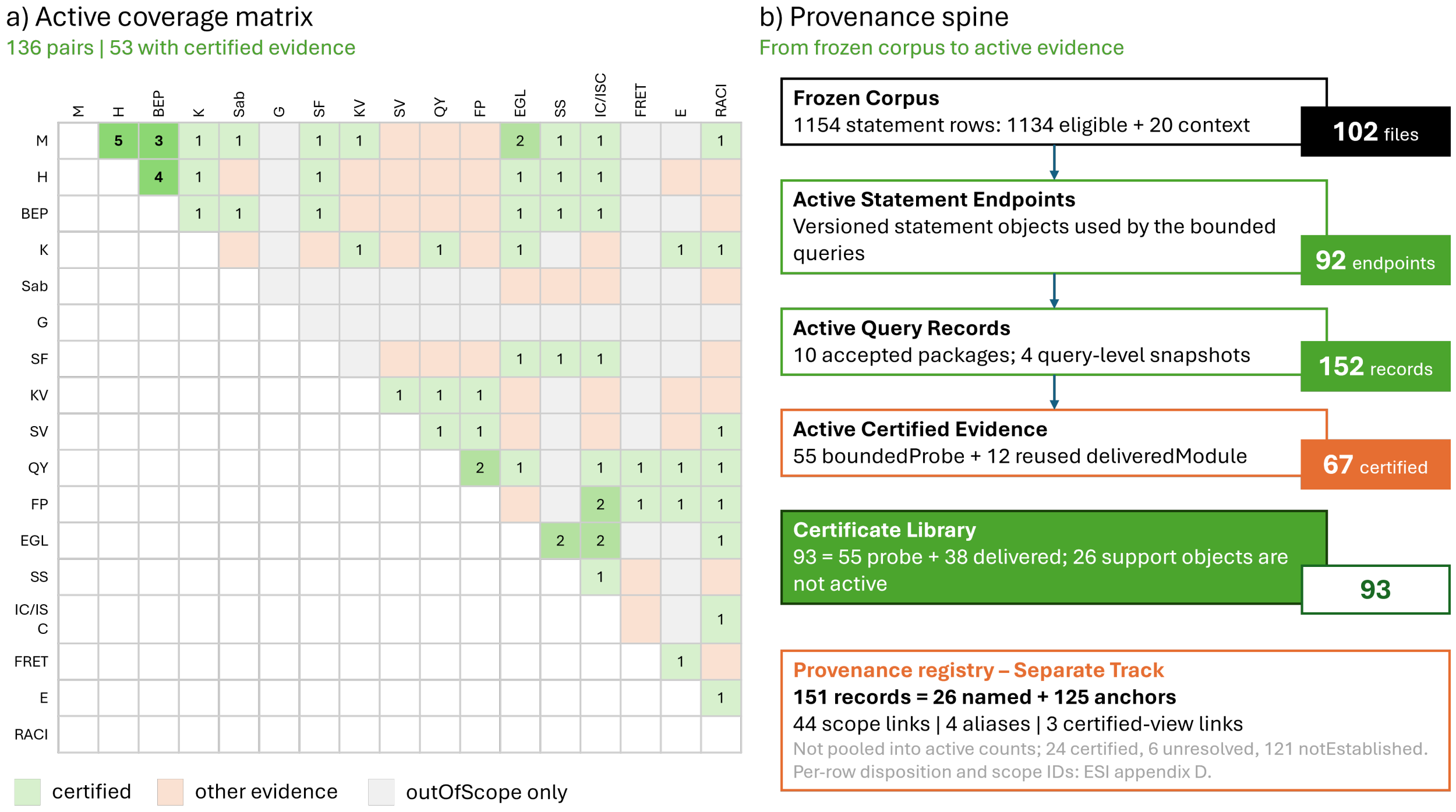

Figure 2: The accepted bounded genealogy of seventeen fixed theory versions and its provenance. **(a),** The 17 × 17 upper-triangular matrix summarizes all 136 theory pairs. Cell numbers count active certified queries; fill distinguishes certified, non-certified and statement-level out-of-scope evidence, and the inner strip reports certified/unresolved/not-established/out-of-scope composition. Multiple queries in one cell are not additional theory pairs. **(b),** Provenance spine from the frozen 102-file corpus and 1,154 statement rows to 92 active statement endpoints, 152 active query records and 67 certified evidence items (55 boundedProbe and 12 reused deliveredModule). The 93-object certificate library and the separate 151-record provenance registry remain distinct and are not pooled into active coverage. Per-record statement, certificate, scope, snapshot and hash provenance is given in SI Appendices A–D.

## 2.4 Adjudication: Where Inherited Chemical Rules Stop

The first use of the genealogy is to replace compressed chemical slogans with statements whose validity and failure boundaries can be inspected. Figure 3 summarizes A1–A3 and, for contrast, the distinct explanatory role of A4 developed below.

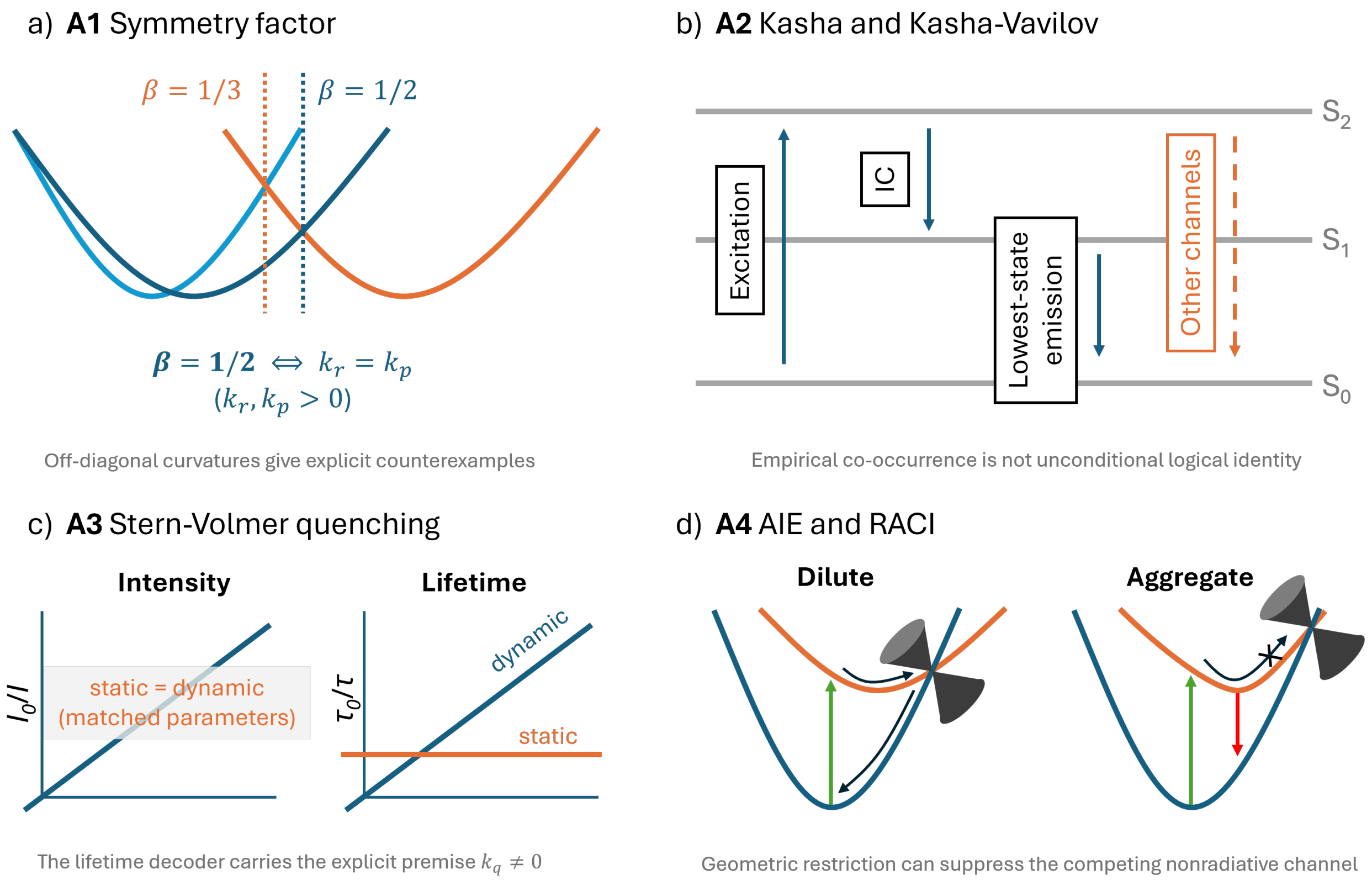

Figure 3: **(a),** A1: in the positive-curvature double-parabola model at zero driving force, $\beta = \frac{1}{2}$ if and only if $k_r = k_p$; unequal-curvature witnesses delimit the equality. **(b),** A2: a retained scoped refutation concerns the unconditional implication from `KashaRule` to `VavilovAt`, whereas the new statement-level certificate `Q-RAD1-KKV-01` establishes `KashaRule` ⇒ `SpecSame` under positive-yield premises. Spectral-shape invariance and quantum-yield independence are different propositions. **(c),** A3: matched static and dynamic quenching models agree in the intensity channel but separate in the joint intensity–lifetime channel, with $k_q \neq 0$ required by the decoder. **(d),** A4 is an explanation rather than a fourth synonymy adjudication: aggregation-induced emission is the outcome, the nonradiative-to-radiative ratio is the kinetic criterion, and restricted conical-intersection access is one conditionally sufficient mechanism.

#### 2.4.1 A1: The Exact Equal-Curvature Boundary of "One-Half"

The symmetry factor is often abbreviated as $\beta \approx \frac{1}{2}$ and treated as a general midpoint property.[1–3] For the positive-curvature double-parabola model at zero driving force, the thermoneutral crossing coordinate has the closed form

$$\beta = \frac{\sqrt{k_p}}{\sqrt{k_r} + \sqrt{k_p}}$$

and the certified equivalence is

$$\beta = \frac{1}{2} \Leftrightarrow k_r = k_p$$

The witnesses $(k_r, k_p) = (1, 4)$ and $(4, 1)$ give $\beta = \frac{2}{3}$ and $\frac{1}{3}$, respectively. Thus one-half is exact on the equal-curvature diagonal rather than a universal definition. The formal result also defines an experimental question: in a system for which this surface model is adequate, independently resolved reactant and product curvatures should determine whether the thermoneutral coordinate lies on or off that diagonal. A measured deviation would constrain the surface parameters, not overturn the formal equivalence.

#### 2.4.2 A2: Kasha and Vavilov Are Related Through Specific Channels, Not Synonyms

Kasha's rule concerns the emitting excited level, whereas Kasha–Vavilov statements concern dependence on excitation.[4–6] Theory-level and provenance evidence distinguishes pointwise independence, conditional closed-form coincidence, separation on the lossless corner, and a refutation of one specified unconditional implication from `KashaRule` to `VavilovAt`. That refutation applies within its declared scope and addresses a different proposition from the active spectral-shape query below.

The active statement-level query `Q-RAD1-KKV-01` certifies a complementary implication: under `RateData` and positive fluorescence-yield premises for every lower level, `KashaRule` implies `SpecSame`. Here `SpecSame` denotes equality of normalized spectral shape, not constancy of total fluorescence quantum yield and not the distinct `VavilovAt` predicate. The implication is one-way and does not establish a general equivalence between Kasha's rule and Vavilov's rule.

These results are complementary rather than contradictory. The genealogy separates two experimental axes that a single slogan conflates: which state emits, and which property—spectral shape or yield—is invariant under excitation. Section 2.6 adds a further premise-gated link from a Marcus internal-conversion window to positive upper-state emission.

#### 2.4.3 A3: Quenching Identifiability Depends on the Observation Channel

Static and dynamic quenching are often described as directly distinguishable by Stern–Volmer experiments.[6,7] Under matched parameters, $k_0 \neq 0$ and $K_{\mathrm{SV}(k_0,k_q)} = K_a$, however, their intensity observations agree pointwise. The certified relation is therefore observational equivalence in the intensity channel, not identity of the physical mechanisms.

The joint intensity–lifetime channel supplies a separating observation: at $k_0 = 2, k_q = 2, K_a = 1$ and $q = 1$, the intensity components agree while the lifetime components differ. A separate decoder recovers the mechanism tag from the complete joint observation profile under $0 < k_0$, $k_q \neq 0$ and $0 < q$. No lifetime separation is claimed at $k_q = 0$.

This case provides a direct experimental design rule: an intensity-only Stern–Volmer plot is not, under the matched conditions, a mechanism discriminator. The certified decoder uses the joint intensity–lifetime channel together with its stated premises; intensity-only matching does not supply that decoder. These theory-level and retained distinguishability results are not silently added to the active 67-query total; their role is to delimit what the available observations can identify.

### 2.5 Explanation: AIE as Outcome, Criterion and Mechanism

Aggregation-induced emission (AIE) is often compressed into a single mechanism narrative: aggregation restricts motion, nonradiative decay decreases, and emission increases.[14,15] The formalization separates three levels that this narrative combines: the observed enhancement, the kinetic criterion that controls it, and a particular structural mechanism capable of satisfying that criterion.

#### 2.5.1 Outcome and Kinetic Criterion

Within the two-channel Jablonski model,

$$\Phi = \frac{k_r}{k_r + k_{\mathrm{nr}}}$$

where $k_r > 0$ and $k_{\mathrm{nr}} \geq 0$. The theory-level equivalence is

$$\Phi_{\mathrm{agg}} > \Phi_{\mathrm{sol}} \Leftrightarrow \frac{k_{\mathrm{nr,agg}}}{k_{\mathrm{r,agg}}} < \frac{k_{\mathrm{nr,sol}}}{k_{\mathrm{r,sol}}}$$

Thus AIE is controlled by the competition ratio $R = \frac{k_{\mathrm{nr}}}{k_r}$, not by a decrease of $k_{\mathrm{nr}}$ alone. Preservation of $k_r$ is a convenient sufficient setting, not a necessary condition. This distinction suggests a direct kinetic analysis: when a single-exponential two-channel model is adequate, use $k_r = \frac{\Phi}{\tau}$ and $k_{\mathrm{nr}} = \frac{1-\Phi}{\tau}$ to test whether the ratio, rather than either rate separately, tracks emission enhancement.

#### 2.5.2 RACI as One Conditionally Sufficient Mechanism

Restricted access to a conical intersection (RACI) supplies one mechanism that can decrease this ratio under explicit assumptions.[16,17] In the declared two-state Hamiltonian, single-torsional-coordinate and barrier-rate model, the chain is

$$\begin{gathered}\text{restricted torsional accessibility}\\ \Rightarrow \text{increased CI accessibility gap}\\ \Rightarrow \text{decreased barrier-controlled } k_{\mathrm{nr}}\\ \Rightarrow \text{decreased } R \Rightarrow \text{enhanced } \Phi\end{gathered}$$

The final steps require the declared rate conditions; the simplified theorem holds $k_r$ fixed, whereas the general version requires only a decrease of the competition ratio. The two-state truncation, torsional coordinate, aggregate accessibility window and barrier-rate law are modelling premises, not consequences of the AIE observation. The named no-CI non-model delimits RACI and is not a counterexample to AIE. A classical double-parabola surface crossing is also not identified with a codimension-two conical intersection.[18–20]

### 2.5.3 Rate-Slot Compatibility Is Not Unique Mechanistic Identification

Three new statement-level observational equivalences extend this explanatory layer. `Q-CI-MR-01`, `Q-CI-EG-01` and `Q-CI-IC-01` show that, at $\beta = \frac{1}{k_B T}$ and with $B$ identified respectively as the Marcus, energy-gap-law or Franck–Condon barrier, the corresponding rate equals the RACI barrier-channel rate in the declared channel. These identities establish modular compatibility of rate laws; they do not identify the physical coordinates, prove CI accessibility or show that ISC is included in the IC-only comparison.

The experimental implication is deliberately limited. A barrier-rate fit alone cannot uniquely identify RACI. Testing the mechanism requires independent evidence such as ultrafast CI dynamics, structural calculations or direct constraints on torsional accessibility, together with measurements of both radiative and nonradiative channels. A decrease of $R$ without such evidence remains consistent with, but does not prove, RACI. The unresolved BEP–RACI query further illustrates why a shared exponential shape cannot by itself identify a chemical barrier.

## 2.6 Generation: New Model-Scoped Chemical Insights

The bounded screening does more than assign trust labels to existing edges: it certifies explicit statement-level connections among theories. We do not interpret all 67 certified queries as independent chemical discoveries. Instead, we select four cases in which a certificate or typed composition yields a clear chemical interpretation and a feasible experimental question. Figure 4 summarizes the model-scoped predictions; these are not experimental data.

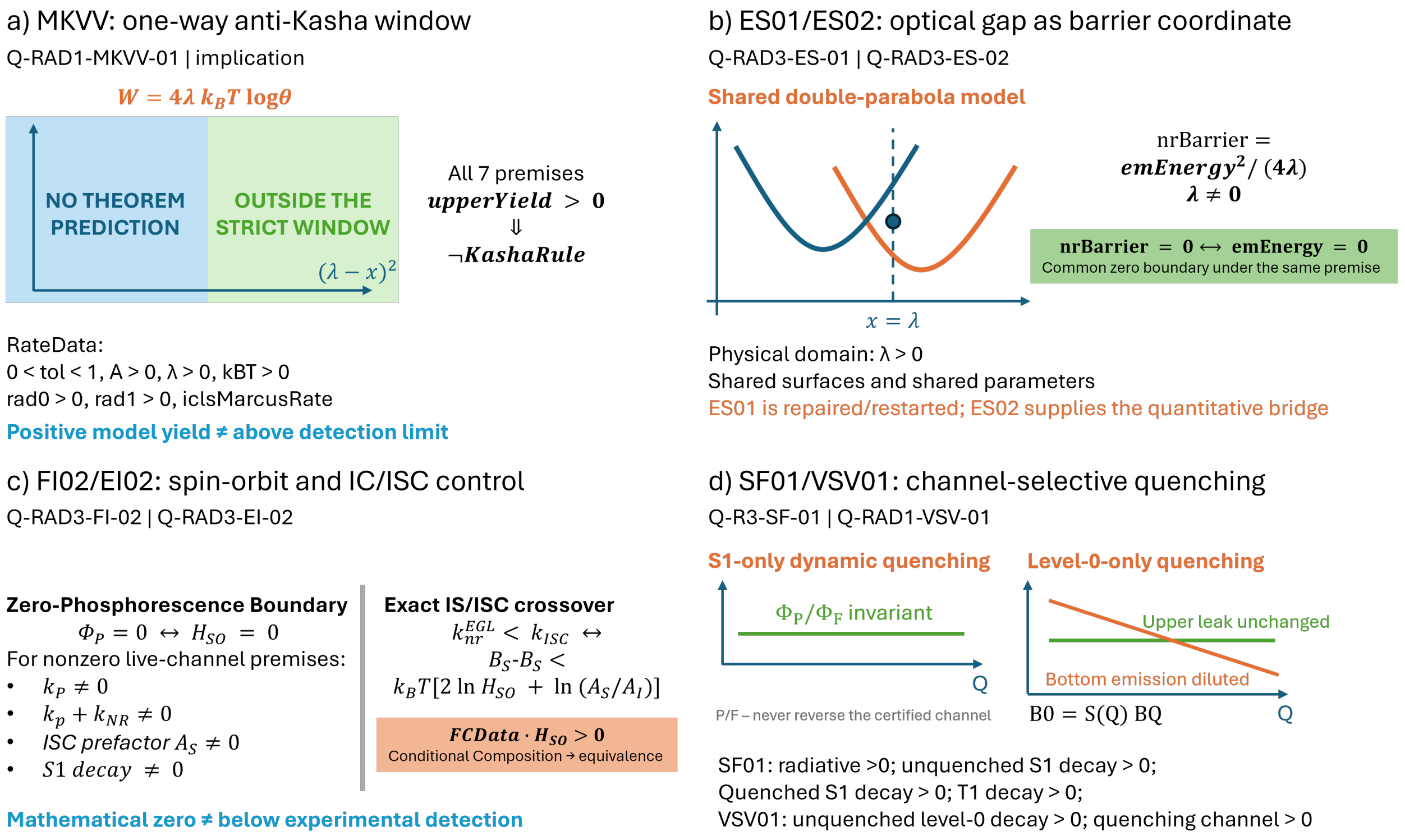

Figure 4: Four generative chemical insights and their model boundaries. **(a),** Under the seven declared premises of `Q-RAD1-MKVV-01`, a squared Marcus gap outside the Kasha tolerance window implies positive upper-state emission; no converse or zero-emission prediction inside the window is made. **(b),** In the shared double-parabola model with $\lambda \neq 0$, `Q-RAD3-ES-02` gives nrBarrier $= \frac{\text{emEnergy}^2}{4\lambda}$, and `Q-RAD3-ES-01` identifies the common zero boundary. **(c),** `Q-RAD3-FI-02` relates the zero-phosphorescence boundary to the spin–orbit ISC channel under four live-channel premises, while `Q-RAD3-EI-02` transfers the IC/ISC crossover to the gap-law IC rate. **(d),** Selective dynamic quenching leaves $\frac{\Phi_P}{\Phi_F}$ invariant in `Q-R3-SF-01`; in `Q-RAD1-VSV-01`, level-0 quenching dilutes bottom emission while leaving the upper-state leak unchanged. All panels are model-scoped predictions, not experimental data.

### 2.6.1 A Marcus Window for Positive Upper-State Emission

The earlier Kasha–Marcus comparison established that a window relation requires an explicit identification of the relevant internal-conversion rate with a Marcus rate. The new implication `Q-RAD1-MKVV-01` carries this identification into the Kasha–Vavilov ladder. Under seven declared premises—`RateData`, $0 < \text{tol} < 1$, positive prefactor, reorganization energy and thermal energy, positive radiative rates at levels zero and one, and `icIsMarcusRate`—a squared-gap condition outside the Kasha tolerance window implies violation of `KashaRule` through a strictly positive upper-state emission yield.

The chemical insight is a conditional route from driving-force and reorganization parameters to an anti-Kasha observable. It is not a universal assertion that every Marcus inverted system emits from an upper state. The implication is one-sided: within the window it makes no zero-emission prediction, and outside the window a positive mathematical yield may still lie below an experimental detection limit. The key experimental commitment is therefore the independent validity of `icIsMarcusRate`, together with the positivity and tolerance premises.

### 2.6.2 An Optical-Gap Coordinate for the Nonradiative Barrier

Two queries connect the energy-gap-law and Stokes-shift surface readings. `Q-RAD3-ES-01` certifies, on the delivered `nrBarrier` and `emEnergy` declarations, the zero-set equivalence

$$\text{nrBarrier}(\lambda, x) = 0 \Leftrightarrow \text{emEnergy}(\lambda, x) = 0 \qquad (\lambda \neq 0)$$

The complementary quantitative observational equivalence `Q-RAD3-ES-02` further gives

$$\text{nrBarrier}(\lambda, x) = \frac{\text{emEnergy}(\lambda, x)^2}{4\lambda} \qquad (\lambda \neq 0)$$

Within the shared double-parabola context, the optical gap is therefore also a coordinate for the nonradiative barrier. The result does not establish this identification for unrelated surfaces or arbitrary experimental activation energies. Experimental tests require an independent estimate of $\lambda$, a justified barrier-rate interpretation of the measured kinetics, and a physical domain with $\lambda > 0$. Negative or zero values of the signed model quantity `emEnergy` are boundaries or extrapolations, not measurable negative-energy photons.

#### 2.6.3 Spin–Orbit and Gap-Law Control of Phosphorescence

`Q-RAD3-FI-02` substitutes the theory-owned ISC rate into the fluorescence–phosphorescence cascade. Under four live-channel premises—nonzero $k_P$, nonzero $k_P + k_{\mathrm{NR}}$, nonzero ISC prefactor and nonzero $S_1$ decay—it proves

$$\Phi_P = 0 \Leftrightarrow H_{\mathrm{SO}} = 0$$

The formal premises are nonzero conditions; their physical rate interpretation uses the appropriate positive domain. This is a boundary for the declared cascade, not a proof that every experimental absence of phosphorescence identifies zero spin–orbit coupling.

The conditional composition `Q-RAD3-EI-02` then transfers the delivered IC/ISC crossover to the energy-gap-law nonradiative rate. With positive Franck–Condon data and $H_{\mathrm{SO}} > 0$,

$$k_{\mathrm{nr}}^{\mathrm{EGL}} < k_{\mathrm{ISC}} \Leftrightarrow B_S - B_I < k_B T \left[ 2 \ln H_{\mathrm{SO}} + \ln\left(\frac{A_S}{A_I}\right) \right]$$

where $B_S - B_I$ is the ISC-minus-IC barrier deficit and $A_S, A_I$ are the corresponding prefactors. Together, the two certificates suggest joint measurements of triplet population, phosphorescence, fluorescence and the competing decay rates. Exact zero is a model boundary; experimental tests should emphasize yield factorization, trends and the crossover threshold with explicit detection limits.

#### 2.6.4 Channel-Selective Quenching and Upper-State Leakage

The observational equivalence `Q-R3-SF-01` considers addition of $k_q[Q]$ to the $S_1$ nonradiative channel. With positive radiative rate, positive unquenched and quenched $S_1$ decay, and positive $T_1$ decay, the observed ratio $\frac{\Phi_P}{\Phi_F}$ is unchanged. The certificate's observation channel is P/F; any use of the reciprocal F/P requires a separate reciprocal statement and the corresponding nonzero conditions.

At the excited-state ladder level, `Q-RAD1-VSV-01` considers a quencher that adds a loss channel only at level zero. The bottom-emission contribution is diluted by the Stern–Volmer factor, whereas the upper-state leak is unchanged. The premises require positive unquenched level-zero decay and a nonnegative quenching channel. These results predict invariant ratios or components only for the declared selective channel. A changing P/F ratio or upper-state leak indicates that additional channels, direct triplet quenching or nonselective quenching must be examined; invariance alone does not prove a unique mechanism.

#### 2.6.5 An Adjacent-Domain Generative Example

Beyond photophysics, `Q-WPS01-BS-01` certifies on the descriptor-axis channel that the linear Sabatier volcano is the pointwise maximum of the two BEP tangent lines.[21–23] The result provides a compact interpretation of the volcano optimum as a switch between two linear-response branches. It is a statement about the linearized model and its barrier channel, not an unconditional derivation of every experimental activity volcano.

### 2.7 An Experimental Agenda Generated by the Genealogy

The generative value of the registry lies not in calling every certificate a discovery, but in making the next measurement and its interpretation explicit. Table 1 prioritizes five hypothesis cards. The order of chemical presentation is not an order of experimental difficulty: channel-selective quenching is comparatively direct, whereas the anti-Kasha and optical-barrier tests require more demanding system selection and independent parameter identification.

Table 1: Certificate-to-hypothesis matrix. Each row combines the formal basis with priority measurements, the model-scoped prediction, controls and failure interpretation. Every prediction is conditional on the stated model and physical parameter identification. No experimental result is asserted here.

| Hypothesis and formal basis | Measurements; model-scoped prediction, controls and failure interpretation |
|---|---|
| **E1: Marcus-window anti-Kasha emission** `Q-RAD1-MKVV-01`; comparison via `Q-RAD1-KKV-01` | **Measure and identify.** Excitation-resolved upper/bottom spectra and absolute yields; ultrafast $S_2$ decay and $S_1$ rise; independently constrain $\lambda$, gap, $A$, $T$, radiative rates, tolerance and `icIsMarcusRate`. <br><br> **Prediction and controls.** Outside the strict window, positive upper yield is implied only under all seven premises. Use direct $S_1$ excitation, matched absorbance, power and photochemistry controls and report detection limits. No inside-window zero prediction is made. Failure localizes the rate/window mapping or missing channels. |
| **E2: Optical-gap/nonradiative-barrier bridge** `Q-RAD3-ES-01`, `Q-RAD3-ES-02` | **Measure and identify.** Emission gap, absolute yield and lifetime for $k_{\mathrm{nr}}$; temperature-dependent kinetics only under an explicit barrier-rate model; independent $\lambda$ and shared-surface checks. <br><br> **Prediction and controls.** Within the shared model and for $\lambda \neq 0$, nrBarrier $= \frac{\mathrm{emEnergy}^2}{4\lambda}$; ES01 gives the common zero boundary under the same premise. Avoid circular fits and do not identify an apparent Arrhenius $E_a$ with `nrBarrier` without justification. Deviation challenges the surface, parameter or rate identification. |
| **E3: Spin–orbit and IC/ISC control** `Q-RAD3-FI-02`, `Q-RAD3-EI-02` | **Measure and identify.** Fluorescence/phosphorescence yields, triplet yield or transient absorption, $S_1/T_1$ lifetimes, $k_{\mathrm{IC}}$, $k_{\mathrm{ISC}}$, barriers and an independently constrained $H_{\mathrm{SO}}$ or proxy. <br><br> **Prediction and controls.** Test the FI02 cascade factorization under its four nonzero live-channel premises and the EI02 exact IC/ISC crossover threshold under `FCData` and $H_{\mathrm{SO}} > 0$. Control oxygen, direct triplet quenching and covariation of other rates. Below-detection emission is not mathematical zero; extra triplet sources or failed channel premises are alternatives. |
| **E4: Channel-selective dynamic quenching** `Q-R3-SF-01`, `Q-RAD1-VSV-01`; joint-channel interpretation from A3 | **Measure and identify.** Quencher titration with intensity, lifetime, $\Phi_F$, $\Phi_P$, $\frac{\Phi_P}{\Phi_F}$ and, where present, spectrally or kinetically resolved upper leak. <br><br> **Prediction and controls.** For the declared selective channels, P/F and upper leak remain invariant while bottom emission dilutes by the Stern–Volmer factor. Control static association, inner filter, oxygen, aggregation and direct $T_1$ quenching. Changes identify additional or nonselective channels; invariance is not mechanism uniqueness. |
| **E5: AIE criterion and RACI mechanism** A4 theory-level certificates; `Q-CI-MR-01`, `Q-CI-EG-01`, `Q-CI-IC-01` | **Measure and identify.** Measure $\Phi$ and $\tau$ in solution and aggregate; only under a valid single-exponential two-channel model extract $k_r = \frac{\Phi}{\tau}$ and $k_{\mathrm{nr}} = \frac{1-\Phi}{\tau}$; independently assess CI dynamics, structure and torsional accessibility. <br><br> **Prediction and controls.** AIE requires a decrease of $R = \frac{k_{\mathrm{nr}}}{k_r}$ within the two-channel model. Control scattering, reabsorption, morphology, multiexponential decay, ISC/FRET and rate covariation. A lower $R$ or a matching rate law does not by itself prove RACI. |

The five hypothesis cards define a decision protocol with feedback rather than a one-shot validation test. A matching observation supports applicability of the declared model and parameter identification within experimental uncertainty. If the predicted observation is not obtained, the first task is to determine whether a premise, channel assignment, parameter identification or rate mapping failed, or whether the predicted signature lies below the detection limit. That diagnosis can motivate a new formal query—for example, adding a direct triplet-quenching channel, replacing a rate identification or changing the observation channel. The resulting model or query revision, rather than an unqualified assertion that the Lean theorem was experimentally refuted, is the scientifically useful output.

### 2.8 Limitations

Machine checking establishes logical validity inside a declared formal model, not the experimental adequacy of that model. Positivity, nonzero rates, closure, shared surfaces, rate identifications and observation maps require chemical and experimental justification. A positive model yield need not exceed an instrument's detection limit, and zero yields, zero rates and zero barriers are ideal mathematical boundaries that experiments can approach only with uncertainties and detection thresholds.

The 136-pair census is bounded registration, not logical completeness. Unresolved, not-established and out-of-scope outcomes are not counterexamples or proofs of nonexistence. Explicitly recorded premises are also not automatically nonvacuous, nonredundant or weakest; those stronger evidence levels require their own certificates. The active outcome counts must not be conflated with retained legacy theory-level evidence, including the scoped A2 refutation and A3 distinguishability.

The workflow is researcher-guided rather than unconstrained automated discovery. Researchers select the chemical questions, models, translations and experimentally meaningful parameters. No new experiments are reported here, and no certified composition is by itself evidence of experimental prevalence, mechanism uniqueness or literature-wide chemical novelty.

### 2.9 Outlook

The same infrastructure can be extended to other layered phenomenological domains: conditional connections among BEP relations, Sabatier criteria, volcano models and kinetic approximations in catalysis; distinctions among thermodynamic potentials, kinetic overpotentials and current responses in electrochemistry; and channel-dependent equivalence or distinguishability in spectroscopy and materials science. Extension should add propositions, contexts and observation channels rather than merely add theory names.

A further direction is to make the experiment-to-query return path systematic. New measurements can justify or reject a proposed parameter identification, motivate additional channels and supply provenance for revised scopes. The aim is a cumulative, auditable loop between theory, proof and experimental design, not an autonomous replacement for chemical judgement.

## 3 Conclusion

A machine-auditable formal genealogy provides more than a graph of theory names. It certifies proposition-level relations with explicit premises and scopes, locates the exact boundaries of inherited rules, separates phenomenological outcomes from kinetic criteria and candidate mechanisms, and converts selected certified combinations into testable experimental agendas. The seventeen-theory census demonstrates both the reach and the sparsity of the current evidence: 67 certified queries over 53 pairs coexist with unresolved and deliberately bounded space. The resulting infrastructure connects theory, proof and prospective measurement while preserving the distinction between model-internal validity and empirical adequacy.

## 4 Methods

### 4.1 Fixed Corpus, Statement Inventory, and Fidelity

The audit fixes seventeen theory-specific versions and 102 frozen `PhotoLean/**/*.lean` files. Its statement-inventory boundary contains 1,154 rows: 1,134 theorem or definition rows are screening-eligible and 20 context-type rows are retained as structure or context carriers but are not screening endpoints. The accepted registry exposes 92 statement objects used by its bounded queries. These counts denote the frozen file set, the inventory boundary and the active-query endpoint set, respectively, and are not interchangeable. The complete theory bibliography and theory-specific statement authorities are supplied in SI Table S2 and Appendix B.

Statement-authority fidelity and definition-body identity are checked separately. Signature-fidelity probes compare declarations through the signature boundary before `:=`; they do not prove equality of definition bodies. Fresh `rfl` certificates pin theory-local copies of shared kernel objects to their kernel definitions, so a relevant body change causes compilation failure. Each theory is represented by declared statement objects and versions rather than by its name or by similar terminology.

### 4.2 Shared Contexts, Translations, and Arithmetic Conventions

Each query declares a shared context $\Delta$, total translations of its endpoints into that context, and a premise predicate $\Pi : \Delta \to \mathrm{Prop}$. Surface comparisons share coordinates, force constants, driving forces and reorgani-zation parameters; rate and observation comparisons share rate data, mechanism tags, observation domains and channel maps. A shared context records the variables needed for a comparison and does not assert unrestricted physical identity.

Lean's division and `Real.sqrt` are totalized. Consequently, substantive chemical statements carry the positivity or nonzero premises required to exclude degenerate totalized corners. Observational channels declare both their maps and admissible domains. Theory-specific conventions, including the Marcus barrier reading, the separate Hammond propositions, the tangent BEP reading, the energy-gap-law statement and the Kasha ladder, remain versioned modelling choices rather than hidden premises.

### 4.3 Typed Certificates and Five Audit Outcomes

For translated propositions $P, Q : \Delta \to \mathrm{Prop}$, an implication certificate establishes

$$\forall \delta, \quad \Pi(\delta) \Rightarrow (P(\delta) \Rightarrow Q(\delta))$$

and an equivalence certificate replaces the inner implication by $P(\delta) \leftrightarrow Q(\delta)$. A refutation of a specified implication requires an explicit witness satisfying $\Pi \wedge P \wedge \neg Q$. Definitional identity uses a fresh `rfl` check on translated terms. Conditional composition consumes typed parents, aligns the mediating proposition, and inherits, bridges or discharges every parent premise. Observational equivalence requires pointwise agreement on the declared channel, while distinguishability requires a separating observation or an identifiability decoder.

Every active query has exactly one outcome: `certified`, `refuted`, `unresolved`, `notEstablished` or `outOfScope`. Only the first attaches an accepted certificate; only `refuted` attaches a counterexample to the named target. Failed searches never become refutations. In the active accepted registry the refuted count is zero; this does not delete the retained legacy A2 refutation of the different unconditional `KashaRule` ⇒ `VavilovAt` target.

### 4.4 Bounded Screening and Evidence Classes

Ten accepted screening packages instantiate queries from the fixed statement inventory. The resulting deterministic registry covers 136 unordered pairs with 152 active queries and records: 67 certified, zero refuted, 39 unresolved, two not-established and 44 out-of-scope outcomes. Certified queries occur over 53 pairs. Ninety-four pairs have at least one non-out-of-scope query and 42 have only a statement-level out-of-scope disposition.

The 67 active certified evidence items consist of 55 `boundedProbe` certificates and 12 reused `deliveredModule` certificates. The certificate library contains 93 objects: 55 bounded probes and 38 delivered-module certificates. Twelve of the latter support active queries; the remaining 26 are retained support or library evidence and do not increase active coverage. The three XD1 candidate declarations are attached only to unresolved queries and never enter the certificate library.

Four query-level snapshots are retained: 105 records in the base snapshot, 17 in `r5rad1`, 16 in `r5rad2`, and 14 in `r4e2`. Full snapshot identifiers, package assignments and per-row provenance are listed in SI Table S1 and Appendix C. A separate 151-record provenance registry is indexed by additive dispositions and contains 26 named queries and 125 pair anchors, including five intra-theory records; none adds active pair or query coverage. Theory-level A1–A4 witnesses and active proposition-level queries remain separate evidence layers.

### 4.5 Certificate-to-Hypothesis Translation

Each generative claim is registered with its query and certificate identifiers, exact statement, shared parameters, load-bearing premises, observation channel and snapshot. A separate hypothesis card records the chemical observables, independently measurable parameters, controls, prediction direction, detection-limit qualifications, adverse-result interpretation and explicit non-claims. Evidence labels distinguish theory-level witnesses, reused delivered-module certificates, bounded-probe certificates and statement-level compositions. These labels describe formal provenance and do not assign experimental validation or literature-wide chemical novelty.

For `Q-RAD3-EI-02`, the conditional composition carries exactly `FCData` and $H_{\mathrm{SO}} > 0$ in a shared context containing the IC and ISC prefactors, reorganization energies, thermal energy, gap coordinates and spin–orbit amplitude. For `Q-R3-SF-01`, the implemented channel map is $\frac{\Phi_P}{\Phi_F}$; no reciprocal channel is claimed without a separate derivation and premises. For `Q-RAD1-VSV-01`, `quenchIC` modifies only ladder level zero.

### 4.6 Declaration-Level Evidence for the Principal Claims

The principal claims are identified in the main text by their A1–A4 labels and by the query and certificate identifiers introduced above and in Table 1. Full Lean declarations, certificate identifiers, module paths and premise records are tabulated in SI Appendices A–D and in the S6–S10 theorem cards. This separation keeps declaration-level provenance in the archival SI while the main text remains at the query and certificate level.

### 4.7 Verification and Model Boundaries

The validation report for the accepted registry records 31,347 checks with zero failures, 17 adversarial mutations rejected fail-closed and five positive controls passed. Nine probe files review 58 declarations: 55 certificate declarations and three unresolved XD1 candidates. Verification combines the whole-tree strict build, deterministic registry regeneration, structure and provenance gates, serial compilation of all nine probes and declaration-level axiom audits. Delivered and audited declarations contain no `sorry` and no custom axioms; the accepted axiom footprint is limited to `propext`, `Classical.choice` and `Quot.sound`. Missing or inconsistent evidence fails closed rather than becoming a successful unresolved result.

These gates audit formal consistency, certificate kinds, premise identifiers, composition structure, source hashes and snapshot discipline. They do not decide whether a selected chemical model adequately represents an experiment. Observational claims remain channel-relative, conditional claims retain their premises, and bounded absence remains distinct from logical nonexistence.

## Author contributions

J.G. designed the formalization programme, the relation calculus and the adjudication protocol, and wrote the manuscript; J.G. and the AI-assisted pipeline developed the Lean development; Z.Q. and B.Z.T. supervised the chemistry and the manuscript. All authors discussed the results and approved the final manuscript.

## Conflicts of interest

The authors declare no competing interests.

## Data availability

The machine-readable audit package accompanying this article contains the fixed theory corpus, statement inventory, accepted registry and schema, the 93-certificate library, nine probe files, the provenance disposition ledger, coverage reports, hypothesis cards and snapshot-specific reproduction metadata. The accepted registry, schema, disposition ledger and validation report are identified by SHA-256 in SI Table S9. Four query-level snapshot groups are retained within the registry rather than assigning one snapshot to every row.

The package is reproduced from the audit repository root using the exact commands and validation record in SI §S13.2. The validator includes the external gates, strict build and scan, nine probe compilations, deterministic isolated rebuild, provenance invariants and adversarial mutations. Exact source and artifact hashes identify the evidence used in this study; source-tree status alone is not used as an integrity identifier.

## Use of AI

AI assistants participated in language drafting and Lean proof development. The authors selected the chemical models, verified the formal statements, and checked the reported results. AI tools are not authors and bear no responsibility for the work. The reliability of the formal claims rests on statement-fidelity checks, compilation, placeholder and custom-axiom scans, and declaration-level axiom audits rather than on the generating process.

## Acknowledgements

This work was supported by the National Natural Science Foundation of China (grant nos. 52003228, 52273197, and 52333007) and by the Shenzhen Key Laboratory of Functional Aggregate Materials (ZDSYS20211021111400001). The authors also thank the Shenzhen Special Position Program for funding support.

## SUPPORTING INFORMATION

# The Same and Not the Same: A Machine-Auditable Genealogy for Chemical Adjudication, Explanation, and Hypothesis Generation

Junyi Gong · Zijie Qiu · Ben Zhong Tang

**Scope of this document.** This Supporting Information presents the accepted formal definitions, certificate semantics, provenance protocol, chemical adjudications, bounded audit registry, model-scoped hypothesis protocols, and reproducibility artifacts. Formal genealogy is synchronic and non-historical: it compares currently stated propositions under explicit contexts, premises, translations, observation channels, scopes, and snapshots. The theorem cards and protocols below generate model-scoped hypotheses, not experimental results; no experimental validation is reported here. Researchers select the chemical questions, theories, statements, models, and scopes. The bounded screen does not perform automatic discovery, and Lean certification does not establish chemical applicability. This file combines the main article and this Supporting Information in a single document, as required by the preprint server: section, page and reference numbering therefore continue the main article, and the SI-only sources appear at the end of the combined reference list that precedes this part.

# S1. Snapshot, evidence layers, and counting conventions

The accepted registry separates pair coverage, proposition-level query outcomes, and certified-relation discovery. There are 136 screened pairs and 152 active queries/records: 67 certified, 0 refuted, 39 unresolved, 2 notEstablished, and 44 outOfScope. The 67 certified queries occur on 53 distinct pairs. They are not 67 new chemical laws, and 136/136 coverage is not 136/136 certification.

Three evidence layers are kept distinct. (i) The active layer contains 152 query rows; its 67 certified rows use 55 bounded-probe certificates and 12 reused delivered-module certificates. (ii) The certificate library contains 93 = 38 deliveredModule + 55 boundedProbe objects; the other 26 deliveredModule objects are retained support/library evidence and cannot be added to active coverage. (iii) A separate provenance layer has 151 records and additive dispositions, plus theory-level A1–A4 witnesses. Provenance dispositions and intra-theory adjudications contribute zero active pair/query coverage.

Ninety-four pairs have at least one non-outOfScope query; 42 have only a statement-level outOfScope record. This is a finite bounded screen of fixed statements, not a logical exhaustion of either theory. Four query-level snapshots coexist in the accepted registry. Every query, record, scope, and premise set retains its own snapshot identifier; Appendix C prints it per row, and S13 records artifact-integrity identifiers.

**Table S1.** Active query partition by snapshot in the accepted registry.

| Snapshot | Queries | Packages |
|---|---|---|
| `snap-bounded-audit-2026-11-04` | 105 | `R3-QUENCH-01, R4-E1-EXCITED-01, RAD3, WP-SURFACE-01, WP-conical-intersection-01, XD1, XD2` |
| `snap-bounded-audit-2026-11-04-r4e2` | 14 | `WP-excited-state-kinetics-02` |
| `snap-bounded-audit-2026-11-04-r5rad1` | 17 | `RAD1` |
| `snap-bounded-audit-2026-11-04-r5rad2` | 16 | `RAD2` |

# S2. Theory versions and statement provenance

The corpus contains seventeen theory-specific versions, not claims about all historical formulations. The shared two-parabola kernel is supporting infrastructure rather than an eighteenth theory. Table S2 gives the complete theory bibliography and the formalized core used in this work.

**Table S2.** The theory inventory and the formalized core used in this work.

| ID | Theory | Formalized core | Primary sources |
|---|---|---|---|
| M | Marcus theory | Quadratic activation barrier, normal and activationless regimes, and the inverted region. | [24,25] |
| H | Hammond postulate | Monotonicity and location of the crossing coordinate relative to the reaction endpoints. | [26] |
| BEP | Bell–Evans–Polanyi relation | Bounds and affine relations connecting activation barriers and reaction energy. | [22,23] |
| K | Kasha's rule | Emission from the lowest excited state of a given multiplicity under an explicit rate ladder. | [4] |
| Sab | Sabatier principle | A volcano-shaped rate–descriptor relation arising from competing adsorption regimes. | [21,27] |
| G | Goldschmidt criteria | Geometric tolerance and substitution predicates over ionic radii and packing ratios. | [28] |
| SF | Symmetry factor | The thermoneutral crossing coordinate and its dependence on reactant and product curvatures. | [1–3] |
| KV | Kasha–Vavilov rule | Excitation-wavelength independence expressed over the same rate data used by the Kasha model. | [5,6] |
| SV | Stern–Volmer relation | Static, dynamic, and mixed quenching in intensity and lifetime observation channels. | [7] |
| QY | Quantum yield | Competitive-channel yields defined by rates normalized by the total parallel decay rate. | [6] |
| FP | Fluorescence–phosphorescence competition | Competition among fluorescence, internal conversion, and intersystem crossing. | [29,30] |
| EGL | Energy-gap law | Affine logarithmic-rate dependence and nonradiative-rate models as functions of an energy gap. | [31] |
| SS | Stokes shift | Absorption–emission separation on relaxed excited-state surfaces. | [32] |
| IC/ISC | Internal conversion vs intersystem crossing | Same- and different-multiplicity decay channels and their rate competition. | [4,33] |
| FRET | Förster resonance energy transfer | Distance-dependent transfer efficiency and the Förster radius. | [34] |
| E | Einstein coefficients | Relations among absorption, stimulated emission, and spontaneous emission coefficients. | [35] |
| RACI | Conical-intersection nonadiabatic kinetics | Two-state degeneracy, constrained accessibility, nonradiative decay, and emission enhancement. | [18–20] |

## S2.1 Inventory denominators and statement fidelity

The accepted statement-inventory boundary contains 1,154 rows: 1,134 theorem/definition rows are screening-eligible and 20 context-type rows are retained but ineligible. The manifest pins 102 files under `PhotoLean/`, with zero file mismatches in validation. Appendix B lists the 92 active statement-endpoint objects. File, inventory, and active-endpoint counts describe different levels of the audit and are not interchangeable.

Each statement record carries a stable identifier, theory attribution, Lean declaration, source artifact, version, and content hash. Statement fidelity compares human-readable Lean statements and signatures after whitespace normalization up to the signature boundary. It is not proof fidelity, definitional equality, historical identity, or chemical equivalence. Fresh `rfl` certificates separately pin definition bodies. Every accepted relation separately pins source hashes, premise IDs, scope, and snapshot. The four query-level snapshots are cumulative frozen views of fixed theory authorities, not experimental replicates.

### S2.2 Human modelling and machine checking

The selection of source statements, comparison queries, translations, observation channels, and scopes is a modelling decision. Lean checks the resulting formal claim; it does not decide whether the chosen model adequately represents an experiment or whether two source passages ought to have been compared. Provenance records keep these responsibilities separate.

## S3. Shared contexts, translations, and theory-specific model choices

### S3.1 Context- and premise-indexed comparisons

A comparison uses a shared context $\Delta$, endpoint predicates $P, Q : \Delta \to \mathrm{Prop}$, and a premise predicate $\Pi : \Delta \to \mathrm{Prop}$. Translations map the source and target statement states into this context and must be total on the declared domain. The same context need not mean the same physical theory: it means that the variables required by this comparison have been made explicit.

Positivity, nonzero rates, matching conditions, continuity, and modelling identifications remain in $\Pi$ rather than being hidden in definitions. A certificate records premise identifiers as well as their formal predicates, allowing the manifest and Lean projections to be checked for equality. Removing a live premise changes the theorem and requires a new query and snapshot.

### S3.2 Definitional identity and totalized arithmetic

A definitional-identity certificate requires the translated terms to elaborate by a fresh `rfl` check. A proved propositional equality is not sufficient for this kind. Conversely, definitional identity of two formal expressions does not establish that their physical interpretations are identical in every context.

Division and `Real.sqrt` have total definitions at degenerate inputs. The substantive adjudications stated here use positive or nonzero premises and do not rely on a degenerate-curvature or dead-channel convention. Totalized rate-function identities likewise do not establish physical applicability at degenerate inputs.

### S3.3 Observation channels

An observation channel declares its name, admissible domain, and observation maps for the two interpreted objects. Intensity, lifetime, and joint intensity–lifetime observations are different comparison settings rather than interchangeable readings of the same experiment. Observational claims are always relative to the declared channels and domains. In particular, the FluorPhos quenching certificate uses $\frac{\Phi_P}{\Phi_F}$, not its inverse.

### S3.4 Theory-specific formalization choices

The shared audit protocol does not impose a single modelling style on all seventeen theories. The following choices define the statement objects compared directly; they are modelling conventions, not claims that each source tradition has only one possible formal reading.

**Kernel and Marcus.** The kernel supplies reactant and product surfaces, forward and reverse barriers, the transition-state coordinate, and the transfer function for the two-parabola family. Marcus theory uses the quadratic barrier and reaction-energy dependence to state the normal, activationless, and inverted regimes. Boundary statements that are often read directly from a surface diagram are represented explicitly.

**Hammond.** The qualitative statement that a transition state resembles the energetically nearer endpoint is represented by separate monotonicity and endpoint-location propositions for `tsCoord`. Distinct readings retain distinct statement identifiers.

**Bell–Evans–Polanyi.** The affine BEP relation is represented through the tangent reading of the quadratic object. Each tangent lies below its corresponding parabola, giving `bepLine_le_eact`. Literature instances are not assumed to realize the same two-parabola model merely because they use similar affine language.

**Symmetry factor.** The quantity $\beta$ is defined as the thermoneutral crossing coordinate rather than as an adjustable constant. This makes $\beta = \frac{1}{2}$ a proposition about reactant and product curvatures.

**Energy-gap law.** The statement authority uses the weak form in which the logarithmic rate is affine in the energy gap, without selecting a particular Franck–Condon integral representation. The barrier-rate logarithmic link `log_barrierRate_eq` carries $A > 0$.

**Stokes shift.** Band-position separation is modelled through relaxation on the $S_0$ and $S_1$ surfaces, with definitional pins for the shared surface objects. The delivered directional statement `emEnergy_pos_iff_inverted` retains its direction and premises.

**Internal conversion and intersystem crossing.** Same- and different-multiplicity channels are represented by rates sharing a barrier object. `icvsisc_barrier_zero_at_crossing` requires no nonzero-curvature premise, whereas `kernel_surfaces_cross_at_tsCoord` carries $\lambda \neq 0$. Similar surface vocabulary therefore does not imply identical premise requirements.

**Kasha ladder and the Kasha-Marcus identification.** Kasha's rule is represented by a finite excited-state ladder with exponential-race branching. The retained Kasha-Marcus comparison additionally requires the modelling identification $h_{\mathrm{ic}}$ between the relevant internal-conversion rate and the Marcus rate. The active MKVV query uses its own seven-premise identification and one-way implication; neither is silently generalized to a universal Kasha-Vavilov equivalence.

### S3.5 Definitional drift and layered regression checks

Fresh `rfl` certificates pin theory-local copies of shared objects to the kernel at the definition-body level. These checks complement, but do not replace, signature-fidelity probes. A changed definition can retain an identical signature while invalidating definitional identity; the nearest fresh-`rfl` certificate then fails during compilation.

The regression protocol checks two distinct failure locations. Perturbing `Kernel.barrier` causes failure inside the kernel at `reverseBarrier_eq_barrier_neg`. Perturbing `Kernel.reactantSurface` causes failure in the theory module retaining the copy, at `cert_s0Surface`. The alarm layers are therefore the kernel algebra, the theory-level definitional pin, and finally the relation audit.

### S3.6 Certificate-to-hypothesis translation

A certified relation enters an experimental hypothesis card only after seven items are stated: query/certificate ID and exact formal proposition; shared parameters and all load-bearing premises; chemical observables; independently measurable parameters and controls; model-internal prediction and direction; the model identification challenged by an adverse experiment; and explicit non-claims. This is a researcher-guided translation protocol, not automatic compilation of Lean into an experiment.

A positive mathematical yield need not exceed an instrument's detection limit. Exact zero is a model boundary rather than an infinitely precise measurement. A mismatch localizes a failed physical mapping, missing channel, or applicability premise; it does not refute the Lean proof.

## S4. Seven relation kinds, five outcomes, and evidence classes

The vocabulary classifies evidence; it does not confer chemical novelty. A record's declared kind is checked against the certificate's computable kind projection.

**Table S3.** Descriptive and formal criteria for the seven certificate kinds.

| Kind | Scientific meaning | Required evidence |
|---|---|---|
| `definitionalIdentity` | The translated objects have the same definition body. This is not propositional equality or unrestricted physical identity. | A fresh `rfl` witness over the declared translations and context. |
| `implication` | Under the shared premises, whenever $P$ holds, $Q$ holds. No converse is implied. | $\forall \delta, \Pi(\delta) \to (P(\delta) \to Q(\delta))$ |
| `equivalence` | Under the shared premises, $P$ and $Q$ hold under exactly the same conditions. | $\forall \delta, \Pi(\delta) \to (P(\delta) \leftrightarrow Q(\delta))$ |
| `conditionalComposition` | A child relation is derived from typed parents through an aligned intermediate statement. | Parent certificates, middle alignment, premise coverage, bridges, discharges, and the computed child form. |
| `observationalEquivalence` | The declared observation channels agree pointwise over their admissible domains. | Channel-indexed equality under the stated premises; no unrestricted propositional equivalence. |
| `distinguishability` | An allowed observation separates the objects, or a decoder recovers their identity from an observation profile. | A separating-observation witness or an identifiability-decoder witness. |
| `refutation` | A concrete counterexample defeats the stated candidate implication. | $\exists \delta, \Pi(\delta) \land P(\delta) \land \neg Q(\delta)$ |

### S4.1 Proof-relevant composition and separating observations

The composition constructor consumes two typed parent certificates, aligns the intermediate proposition, and checks that every parent premise is inherited, bridged, or discharged. A list of parent names beside an independent child proof is not a composition certificate. Distinguishability instead requires a separating observation or an identifiability decoder; neither form is promoted to observational equivalence or mechanism uniqueness.

### S4.2 Provenance and premise evidence

An audit record identifies its query, statement versions, scope, snapshot, status, evidence reference, and source provenance. The conflict key is $(\mathrm{queryId}, \mathrm{statementVersion}, \mathrm{scope})$, not a theory-pair name. Different relation kinds may coexist for one pair because they answer different queries or use different channels.

**Table S4.** Premise descriptors and the evidence they require.

| Descriptor | Meaning | Evidence standard |
|---|---|---|
| `explicit` | The premise occurs in the certificate signature and manifest projection. | Identifier and premise-projection equality. |
| `nonvacuous` | The premise can be instantiated in the declared context. | A typed inhabitation or model witness. |
| `nonredundant` | Removing the premise changes provability in the declared comparison. | A typed removal/comparison witness. |
| `weakest` | The premise is minimal relative to a declared alternative set and preorder. | Typed minimality evidence over that collection. |

The registry checks declared premise identifiers and their Lean projections. It does not award nonvacuity, nonredundancy, or weakest-premise badges without corresponding typed evidence. Explicit, inherit, derivable, premise-of, trivial, heuristic, and shape-only labels describe premise or prose support only; they never promote a query to certified.

### S4.3 Five audit outcomes and two evidence classes

**Table S5.** The five active audit outcomes.

| Outcome | Interpretation |
|---|---|
| certified | The query references an accepted certificate whose kind, premises, provenance, and snapshot pass the applicable checks. |
| refuted | An accepted scoped counterexample refutes the exact named proposition. The active accepted registry has zero such rows; retained provenance refutation evidence is separate. |
| unresolved | A candidate or required evidence remains open. This is neither a certificate nor a negative theorem. |
| notEstablished | No direct certificate is established in the named snapshot and scope, with a reproducible dependency. This is not logical nonexistence. |
| outOfScope | The query lies outside the declared audit jurisdiction. This is not a claim of chemical irrelevance. |

A `deliveredModule` certificate is in the frozen `PhotoLean/**/*.lean` tree. A `boundedProbe` certificate is package-local accepted evidence, with source hash, compile log, declaration, kind pin, premise set, and axiom footprint; it is not an addition to the frozen delivered tree. Failed search never becomes refutation, and no unresolved, notEstablished, or outOfScope row asserts logical or chemical non-existence.

# S5. Complete bounded screening

For seventeen theories, the unordered pair universe contains $\binom{17}{2} = 136$ pairs. Registry coverage asks whether every pair has a named statement-level query or statement-level outOfScope record. Certified coverage counts pairs with at least one active certified query. Neither measure is logical completeness.

The registry integrates ten accepted packages: one R2, one R3, three R4, and five R5 packages. All ten are required for 136 pairs and 152 queries. The three XD1 declarations remain unresolved candidate evidence and are excluded from the 93-certificate library.

**Table S6.** Certified active queries and the full certificate library use different denominators.

| Kind | Active certified queries | Library objects |
|---|---|---|
| definitionalIdentity | 12 | 28 |
| implication | 4 | 5 |
| equivalence | 13 | 18 |
| conditionalComposition | 2 | 2 |
| observationalEquivalence | 36 | 37 |
| distinguishability | 0 | 2 |
| refutation | 0 | 1 |

**Table S7.** Accepted screening packages.

| Package | Round | Acceptance | Pairs | Queries | Cert. | Probe | Legacy |
|---|---|---|---|---|---|---|---|
| WP-SURFACE-01 | R2 | accepted-R2 | 10 | 19 | 16 | 4 | 12 |
| R3-QUENCH-01 | R3 | accepted-R3 | 11 | 12 | 7 | 7 | 0 |
| R4-E1-EXCITED-01 | R4 | accepted-R4 | 16 | 16 | 6 | 6 | 0 |
| WP-excited-state-kinetics-02 | R4 | accepted-R4 | 14 | 14 | 5 | 5 | 0 |
| WP-conical-intersection-01 | R4 | accepted-R4 | 16 | 16 | 8 | 8 | 0 |
| RAD1 | R5 | accepted-R5 | 16 | 17 | 11 | 11 | 0 |
| RAD2 | R5 | accepted-R5 | 16 | 16 | 4 | 4 | 0 |
| RAD3 | R5 | accepted-R5 | 13 | 16 | 10 | 10 | 0 |
| XD1 | R5 | accepted-R5 | 15 | 15 | 0 | 0 | 0 |
| XD2 | R5 | accepted-R5 | 9 | 11 | 0 | 0 | 0 |

**Table S8.** Complete 136-pair screening index; C/R/U/N/O = certified/refuted/unresolved/notEstablished/outOfScope.

| Pair ID | Theories | Q | C | R | U | N | O |
|---|---|---|---|---|---|---|---|
| P-Marcus-Hammond | Marcus / Hammond | 5 | 5 | 0 | 0 | 0 | 0 |
| P-Marcus-BEP | Marcus / BEP | 3 | 3 | 0 | 0 | 0 | 0 |
| P-Marcus-Kasha | Marcus / Kasha | 1 | 1 | 0 | 0 | 0 | 0 |
| P-Marcus-Sabatier | Marcus / Sabatier | 1 | 1 | 0 | 0 | 0 | 0 |
| P-Marcus-Goldschmidt | Marcus / Goldschmidt | 1 | 0 | 0 | 0 | 0 | 1 |
| P-Marcus-SymmetryFactor | Marcus / SymmetryFactor | 1 | 1 | 0 | 0 | 0 | 0 |
| P-Marcus-KashaVavilov | Marcus / KashaVavilov | 1 | 1 | 0 | 0 | 0 | 0 |
| P-Marcus-SternVolmer | Marcus / SternVolmer | 1 | 0 | 0 | 1 | 0 | 0 |
| P-Marcus-QuantumYield | Marcus / QuantumYield | 1 | 0 | 0 | 1 | 0 | 0 |
| P-Marcus-FluorPhos | Marcus / FluorPhos | 1 | 0 | 0 | 1 | 0 | 0 |
| P-Marcus-EnergyGapLaw | Marcus / EnergyGapLaw | 2 | 2 | 0 | 0 | 0 | 0 |
| P-Marcus-StokesShift | Marcus / StokesShift | 1 | 1 | 0 | 0 | 0 | 0 |
| P-Marcus-ICvsISC | Marcus / ICvsISC | 1 | 1 | 0 | 0 | 0 | 0 |
| P-Marcus-Forster | Marcus / Forster | 1 | 0 | 0 | 0 | 0 | 1 |
| P-Marcus-Einstein | Marcus / Einstein | 1 | 0 | 0 | 0 | 0 | 1 |
| P-Marcus-RACI | Marcus / RACI | 1 | 1 | 0 | 0 | 0 | 0 |
| P-Hammond-BEP | Hammond / BEP | 4 | 4 | 0 | 0 | 0 | 0 |
| P-Hammond-Kasha | Hammond / Kasha | 1 | 1 | 0 | 0 | 0 | 0 |

| Pair ID | Theories | Q | C | R | U | N | O |
|---|---|---|---|---|---|---|---|
| P-Hammond-Sabatier | Hammond / Sabatier | 1 | 0 | 0 | 1 | 0 | 0 |
| P-Hammond-Goldschmidt | Hammond / Goldschmidt | 1 | 0 | 0 | 0 | 0 | 1 |
| P-Hammond-SymmetryFactor | Hammond / SymmetryFactor | 1 | 1 | 0 | 0 | 0 | 0 |
| P-Hammond-KashaVavilov | Hammond / KashaVavilov | 1 | 0 | 0 | 1 | 0 | 0 |
| P-Hammond-SternVolmer | Hammond / SternVolmer | 1 | 0 | 0 | 1 | 0 | 0 |
| P-Hammond-QuantumYield | Hammond / QuantumYield | 1 | 0 | 0 | 1 | 0 | 0 |
| P-Hammond-FluorPhos | Hammond / FluorPhos | 1 | 0 | 0 | 1 | 0 | 0 |
| P-Hammond-EnergyGapLaw | Hammond / EnergyGapLaw | 1 | 1 | 0 | 0 | 0 | 0 |
| P-Hammond-StokesShift | Hammond / StokesShift | 1 | 1 | 0 | 0 | 0 | 0 |
| P-Hammond-ICvsISC | Hammond / ICvsISC | 1 | 1 | 0 | 0 | 0 | 0 |
| P-Hammond-Forster | Hammond / Forster | 1 | 0 | 0 | 0 | 0 | 1 |
| P-Hammond-Einstein | Hammond / Einstein | 1 | 0 | 0 | 1 | 0 | 0 |
| P-Hammond-RACI | Hammond / RACI | 1 | 0 | 0 | 1 | 0 | 0 |
| P-BEP-Kasha | BEP / Kasha | 1 | 1 | 0 | 0 | 0 | 0 |
| P-BEP-Sabatier | BEP / Sabatier | 1 | 1 | 0 | 0 | 0 | 0 |
| P-BEP-Goldschmidt | BEP / Goldschmidt | 1 | 0 | 0 | 0 | 0 | 1 |
| P-BEP-SymmetryFactor | BEP / SymmetryFactor | 1 | 1 | 0 | 0 | 0 | 0 |
| P-BEP-KashaVavilov | BEP / KashaVavilov | 1 | 0 | 0 | 1 | 0 | 0 |
| P-BEP-SternVolmer | BEP / SternVolmer | 1 | 0 | 0 | 1 | 0 | 0 |
| P-BEP-QuantumYield | BEP / QuantumYield | 1 | 0 | 0 | 1 | 0 | 0 |
| P-BEP-FluorPhos | BEP / FluorPhos | 1 | 0 | 0 | 1 | 0 | 0 |
| P-BEP-EnergyGapLaw | BEP / EnergyGapLaw | 1 | 1 | 0 | 0 | 0 | 0 |
| P-BEP-StokesShift | BEP / StokesShift | 1 | 1 | 0 | 0 | 0 | 0 |
| P-BEP-ICvsISC | BEP / ICvsISC | 1 | 1 | 0 | 0 | 0 | 0 |
| P-BEP-Forster | BEP / Forster | 1 | 0 | 0 | 0 | 0 | 1 |
| P-BEP-Einstein | BEP / Einstein | 1 | 0 | 0 | 0 | 0 | 1 |
| P-BEP-RACI | BEP / RACI | 1 | 0 | 0 | 1 | 0 | 0 |
| P-Kasha-Sabatier | Kasha / Sabatier | 1 | 0 | 0 | 1 | 0 | 0 |
| P-Kasha-Goldschmidt | Kasha / Goldschmidt | 1 | 0 | 0 | 0 | 0 | 1 |
| P-Kasha-SymmetryFactor | Kasha / SymmetryFactor | 1 | 0 | 0 | 1 | 0 | 0 |
| P-Kasha-KashaVavilov | Kasha / KashaVavilov | 1 | 1 | 0 | 0 | 0 | 0 |
| P-Kasha-SternVolmer | Kasha / SternVolmer | 1 | 0 | 0 | 1 | 0 | 0 |
| P-Kasha-QuantumYield | Kasha / QuantumYield | 1 | 1 | 0 | 0 | 0 | 0 |
| P-Kasha-FluorPhos | Kasha / FluorPhos | 1 | 0 | 0 | 1 | 0 | 0 |
| P-Kasha-EnergyGapLaw | Kasha / EnergyGapLaw | 1 | 1 | 0 | 0 | 0 | 0 |
| P-Kasha-StokesShift | Kasha / StokesShift | 1 | 0 | 0 | 0 | 0 | 1 |
| P-Kasha-ICvsISC | Kasha / ICvsISC | 1 | 0 | 0 | 1 | 0 | 0 |
| P-Kasha-Forster | Kasha / Forster | 1 | 0 | 0 | 0 | 0 | 1 |
| P-Kasha-Einstein | Kasha / Einstein | 1 | 1 | 0 | 0 | 0 | 0 |
| P-Kasha-RACI | Kasha / RACI | 1 | 1 | 0 | 0 | 0 | 0 |
| P-Sabatier-Goldschmidt | Sabatier / Goldschmidt | 1 | 0 | 0 | 0 | 0 | 1 |
| P-Sabatier-SymmetryFactor | Sabatier / SymmetryFactor | 1 | 0 | 0 | 1 | 0 | 0 |
| P-Sabatier-KashaVavilov | Sabatier / KashaVavilov | 1 | 0 | 0 | 0 | 0 | 1 |
| P-Sabatier-SternVolmer | Sabatier / SternVolmer | 1 | 0 | 0 | 0 | 0 | 1 |
| P-Sabatier-QuantumYield | Sabatier / QuantumYield | 1 | 0 | 0 | 0 | 0 | 1 |
| P-Sabatier-FluorPhos | Sabatier / FluorPhos | 1 | 0 | 0 | 0 | 0 | 1 |
| P-Sabatier-EnergyGapLaw | Sabatier / EnergyGapLaw | 1 | 0 | 0 | 1 | 0 | 0 |
| P-Sabatier-StokesShift | Sabatier / StokesShift | 1 | 0 | 0 | 1 | 0 | 0 |
| P-Sabatier-ICvsISC | Sabatier / ICvsISC | 1 | 0 | 0 | 1 | 0 | 0 |
| P-Sabatier-Forster | Sabatier / Forster | 1 | 0 | 0 | 0 | 0 | 1 |
| P-Sabatier-Einstein | Sabatier / Einstein | 1 | 0 | 0 | 0 | 0 | 1 |
| P-Sabatier-RACI | Sabatier / RACI | 1 | 0 | 0 | 1 | 0 | 0 |
| P-Goldschmidt-SymmetryFactor | Goldschmidt / SymmetryFactor | 1 | 0 | 0 | 0 | 0 | 1 |
| P-Goldschmidt-KashaVavilov | Goldschmidt / KashaVavilov | 1 | 0 | 0 | 0 | 0 | 1 |
| P-Goldschmidt-SternVolmer | Goldschmidt / SternVolmer | 1 | 0 | 0 | 0 | 0 | 1 |
| P-Goldschmidt-QuantumYield | Goldschmidt / QuantumYield | 1 | 0 | 0 | 0 | 0 | 1 |
| P-Goldschmidt-FluorPhos | Goldschmidt / FluorPhos | 1 | 0 | 0 | 0 | 0 | 1 |
| P-Goldschmidt-EnergyGapLaw | Goldschmidt / EnergyGapLaw | 2 | 0 | 0 | 0 | 0 | 2 |
| P-Goldschmidt-StokesShift | Goldschmidt / StokesShift | 1 | 0 | 0 | 0 | 0 | 1 |
| P-Goldschmidt-ICvsISC | Goldschmidt / ICvsISC | 1 | 0 | 0 | 0 | 0 | 1 |
| P-Goldschmidt-Forster | Goldschmidt / Forster | 2 | 0 | 0 | 0 | 0 | 2 |
| P-Goldschmidt-Einstein | Goldschmidt / Einstein | 1 | 0 | 0 | 0 | 0 | 1 |
| P-Goldschmidt-RACI | Goldschmidt / RACI | 1 | 0 | 0 | 0 | 0 | 1 |
| P-SymmetryFactor-KashaVavilov | SymmetryFactor / KashaVavilov | 1 | 0 | 0 | 0 | 0 | 1 |
| P-SymmetryFactor-SternVolmer | SymmetryFactor / SternVolmer | 1 | 0 | 0 | 1 | 0 | 0 |
| P-SymmetryFactor-QuantumYield | SymmetryFactor / QuantumYield | 1 | 0 | 0 | 1 | 0 | 0 |

| Pair ID | Theories | Q | C | R | U | N | O |
|---|---|---|---|---|---|---|---|
| P-SymmetryFactor-FluorPhos | SymmetryFactor / FluorPhos | 1 | 0 | 0 | 1 | 0 | 0 |
| P-SymmetryFactor-EnergyGapLaw | SymmetryFactor / EnergyGapLaw | 1 | 1 | 0 | 0 | 0 | 0 |
| P-SymmetryFactor-StokesShift | SymmetryFactor / StokesShift | 1 | 1 | 0 | 0 | 0 | 0 |
| P-SymmetryFactor-ICvsISC | SymmetryFactor / ICvsISC | 1 | 1 | 0 | 0 | 0 | 0 |
| P-SymmetryFactor-Forster | SymmetryFactor / Forster | 1 | 0 | 0 | 0 | 0 | 1 |
| P-SymmetryFactor-Einstein | SymmetryFactor / Einstein | 1 | 0 | 0 | 0 | 0 | 1 |
| P-SymmetryFactor-RACI | SymmetryFactor / RACI | 1 | 0 | 0 | 1 | 0 | 0 |
| P-KashaVavilov-SternVolmer | KashaVavilov / SternVolmer | 1 | 1 | 0 | 0 | 0 | 0 |
| P-KashaVavilov-QuantumYield | KashaVavilov / QuantumYield | 1 | 1 | 0 | 0 | 0 | 0 |
| P-KashaVavilov-FluorPhos | KashaVavilov / FluorPhos | 1 | 1 | 0 | 0 | 0 | 0 |
| P-KashaVavilov-EnergyGapLaw | KashaVavilov / EnergyGapLaw | 1 | 0 | 0 | 1 | 0 | 0 |
| P-KashaVavilov-StokesShift | KashaVavilov / StokesShift | 1 | 0 | 0 | 0 | 0 | 1 |
| P-KashaVavilov-ICvsISC | KashaVavilov / ICvsISC | 1 | 0 | 0 | 1 | 0 | 0 |
| P-KashaVavilov-Forster | KashaVavilov / Forster | 1 | 0 | 0 | 0 | 0 | 1 |
| P-KashaVavilov-Einstein | KashaVavilov / Einstein | 1 | 0 | 0 | 1 | 0 | 0 |
| P-KashaVavilov-RACI | KashaVavilov / RACI | 1 | 0 | 0 | 0 | 1 | 0 |
| P-SternVolmer-QuantumYield | SternVolmer / QuantumYield | 1 | 1 | 0 | 0 | 0 | 0 |
| P-SternVolmer-FluorPhos | SternVolmer / FluorPhos | 1 | 1 | 0 | 0 | 0 | 0 |
| P-SternVolmer-EnergyGapLaw | SternVolmer / EnergyGapLaw | 1 | 0 | 0 | 1 | 0 | 0 |
| P-SternVolmer-StokesShift | SternVolmer / StokesShift | 1 | 0 | 0 | 0 | 0 | 1 |
| P-SternVolmer-ICvsISC | SternVolmer / ICvsISC | 1 | 0 | 0 | 1 | 0 | 0 |
| P-SternVolmer-Forster | SternVolmer / Forster | 1 | 0 | 0 | 0 | 0 | 1 |
| P-SternVolmer-Einstein | SternVolmer / Einstein | 1 | 0 | 0 | 1 | 0 | 0 |
| P-SternVolmer-RACI | SternVolmer / RACI | 1 | 1 | 0 | 0 | 0 | 0 |
| P-QuantumYield-FluorPhos | QuantumYield / FluorPhos | 2 | 2 | 0 | 0 | 0 | 0 |
| P-QuantumYield-EnergyGapLaw | QuantumYield / EnergyGapLaw | 1 | 1 | 0 | 0 | 0 | 0 |
| P-QuantumYield-StokesShift | QuantumYield / StokesShift | 1 | 0 | 0 | 0 | 0 | 1 |
| P-QuantumYield-ICvsISC | QuantumYield / ICvsISC | 1 | 1 | 0 | 0 | 0 | 0 |
| P-QuantumYield-Forster | QuantumYield / Forster | 1 | 1 | 0 | 0 | 0 | 0 |
| P-QuantumYield-Einstein | QuantumYield / Einstein | 1 | 1 | 0 | 0 | 0 | 0 |
| P-QuantumYield-RACI | QuantumYield / RACI | 1 | 1 | 0 | 0 | 0 | 0 |
| P-FluorPhos-EnergyGapLaw | FluorPhos / EnergyGapLaw | 1 | 0 | 0 | 1 | 0 | 0 |
| P-FluorPhos-StokesShift | FluorPhos / StokesShift | 1 | 0 | 0 | 0 | 0 | 1 |
| P-FluorPhos-ICvsISC | FluorPhos / ICvsISC | 2 | 2 | 0 | 0 | 0 | 0 |
| P-FluorPhos-Forster | FluorPhos / Forster | 1 | 1 | 0 | 0 | 0 | 0 |
| P-FluorPhos-Einstein | FluorPhos / Einstein | 1 | 1 | 0 | 0 | 0 | 0 |
| P-FluorPhos-RACI | FluorPhos / RACI | 1 | 1 | 0 | 0 | 0 | 0 |
| P-EnergyGapLaw-StokesShift | EnergyGapLaw / StokesShift | 2 | 2 | 0 | 0 | 0 | 0 |
| P-EnergyGapLaw-ICvsISC | EnergyGapLaw / ICvsISC | 2 | 2 | 0 | 0 | 0 | 0 |
| P-EnergyGapLaw-Forster | EnergyGapLaw / Forster | 1 | 0 | 0 | 0 | 0 | 1 |
| P-EnergyGapLaw-Einstein | EnergyGapLaw / Einstein | 1 | 0 | 0 | 0 | 0 | 1 |
| P-EnergyGapLaw-RACI | EnergyGapLaw / RACI | 1 | 1 | 0 | 0 | 0 | 0 |
| P-StokesShift-ICvsISC | StokesShift / ICvsISC | 1 | 1 | 0 | 0 | 0 | 0 |
| P-StokesShift-Forster | StokesShift / Forster | 1 | 0 | 0 | 0 | 1 | 0 |
| P-StokesShift-Einstein | StokesShift / Einstein | 1 | 0 | 0 | 0 | 0 | 1 |
| P-StokesShift-RACI | StokesShift / RACI | 1 | 0 | 0 | 1 | 0 | 0 |
| P-ICvsISC-Forster | ICvsISC / Forster | 1 | 0 | 0 | 1 | 0 | 0 |
| P-ICvsISC-Einstein | ICvsISC / Einstein | 1 | 0 | 0 | 0 | 0 | 1 |
| P-ICvsISC-RACI | ICvsISC / RACI | 1 | 1 | 0 | 0 | 0 | 0 |
| P-Forster-Einstein | Forster / Einstein | 1 | 1 | 0 | 0 | 0 | 0 |
| P-Forster-RACI | Forster / RACI | 1 | 0 | 0 | 1 | 0 | 0 |
| P-Einstein-RACI | Einstein / RACI | 1 | 1 | 0 | 0 | 0 | 0 |

# S6. Adjudication A1: the symmetry-factor boundary

## S6.1 Formal statement

At zero driving force, the crossing coordinate for two parabolic wells has the closed form

$$\beta = \frac{\sqrt{k_p}}{\sqrt{k_r} + \sqrt{k_p}}.$$

For positive force constants, `PhotoLean.RelationAudit.betaHalfEquivalence` exposes the certified equivalence

$$\beta = \frac{1}{2} \leftrightarrow k_r = k_p.$$

The retained certificate is `cert-a1`, record `rec-a1`; its premise identifier `symmetryFactor.positiveForceConstants` represents $0 < k_r \wedge 0 < k_p$. This is theory-level intra-theory evidence and adds no active pair/query coverage.

The result concerns the thermoneutral crossing coordinate. It does not make $\beta = \frac{1}{2}$ a definition, a universal transfer coefficient, or structural identity.

### S6.2 Off-diagonal witnesses

For $(k_r, k_p) = (1, 4)$, the closed form gives $\beta = \frac{2}{3}$, whereas $(4, 1)$ gives $\beta = \frac{1}{3}$. These witnesses show failure away from the equal-curvature diagonal and direction asymmetry under curvature exchange. They are theory-level witnesses, not additional active pair relations.

## S7. Adjudication A2: legacy refutation and two directional active bridges

### S7.1 Pointwise independence and the loss boundary

The two rules are represented by distinct predicates over shared rate data. The theory-level `kv_d2_verdict` supplies witnesses in both directions within positive excitation and loss: one in which Kasha holds while the Kasha–Vavilov condition fails at level 1, and one in which the Kasha–Vavilov condition holds while Kasha fails. Under closure quantification with lossy channels, the closed formulas coincide for rate data with positive internal conversion; at the lossless corner, they separate. The companion `kv_antiKasha_boundary` relates a violation of Kasha's rule to observable anti-Kasha emission through the maximal-emitter argument.

### S7.2 Retained scoped refutation

The retained provenance record `rec-kasha-vavilov` and certificate `cert-kv-refutation/PhotoLean.RelationAudit.kvRefutation` refute the stronger inference KashaRule $\Rightarrow$ `VavilovAt` at level 1. Its premises are `Kasha.RateData rad ic 2` and `0 < ic 0`; the lossy scope matters. This scoped refutation does not contradict active `refuted = 0`, because the active queries below address different statement-level propositions.

### S7.3 KKV and MKVV address distinct targets

Active query `Q-RAD1-KKV-01` certifies only the one-way implication `KashaRule rad ic N` $\Rightarrow$ `for all M <= N, SpecSame rad ic M N`, under `Kasha.RateData rad ic N` and positive `Kasha.fluoYield rad ic M` for every `M <= N`. Its consequent is `SpecSame`, not `VavilovAt`, and it does not certify yield independence. Full certificate details are in S10.1.

Active query `Q-RAD1-MKVV-01` certifies only that the strict outside-window inequality, together with its seven premises, implies `not KashaRule rad ic 1` via positive `upperYield`. It does not give an inside-window zero-emission prediction, a converse, a complete inside/outside criterion, or an experimental detectability guarantee. The exact antecedent and premises are in S10.2.

## S8. Adjudication A3: quenching, VSV, and the P/F direction

### S8.1 Intensity-only observational equivalence

The intensity comparison uses context $(k_0, k_q, K_a)$ and premises

$$k_0 \neq 0, \quad K_{\mathrm{SV}}(k_0, k_q) = K_a.$$

Under these premises, `cert-sv-intensity/PhotoLean.RelationAudit.svIntensityObservationalEquivalence` proves pointwise agreement of the dynamic and static intensity observations over the declared concentration domain. This is channel-indexed observational equivalence, not identity of mechanisms.

### S8.2 Joint intensity–lifetime separation and decoder

The joint channel observes an intensity ratio and a lifetime ratio together. The separating witness uses $k_0 = 2$, $k_q = 2$, and $K_a = 1$ at concentration $q = 1$. The intensity components agree, while the lifetime components differ. `cert-sv-joint/svJointDistinguishability` uses the same matched premises as S8.1.

`cert-sv-decoder/svJointIdentifiability` instead requires $0 < k_0$, $k_q \neq 0$, and $0 < q$. `svDecoder_correct` proves recovery of the mechanism tag from the joint observation profile. No lifetime separation is claimed at $k_q = 0$. These are retained theory-level provenance witnesses, not active pair coverage.

### S8.3 Level-selective VSV and P/F-invariant SF certificates

`Q-RAD1-VSV-01` adds a collisional loss only at level 0. It leaves `upperYield` and the funnel cascade unchanged and equates the unquenched bottom-emission channel to the quenched bottom channel multiplied by `svRatioDyn` evaluated at the unquenched level-0 decay. Its two premises are `0 < rad 0 + ic 0` and `0 <= kq*q`. S10.5 preserves the equation's orientation.

`Q-R3-SF-01` proves invariance of the actual channel $\frac{\Phi_P}{\Phi_F}$ under an added S1 internal-conversion channel. The live premises are `0 < kF`, positive unquenched S1 total decay, positive quenched S1 total decay, and `0 < kP + kNR`. Do not replace this by $\frac{\Phi_F}{\Phi_P}$ without a new derivation. Invariance does not identify quenching mechanism or establish collisional versus other loss models. Full details are in S10.6.

# S9. Explanation A4: layered RACI mechanism and rate slots

## S9.1 Phenomenon, kinetic criterion, and mechanism

Aggregation-induced emission (AIE) is treated as the observable inequality $\Phi_{\mathrm{agg}} > \Phi_{\mathrm{sol}}$. Restricted access to a conical intersection (RACI) is a mechanism that can produce this inequality in a declared model class. The two are not definitionally identical.

In the two-channel Jablonski model,

$$\Phi = \frac{k_r}{k_r + k_{\mathrm{nr}}}.$$

For positive radiative rates and nonnegative nonradiative rates, `quantumYield_gt_iff_ratio_lt` proves

$$\Phi_{\mathrm{agg}} > \Phi_{\mathrm{sol}} \leftrightarrow \frac{k_{\mathrm{nr,agg}}}{k_{r,\mathrm{agg}}} < \frac{k_{\mathrm{nr,sol}}}{k_{r,\mathrm{sol}}}.$$

Preservation of $k_r$ is a convenient sufficient setting, not a necessary condition. `raci_emission_enhancement` treats the simplified fixed-$k_r$ case; `raci_emission_enhancement_general` uses only the decrease of the competition ratio.

## S9.2 Geometric and dynamical chain

For a real symmetric two-state Hamiltonian with entries $(a, b, d)$, the conical-intersection condition is

$$(a - d)^2 + 4b^2 = 0,$$

equivalently $a = d$ and $b = 0$. In the single-torsion-coordinate model, the free phase can access the conical point at $\theta = 0$. The aggregate allows only $|\theta| \geq \delta_{\mathrm{agg}}$ with $\delta_{\mathrm{agg}} > 0$, and the minimal accessibility gap increases from 0 to $2\delta_{\mathrm{agg}}$.

For the barrier-rate model $k_{\mathrm{nr}} = A\exp(-\gamma B)$ with $A > 0$ and $\gamma > 0$, the increased gap strictly decreases the nonradiative rate. `torsion_raci_emission_enhancement` combines this chain with an unchanged positive radiative rate. The named `nonModelNoCI` has an everywhere-empty conical set, so the RACI premise chain cannot be instantiated there. This non-model delimits RACI; it is not a counterexample to AIE. A classical double-parabola crossing is not identified with a codimension-two conical intersection.

## S9.3 Active RACI rate-slot identities

Three active observational-equivalence certificates make the rate slots explicit. `Q-CI-MR-01` identifies `RACI.barrierRate A (1/(kB*T)) (Marcus.barrier lam x)` with `Marcus.rate A lam kB T x`; `Q-CI-EG-01` identifies the same slot at `EnergyGapLaw.nrBarrier lam x` with `EnergyGapLaw.nrRate A lam kB T x`; `Q-CI-IC-01` identifies it at `ICvsISC.fcBarrier lam x` with `ICvsISC.icRate A lam kB T x`.

Their declarations are respectively `PhotoLean.RelationAudit.BoundedAudit.CI.marcusRateBridge`, `eglRateBridge`, and `icRateBridge`. All use `scope-CI-rate-law`, `snap-bounded-audit-2026-11-04`, probe SHA-256 `33cf7fe66d3c3cee3431f9ace125259bca3c36ef798d9f5324154b817ce80572`, and empty premise lists (`True`). They are totalized rate-function identities, not evidence of physical coordinate identity, CI accessibility, ISC inclusion, or a unique RACI mechanism.

The stronger BEP–RACI barrier identification `Q-CI-BP-01` remains unresolved: shared Arrhenius shape alone does not identify reactive activation energy with the CI accessibility gap. Algebraic substitutability is therefore separated from mechanistic identification.

# S10. New model-scoped insight theorem cards

Each card states a formal result, not an experimental result. Query IDs identify statement-level questions; certificate IDs and declarations identify evidence. No card claims automatic discovery. Artifact hashes identify the accepted evidence package, and S13 gives its validation and reproduction record.

## S10.1 KKV: exact Kasha rule to spectral sameness

**Query/certificate/declaration/kind.** `Q-RAD1-KKV-01`; `RC-RAD1-KKV-01`; `PhotoLean.RelationAudit.BoundedAudit.RAD1.kashaVavBridge`; `implication`.

**Formal statement.** Under `Kasha.RateData rad ic N` and `for all M <= N, 0 < Kasha.fluoYield rad ic M`, `Kasha.KashaRule rad ic N` implies `for all M <= N, KashaVavilov.SpecSame rad ic M N`. There is no numerical observation channel: this is a proposition-level one-way implication.

**Scope/snapshot/hash.** `scope-RAD1-kasha-vav-spec`; `snap-bounded-audit-2026-11-04-r5rad1`; RAD1 probe SHA-256 `38cc6056139db6890175441cf029bf9e5c01ad6e1b63c88e2d2f9ad5b388cfeb`.

**Caveat/falsification boundary.** The conclusion is neither `VavilovAt` nor yield independence. A falsifying context must satisfy both premises and the exact Kasha rule while violating `SpecSame` for some `M <= N`; failure of the stronger legacy inference is not a counterexample to this query.

## S10.2 MKVV: Marcus-window exit to anti-Kasha observability

**Query/certificate/declaration/kind.** `Q-RAD1-MKVV-01`; `RC-RAD1-MKVV-01`; `PhotoLean.RelationAudit.BoundedAudit.RAD1.mkvBridge`; `implication`.

**Formal statement and channel.** There is no numeric channel. With the seven premises below, the antecedent and consequent are exactly:

```
4*lam*(kB*T)*log(Kasha.kashaGapThreshold A (rad 0)
  (Kasha.decay rad ic 0) (rad 1) tol) < (lam-x)^2
  -> ¬ Kasha.KashaRule rad ic 1
```

The proof obtains `0 < upperYield rad ic 1` and then uses `antiKasha_observable_iff`.

**All load-bearing premises.** `mkv.rateData`: `Kasha.RateData rad ic 1`; `mkv.tolerance`: `0 < tol` and `tol < 1`; `mkv.prefactorPositive`: `0 < A`; `mkv.lamPositive`: `0 < lam`; `mkv.thermalPositive`: `0 < kB*T`; `mkv.ratesPositive`: `0 < rad 0` and `0 < rad 1`; `mkv.icIsMarcusRate`: `ic 1 = Kasha.marcusIC A lam kB T x`.

**Scope/snapshot/hash.** `scope-RAD1-marcus-kv-window`; `snap-bounded-audit-2026-11-04-r5rad1`; probe SHA-256 `38cc6056139db6890175441cf029bf9e5c01ad6e1b63c88e2d2f9ad5b388cfeb`.

**Caveat/falsification boundary.** This is a one-way model-scoped hypothesis. No inside-window zero-yield result or converse is proved. A counterexample must satisfy all seven premises and the strict inequality while the Kasha rule still holds; an intensity below a detection limit is not a Lean counterexample.

## S10.3 ES01/ES02: the same gap-window coordinate

**Evidence.** `Q-RAD3-ES-01`/certificate `Q-RAD3-ES-01/PhotoLean.RelationAudit.BoundedAudit.Rad3.esZeroBarrierBridge/equivalence`; `Q-RAD3-ES-02`/certificate `Q-RAD3-ES-02/PhotoLean.RelationAudit.BoundedAudit.Rad3.esBarrierEmissionBridge/observationalEquivalence`.

**Formal statements.** ES01 proves

```
EnergyGapLaw.nrBarrier lam x = 0 <-> StokesShift.emEnergy lam x = 0
```

ES02 declares the `barrier-height` channel on domain `Unit` and equates

```
EnergyGapLaw.nrBarrier lam x
  = StokesShift.emEnergy lam x ^ 2 / (4 * lam)
```

**Actual premise.** Both consume exactly `RAD3-ES-lam-nonzero`: `lam != 0`, not an added `lam > 0`. At `lam = 0`, totalized `nrBarrier` is zero for every `x` while `emEnergy lam x = x-lam`; hence `x != 0` falsifies the unpremised ES01 equivalence.

**Scope/snapshot/hash.** `scope-RAD3-gap-window`; `snap-bounded-audit-2026-11-04`; probe SHA-256 `c693f19830aeb8ce7c01f43f1c2821bc57920f98d0d03f3cdca5dbd1a3f95f09`.

**Scope and caveat.** ES01 certifies the zero-set boundary and ES02 supplies the quantitative barrier-coordinate relation. The context is one equal-curvature two-parabola model with shared `lam` and `x`. Neither certificate equates barriers or emission windows across different surface parameters or extends ES01 to `lam = 0`.

## S10.4 FI02/EI02: live-channel phosphorescence and ISC crossover

**FI02 evidence.** Query/certificate `Q-RAD3-FI-02`; declaration `PhotoLean.RelationAudit.BoundedAudit.Rad3.fiElSayedBridge`; kind `equivalence`.

**FI02 statement and premise direction.** With the theory-owned `FluorPhos.kISC` identified as `ICvsISC.iscRate`, the certificate yields zero `FluorPhos.phiP` iff `HSO = 0`. Its four actual premises are nonzero, not merely positive: `RAD3-FIES-kP-nonzero` (`kP != 0`); `RAD3-FIES-triplet-total-nonzero` (`kP+kNR != 0`); `RAD3-FIES-AS-nonzero` (`AS != 0`); and `RAD3-FIES-s1decay-nonzero` (the S1 total decay containing the ISC rate is nonzero). No numerical observation channel is declared; this is a proposition equivalence.

**FI02 caveat.** Mathematical zero is not an experimental detection threshold. Removing a live-channel premise changes the theorem; a measured undetectable phosphorescence signal alone does not refute Lean.

**EI02 evidence.** Query/certificate `Q-RAD3-EI-02`; declaration `PhotoLean.RelationAudit.BoundedAudit.Rad3.eiCrossoverComposition`; kind `conditionalComposition`; computed child form `equivalence`; parents `RAD3-EI-rate-identification` and `RAD3-EI-isc-dominance`; bridge list empty.

**EI02 exact statement.** Under `ICvsISC.FCData AI AS lamI lamS kB T` and `0 < HSO`, the child proves

```
EnergyGapLaw.nrRate AI lamI kB T xI
    < ICvsISC.iscRate HSO AS lamS kB T xS
  <->
ICvsISC.fcBarrier lamS xS - ICvsISC.fcBarrier lamI xI
    < (kB*T)*(2*log HSO + log(AS/AI))
```

The child premise list is exactly `RAD3-EI-fcdata`, `RAD3-EI-hso-positive`; the first parent has no premises and the second supplies both.

**Shared FI02/EI02 provenance.** `snap-bounded-audit-2026-11-04`; probe SHA-256 `c693f19830aeb8ce7c01f43f1c2821bc57920f98d0d03f3cdca5dbd1a3f95f09`; FI02 scope `scope-RAD3-el-sayed`.

**EI02 formal context.** The shared two-channel context contains the IC and ISC prefactors, reorganization energies, thermal energy, gap coordinates, and spin–orbit amplitude. The child certificate consumes exactly `FCData` and `HSO > 0` within this context; Appendix C records the query-level snapshot.

### S10.5 VSV: channel-selective dilution

**Query/certificate/declaration/kind.** `Q-RAD1-VSV-01`; `RC-RAD1-VSV-01`; `PhotoLean.RelationAudit.BoundedAudit.RAD1.kvSvBridge`; `observationalEquivalence`.

**Formal statement and direction.** Define `quenchIC ic kq q n = ic n + kq*q` if `n=0`, otherwise `ic n`. With `0 < rad 0 + ic 0` and `0 <= kq*q`, the `bottomEmissionChannelUnderQuench` domain is `Unit`, and the channel equation is

```
fluoYield rad ic N - upperYield rad ic N
  = (fluoYield rad (quenchIC ic kq q) N - upperYield rad ic N)
    * SternVolmer.svRatioDyn (rad 0 + ic 0) kq q
```

The ratio is evaluated at the unquenched decay. Auxiliary certified steps leave `upperYield` and the funnel cascade unchanged.

**Scope/snapshot/hash/caveat.** `scope-RAD1-kv-sv-quench`; `snap-bounded-audit-2026-11-04-r5rad1`; probe SHA-256 `38cc6056139db6890175441cf029bf9e5c01ad6e1b63c88e2d2f9ad5b388cfeb`. This equation does not reverse the dilution direction and does not prove that an observed quencher acts only at level 0.

### S10.6 SF: phosphorescence/fluorescence-ratio invariance

**Query/certificate/declaration/kind.** `Q-R3-SF-01`; `Q-R3-SF-01`; `PhotoLean.RelationAudit.BoundedAudit.R3.fpRatioInvariantBridge`; `observationalEquivalence`.

**Formal channel.** The declared channel is exactly

```
FluorPhos.phiP kF kISC kIC kP kNR / FluorPhos.phiF kF kISC kIC
  = FluorPhos.phiP kF kISC (kIC+kq*q) kP kNR /
    FluorPhos.phiF kF kISC (kIC+kq*q)
```

on domain `Unit`. Thus the direction is $\frac{\Phi_P}{\Phi_F}$, not $\frac{\Phi_F}{\Phi_P}$.

**All premises.** `0 < kF`; `0 < s1Decay kF kISC kIC`; `0 < s1Decay kF kISC (kIC+kq*q)`; `0 < kP+kNR`.

**Scope/snapshot/hash.** `scope-R3-sv-fp-invariance`; `snap-bounded-audit-2026-11-04`; probe SHA-256 `16d5b8424d17e1e608c9780a0e6fbebb5838a80e59fb57b6d52fe3e70e7fb368`.

**Channel and falsification boundary.** The certified channel is phosphorescence/fluorescence, P/F. A counterexample must satisfy all four positivity premises and make the two P/F ratios differ. The theorem does not establish mechanism uniqueness, infinite experimental precision, or invariance of either yield separately.

### S10.7 Adjacent illustration: the linear-volcano descriptor axis

`Q-WPS01-BS-01`, certificate `QP-bepSabatierLinearVolcanoBridge`, declaration `PhotoLean.RelationAudit.BoundedAudit.Surface.bepSabatierLinearVolcanoBridge`, is a premise-free observational equivalence on the descriptor-axis channel:

```
Sabatier.volcanoBarrier (1/2) (lam1/4) (1/2) (lam2/4) dE
  = max (BEP.bepLine lam1 (-dE)) (BEP.bepLine lam2 dE)
```

Its scope is `scope-WPS01-descriptor-axis`, snapshot `snap-bounded-audit-2026-11-04`, probe SHA-256 `db1975be8ebe38d20b05252a9443faf0dcebd7803d97c90cca5d2ca942ba755c`. This is a model construction, not a claim that a catalytic descriptor is an optical gap.

## S11. Experimental hypothesis protocols E1–E5

All five protocols are researcher-designed, model-scoped hypotheses and measurement plans, not experimental results. A mismatch localizes a failed model identification or applicability premise; it does not refute a Lean proof.

### E1. Anti-Kasha emission from a Marcus-controlled upper channel

**Basis.** S10.2. **Measure/independently constrain.** Resolve bottom and upper emission channels and absolute yields; measure ultrafast upper-state decay and lowest-state growth; independently estimate `lam`, `x`, `A`, `kB*T`, the bottom and funnel radiative rates, and the tolerance; test `ic 1 = Kasha.marcusIC A lam kB T x`. **Prediction.** Only under all seven premises and the strict outside-window inequality does the model imply positive upper yield and KashaRule failure. **Controls.** Include an inside-window negative control, direct S1 excitation, matched absorbance, spectral channel assignment, excitation-power and photochemistry checks, and an explicit detection limit. **Failure/nonclaim.** No detected upper emission can reflect the one-way direction, failed parameter identification, additional loss channels, or detection limit; inside-window zero emission is not a theorem prediction.

### E2. Coupled gap-law barrier and Stokes emission window

**Basis.** S10.3. **Measure/independently constrain.** Determine the emission gap and nonradiative barrier in the same sample/temperature series; independently constrain the shared equal-curvature surfaces and `lam`; check `lam != 0`. Use temperature-dependent kinetics only after an explicit barrier-rate identification. **Prediction.** Within this model, the zero boundaries coincide and `nrBarrier = emEnergy^2/(4*lam)`. **Controls/failure.** Use independent optical and kinetic measurements rather than circularly fitting the same data. An Arrhenius apparent activation energy is not automatically `nrBarrier`; different surface parameters, temperature dependence, or a failed shared-gap identification defeat applicability, not the certificate.

### E3. Simultaneous ISC/fluorescence and phosphorescence competition

**Basis.** S10.4. **Measure/independently constrain.** Measure absolute fluorescence and phosphorescence yields, triplet yield or transient absorption, S1/T1 lifetimes, and competing rates with live S1 and triplet channels; independently constrain `HSO` or a justified proxy, `AS/AI`, both barriers, and `kB*T`. **Prediction.** In the FI model, mathematical phosphorescence zero corresponds to `HSO=0`; in the EI model, the stated ISC-rate inequality is equivalent to the exact barrier/log-prefactor threshold under `FCData` and `HSO>0`. **Controls/failure.** Control oxygen and direct triplet quenching, record detection limits, and track substitution-induced changes in other rates. Simultaneous variation of both rates is admissible; a single undetectable yield is not a proof refutation. Every EI test must instantiate `FCData`, `HSO>0`, and the shared two-channel parameter identification.

### E4. Channel-selective quenching

**Basis.** S10.5–S10.6. **Measure/independently constrain.** Separate upper and bottom emission, measure intensity, lifetime, $\Phi_F$, $\Phi_P$, and $\frac{\Phi_P}{\Phi_F}$, and independently establish concentration and `kq*q`; test the level-0-only channel identification and all positivity premises. **Prediction.** VSV gives the oriented bottom-channel equation with invariant upper leak; SF gives invariant $\frac{\Phi_P}{\Phi_F}$, not its inverse and not separate yield invariance. **Controls/failure.** Include static-association, inner-filter, oxygen, aggregation, spectral-reabsorption, direct-triplet-quenching, and nonselective-channel controls. Violations locate an incorrect channel map, mechanism, or premise; invariance itself never proves mechanism uniqueness.

### E5. RACI slot discrimination and layer controls

**Basis.** S9. **Measure/independently constrain.** Measure aggregate/monomer quantum yields and lifetimes; only under an applicable single-exponential two-channel model extract $k_r = \frac{\Phi}{\tau}$ and $k_{\mathrm{nr}} = \frac{1-\Phi}{\tau}$; independently test which, if any, of Marcus, EnergyGapLaw, or ICvsISC barriers is the CI accessibility gap using structural, ultrafast, or other dynamical evidence. **Prediction.** AIE corresponds to decreased $R = \frac{k_{\mathrm{nr}}}{k_r}$ in the two-channel model; only after a barrier identification may a rate-slot identity be used, and the RACI geometric/rate premises remain separate. **Controls/failure.** Include aggregate-morphology, scattering, reabsorption, multiexponential-decay, ISC/FRET, radiative-rate, viscosity/restriction, and alternative-mechanism controls. A rate-law coincidence is not mechanistic uniqueness; `Q-CI-BP-01` remains unresolved and cannot supply the missing bridge.

# S12. Provenance reconciliation

The 151-record provenance registry has SHA-256 `25d38791ce86f9030598eb68b6894a886513164ddb9165e0fff371883c1ecbf0`. Its records comprise 24 certified, 6 unresolved, and 121 notEstablished outcomes, indexed as 26 named queries and 125 pair anchors. Five intra-theory records and all dispositions contribute zero active pair/query coverage. The certificate library retains 38 delivered-Module objects: 12 support active queries and 26 provide supporting evidence. Active certified evidence therefore remains 55+12, distinct from the 93-object library and from provenance dispositions.

The derived provenance view resolves four declaration aliases: `PhotoLean.BEP.rateViaEact` to `PhotoLean.BEP.rate_eq_exp_neg_eact`; `PhotoLean.Marcus.rateDescriptor` to `PhotoLean.Marcus.descriptor_sharp`; `PhotoLean.EnergyGapLaw.nrBarrier_eq_zero` to `PhotoLean.EnergyGapLaw.nrBarrier`; and `PhotoLean.StokesShift.emEnergy_eq_zero` to `PhotoLean.StokesShift.emEnergy`. It also records three certified-view links for `rec-D-S05-bep-rate-eq-exp-neg-eact`, `rec-D-S06-hammond-sharp-iff-marcus-sharp`, and `rec-D-S15-egBarrier-zero-iff-emEnergy-zero`.

Forty-four scope relationships are classified as 41 `superseded-in-named-scope` and 3 `coexists-different-scope`. The A2 refutation and A1/A3/A4 theory-level evidence remain separate from active `refuted=0`. Appendix D lists all 151 dispositions. These classifications preserve scope and evidence provenance without adding active coverage or treating scoped absence as logical nonexistence.

# S13. Verification and reproduction

## S13.1 Validation and artifact integrity

Validation of the accepted registry is PASS: 31,347 checks, 0 failures; all 17 adversarial mutations failed closed and all 5 positive controls passed. Nine probe files review 58 declarations = 55 certificates + 3 unresolved XD1 candidates. The validator runs the strict build/scan, the nine probe compilations, deterministic isolated rebuild, and provenance invariants. The delivered tree has 102 files and 0 mismatches; rebuild is byte-identical. Allowed axiom footprints are `propext`, `Classical.choice`, and `Quot.sound`; `sorryAx` and custom axioms are forbidden.

Table S9 identifies the exact artifacts used for the counts, evidence, and validation reported here. The provenance registry remains a separate input and is not merged into active totals.

**Table S9.** SHA-256 identifiers for the accepted artifacts analyzed in this work.

| Artifact | SHA-256 |
|---|---|
| Accepted registry | `894fa88a56de663e5447c9ec9667dc3461531535427ba5c492df9a8ba972c1ba` |
| Accepted schema | `95f60be6f1f0aa96d454bfb77fd5628c7e3da620ca018a13142db632ef197e1d` |
| Provenance disposition ledger | `98f0a4e1b384089acb9cf8ac6d51f453f017af820e2d5998ee2bbb933d379c1d` |
| Validation report | `e48b2110f4ca9125a00595ac11ae0300afea1bedc588ae4003ea3936db329b2a` |

| Artifact | SHA-256 |
|---|---|
| Provenance registry | 25d38791ce86f9030598eb68b6894a886513164ddb9165e0fff371883c1ecbf0 |

## S13.2 Reproduction commands

Reproduce the accepted registry from the audit repository root (`bounded-relation-audit-2026-11-04/photolean`):

```
proofs/scripts/check.sh --strict
python3 tools/relation_audit_bounded/build.py --acceptance accepted-R8
python3 tools/relation_audit_bounded/validate.py --write-report
python3 tools/relation_audit_bounded/tests/test_adversarial.py
```

The unskipped validator includes the external gates, strict build/scan, probe recompilation, deterministic rebuild, provenance invariants, and adversarial tests. For separate manual inspection, the nine registry-recorded probes are:

```
for p in \
  bounded_audit/r3/probes/R3-probe.lean \
  bounded_audit/r4/conical-01/probes/R4-CI-probe.lean \
  bounded_audit/r4/excited-01/probes/E1-probe.lean \
  bounded_audit/r4/excited-02/probes/E2-probe.lean \
  bounded_audit/r5/cross-domain-01/probes/XD1-probe.lean \
  bounded_audit/r5/radiative-01/probes/RAD1-probe.lean \
  bounded_audit/r5/radiative-02/probes/RAD2-probe.lean \
  bounded_audit/r5/radiative-03/probes/RAD3-probe.lean \
  bounded_audit/wp-surface-01/probes/Surface-probe.lean
do
  proofs/scripts/lake env lean "$p"
done
```

No run that skips external gates, probe compilation, deterministic rebuild, or mutation tests is described as full validation. The dirty/clean state of any source tree is reported only from its own snapshot metadata; the legacy registry is never rewritten during reproduction.

# Appendix A. Complete certificate index

This flattened library lists all 93 certificates: 38 deliveredModule and 55 boundedProbe. `Premises` counts the normalized premise array; an empty array is a theorem fact, not missing metadata. Library role distinguishes 55 active probes, 12 reused active delivered certificates, and 26 retained support objects. Presence in this table does not itself confer active coverage.

**Table S10.** Complete certificate index.

| Certificate ID | Kind | Lean declaration | Premises | Evidence | Library role |
|---|---|---|---|---|---|
| Q-CI-EB-01 | observationalEquivalence | PhotoLean.RelationAudit.BoundedAudit.CI.einsteinYieldBridge | 0 | boundedProbe | active-certificate |
| Q-CI-EG-01 | observationalEquivalence | PhotoLean.RelationAudit.BoundedAudit.CI.eglRateBridge | 0 | boundedProbe | active-certificate |
| Q-CI-FP-01 | observationalEquivalence | PhotoLean.RelationAudit.BoundedAudit.CI.fluorPhosS1Bridge | 0 | boundedProbe | active-certificate |
| Q-CI-IC-01 | observationalEquivalence | PhotoLean.RelationAudit.BoundedAudit.CI.icRateBridge | 0 | boundedProbe | active-certificate |
| Q-CI-KS-01 | definitionalIdentity | PhotoLean.RelationAudit.BoundedAudit.CI.kashaRadiativeBranchPin | 0 | boundedProbe | active-certificate |
| Q-CI-MR-01 | observationalEquivalence | PhotoLean.RelationAudit.BoundedAudit.CI.marcusRateBridge | 0 | boundedProbe | active-certificate |
| Q-CI-QY-01 | observationalEquivalence | PhotoLean.RelationAudit.BoundedAudit.CI.qyTwoChannelBridge | 0 | boundedProbe | active-certificate |
| Q-CI-SV-01 | observationalEquivalence | PhotoLean.RelationAudit.BoundedAudit.CI.svQuenchDilutionBridge | 2 | boundedProbe | active-certificate |
| EB2-BI-01 | definitionalIdentity | PhotoLean.RelationAudit.BoundedAudit.E2.bepICBridge | 0 | boundedProbe | active-certificate |
| EB2-BS-01 | observationalEquivalence | PhotoLean.RelationAudit.BoundedAudit.E2.bepStokesBridge | 0 | boundedProbe | active-certificate |
| EB2-SE-01 | observationalEquivalence | PhotoLean.RelationAudit.BoundedAudit.E2.scEgBridge | 1 | boundedProbe | active-certificate |
| EB2-SI-01 | observationalEquivalence | PhotoLean.RelationAudit.BoundedAudit.E2.scICBridge | 1 | boundedProbe | active-certificate |
| EB2-SS-01 | definitionalIdentity | PhotoLean.RelationAudit.BoundedAudit.E2.scStokesBridge | 0 | boundedProbe | active-certificate |
| Q-R3-KQ-01 | observationalEquivalence | PhotoLean.RelationAudit.BoundedAudit.R3.kashaRadiativeBranchYieldBridge | 0 | boundedProbe | active-certificate |
| Q-R3-MK-01 | equivalence | PhotoLean.RelationAudit.BoundedAudit.R3.marcusKashaWindowBridge | 9 | boundedProbe | active-certificate |
| Q-R3-MS-01 | observationalEquivalence | PhotoLean.RelationAudit.BoundedAudit.R3.marcusSymFactorBarrierBridge | 1 | boundedProbe | active-certificate |
| Q-R3-QF-01 | observationalEquivalence | PhotoLean.RelationAudit.BoundedAudit.R3.fluorPhosPhiFYieldBridge | 0 | boundedProbe | active-certificate |
| Q-R3-QF-02 | observationalEquivalence | PhotoLean.RelationAudit.BoundedAudit.R3.fluorPhosPhiPYieldCascadeBridge | 0 | boundedProbe | active-certificate |
| Q-R3-SF-01 | observationalEquivalence | PhotoLean.RelationAudit.BoundedAudit.R3.fpRatioInvariantBridge | 4 | boundedProbe | active-certificate |
| Q-R3-SQ-01 | observationalEquivalence | PhotoLean.RelationAudit.BoundedAudit.R3.svQuenchDilutesYieldBridge | 2 | boundedProbe | active-certificate |
| Q-R4-E1-BEG-01 | equivalence | PhotoLean.RelationAudit.BoundedAudit.R4Excited01.bepEnergyGapInvertedEquiv | 1 | boundedProbe | active-certificate |
| Q-R4-E1-BK-01 | equivalence | PhotoLean.RelationAudit.BoundedAudit.R4Excited01.bepKashaTransferZeroEquiv | 3 | boundedProbe | active-certificate |
| Q-R4-E1-HEG-01 | equivalence | PhotoLean.RelationAudit.BoundedAudit.R4Excited01.hammondEnergyGapBarrierZeroEquiv | 1 | boundedProbe | active-certificate |
| Q-R4-E1-HIC-01 | definitionalIdentity | PhotoLean.RelationAudit.BoundedAudit.R4Excited01.hammondICvsISCBarrierPin | 0 | boundedProbe | active-certificate |
| Q-R4-E1-HK-01 | equivalence | PhotoLean.RelationAudit.BoundedAudit.R4Excited01.hammondKashaICMaxEquiv | 3 | boundedProbe | active-certificate |
| Q-R4-E1-HSS-01 | observationalEquivalence | PhotoLean.RelationAudit.BoundedAudit.R4Excited01.hammondStokesAbsorptionBridge | 0 | boundedProbe | active-certificate |
| RC-RAD1-KEB-01 | observationalEquivalence | PhotoLean.RelationAudit.BoundedAudit.RAD1.kashaEinsteinBridge | 0 | boundedProbe | active-certificate |
| RC-RAD1-KEG-01 | definitionalIdentity | PhotoLean.RelationAudit.BoundedAudit.RAD1.kashaEglRateBridge | 0 | boundedProbe | active-certificate |
| RC-RAD1-KKV-01 | implication | PhotoLean.RelationAudit.BoundedAudit.RAD1.kashaVavBridge | 2 | boundedProbe | active-certificate |
| RC-RAD1-MEG-01 | definitionalIdentity | PhotoLean.RelationAudit.BoundedAudit.RAD1.marcusEglRateBridge | 0 | boundedProbe | active-certificate |
| RC-RAD1-MEG-02 | equivalence | PhotoLean.RelationAudit.BoundedAudit.RAD1.marcusEglZoneBridge | 0 | boundedProbe | active-certificate |
| RC-RAD1-MIC-01 | definitionalIdentity | PhotoLean.RelationAudit.BoundedAudit.RAD1.marcusICRateBridge | 0 | boundedProbe | active-certificate |
| RC-RAD1-MKVV-01 | implication | PhotoLean.RelationAudit.BoundedAudit.RAD1.mkvBridge | 7 | boundedProbe | active-certificate |
| RC-RAD1-MSS-01 | equivalence | PhotoLean.RelationAudit.BoundedAudit.RAD1.marcusSSBridge | 0 | boundedProbe | active-certificate |
| RC-RAD1-VFP-01 | observationalEquivalence | PhotoLean.RelationAudit.BoundedAudit.RAD1.kvFpBridge | 0 | boundedProbe | active-certificate |
| RC-RAD1-VQY-01 | observationalEquivalence | PhotoLean.RelationAudit.BoundedAudit.RAD1.kvQyBridge | 0 | boundedProbe | active-certificate |

| Certificate ID | Kind | Lean declaration | Premises | Evidence | Library role |
|---|---|---|---|---|---|
| RC-RAD1-VSV-01 | observationalEquivalence | PhotoLean.RelationAudit.BoundedAudit.RAD1.kvSvBridge | 2 | boundedProbe | active-certificate |
| RAD2-QYE-01 | observationalEquivalence | PhotoLean.RelationAudit.BoundedAudit.RAD2.qyEgBridge | 2 | boundedProbe | active-certificate |
| RAD2-QYEB-01 | observationalEquivalence | PhotoLean.RelationAudit.BoundedAudit.RAD2.qyEbBridge | 0 | boundedProbe | active-certificate |
| RAD2-QYF-01 | observationalEquivalence | PhotoLean.RelationAudit.BoundedAudit.RAD2.qy-FoBridge | 3 | boundedProbe | active-certificate |
| RAD2-QYI-01 | observationalEquivalence | PhotoLean.RelationAudit.BoundedAudit.RAD2.qy-IcBridge | 2 | boundedProbe | active-certificate |
| Q-RAD3-EI-01 | definitionalIdentity | PhotoLean.RelationAudit.BoundedAudit.Rad3.eiRateDefinition | 0 | boundedProbe | active-certificate |
| Q-RAD3-EI-02 | conditionalComposition | PhotoLean.RelationAudit.BoundedAudit.Rad3.eiCrossoverComposition | 2 | boundedProbe | active-certificate |
| Q-RAD3-ES-01 | equivalence | PhotoLean.RelationAudit.BoundedAudit.Rad3.es-ZeroBarrierBridge | 1 | boundedProbe | active-certificate |
| Q-RAD3-ES-02 | observationalEquivalence | PhotoLean.RelationAudit.BoundedAudit.Rad3.es-BarrierEmissionBridge | 1 | boundedProbe | active-certificate |
| Q-RAD3-FE-01 | observationalEquivalence | PhotoLean.RelationAudit.BoundedAudit.Rad3.fpEinsBridge | 0 | boundedProbe | active-certificate |
| Q-RAD3-FF-01 | observationalEquivalence | PhotoLean.RelationAudit.BoundedAudit.Rad3.ff-TransferBridge | 2 | boundedProbe | active-certificate |
| Q-RAD3-FI-01 | observationalEquivalence | PhotoLean.RelationAudit.BoundedAudit.Rad3.fi-BranchBridge | 0 | boundedProbe | active-certificate |
| Q-RAD3-FI-02 | equivalence | PhotoLean.RelationAudit.BoundedAudit.Rad3.fiElSayedBridge | 4 | boundedProbe | active-certificate |
| Q-RAD3-FoE-01 | observationalEquivalence | PhotoLean.RelationAudit.BoundedAudit.Rad3.foEinBridge | 0 | boundedProbe | active-certificate |
| Q-RAD3-SI-01 | observationalEquivalence | PhotoLean.RelationAudit.BoundedAudit.Rad3.siSurfaceBridge | 0 | boundedProbe | active-certificate |
| QP-symmetryFactorBepThermoneu-tralBridge | observationalEquivalence | PhotoLean.RelationAudit.BoundedAudit.Surface.symmetryFactorBepThermoneutralBridge | 1 | boundedProbe | active-certificate |
| QP-bepSabatierLinearVolcanoBridge | observationalEquivalence | PhotoLean.RelationAudit.BoundedAudit.Surface.bepSabatierLinearVolcanoBridge | 0 | boundedProbe | active-certificate |
| QP-symmetryFactorHammondTher-moneutralBridge | observationalEquivalence | PhotoLean.RelationAudit.BoundedAudit.Surface.symmetryFactorHammondThermoneutralBridge | 1 | boundedProbe | active-certificate |
| QP-marcusRateSabatierActivityPin | definitionalIdentity | PhotoLean.RelationAudit.BoundedAudit.Surface.marcusRateSabatierActivityPin | 0 | boundedProbe | active-certificate |
| cert-D-S01-kernel-barrier-eq-bep | definitionalIdentity | PhotoLean.RelationAudit.Full.Sections01_05.kernelBarrierBepPin | 0 | deliveredModule | retained-legacy-support |
| cert-D-S01-kernel-barrier-eq-ham-mond | definitionalIdentity | PhotoLean.RelationAudit.Full.Sections01_05.kernelBarrierHammondPin | 0 | deliveredModule | retained-legacy-support |
| cert-D-S01-kernel-productSurface-eq-hammond | definitionalIdentity | PhotoLean.RelationAudit.Full.Sections01_05.kernelProductSurfaceHammondPin | 0 | deliveredModule | retained-legacy-support |
| cert-D-S01-kernel-reactantSurface-eq-hammond | definitionalIdentity | PhotoLean.RelationAudit.Full.Sections01_05.kernelReactantSurfaceHammondPin | 0 | deliveredModule | retained-legacy-support |
| cert-D-S01-kernel-reverseBarrier-eq-hammond | definitionalIdentity | PhotoLean.RelationAudit.Full.Sections01_05.kernelReverseBarrierHammondPin | 0 | deliveredModule | retained-legacy-support |
| cert-D-S01-kernel-transfer-eq-bep | definitionalIdentity | PhotoLean.RelationAudit.Full.Sections01_05.kernelTransferBepPin | 0 | deliveredModule | retained-legacy-support |
| cert-D-S01-kernel-tsCoord-eq-ham-mond | definitionalIdentity | PhotoLean.RelationAudit.Full.Sections01_05.kernelTsCoordHammondPin | 0 | deliveredModule | retained-legacy-support |
| cert-D-S02-hammondZoneQ-beyon-dReactant-iff-inverted | equivalence | PhotoLean.RelationAudit.Full.Sections01_05.hammondZoneQInvertedEquivalence | 1 | deliveredModule | reused-active-certificate |
| cert-D-S02-lefflerSecant-neg-iff-inverted | equivalence | PhotoLean.RelationAudit.Full.Sections01_05.lefflerSecantNegInvertedEquivalence | 2 | deliveredModule | reused-active-certificate |
| cert-D-S02-secSlope-eq-lefflerSe-cant | observationalEquivalence | PhotoLean.RelationAudit.Full.Sections01_05.secSlopeLefflerSecantBridge | 0 | deliveredModule | reused-active-certificate |
| cert-D-S02-transfer-eq-tsCoord-bridge | observationalEquivalence | PhotoLean.RelationAudit.Full.Sections01_05.transferTsCoordBridge | 1 | deliveredModule | reused-active-certificate |
| cert-D-S02-tsCoord-lt-zero-iff-inverted | equivalence | PhotoLean.RelationAudit.Full.Sections01_05.tsCoordLtZeroInvertedEquivalence | 1 | deliveredModule | reused-active-certificate |
| cert-D-S03-hammondDescriptor-of-epConformsOnWindow | implication | PhotoLean.RelationAudit.Full.Sections01_05.windowConformanceDescriptorImplication | 0 | deliveredModule | reused-active-certificate |
| cert-D-S05-bep-eact-eq-marcus-barrier | definitionalIdentity | PhotoLean.RelationAudit.Full.Sections01_05.be-pEactMarcusBarrierPin | 0 | deliveredModule | reused-active-certificate |
| cert-D-S05-bep-rate-eq-exp-neg-eact | definitionalIdentity | PhotoLean.RelationAudit.Full.Sections01_05.bepRateExpNegEactPin | 0 | deliveredModule | reused-active-certificate |
| cert-D-S05-hammond-barrier-eq-gapReactant | definitionalIdentity | PhotoLean.RelationAudit.Full.Sections01_05.hammondBarrierGapReactantPin | 0 | deliveredModule | reused-active-certificate |
| cert-D-S06-hammond-sharp-iff-mar-cus-sharp | equivalence | PhotoLean.RelationAudit.Full.Sections06_10.hammondMarcusSharpCertificate | 3 | deliveredModule | reused-active-certificate |
| cert-D-S07-kernel-marcusIC | definitionalIdentity | PhotoLean.RelationAudit.Full.Sections06_10.kernelMarcusICPin | 0 | deliveredModule | retained-legacy-support |
| cert-D-S12-eg-barrier-eq-kernel | definitionalIdentity | PhotoLean.RelationAudit.Full.Sections11_17.eg-BarrierEqKernelCertificate | 0 | deliveredModule | retained-legacy-support |
| cert-D-S12-eg-invertedGap-iff-marcus | equivalence | PhotoLean.RelationAudit.Full.Sections11_17.egInvertedGapEquivalenceCertificate | 0 | deliveredModule | retained-legacy-support |

| Certificate ID | Kind | Lean declaration | Premises | Evidence | Library role |
|---|---|---|---|---|---|
| cert-D-S12-eg-rate-eq-marcus | definitionalIdentity | PhotoLean.RelationAudit.Full.Sections11_17.egRateEqMarcusCertificate | 0 | deliveredModule | retained-legacy-support |
| cert-D-S12-icvscic-barrier-eq-kernel | definitionalIdentity | PhotoLean.RelationAudit.Full.Sections11_17.icvscicBarrierEqKernelCertificate | 0 | deliveredModule | retained-legacy-support |
| cert-D-S12-icvscic-icRate-eq-marcus | definitionalIdentity | PhotoLean.RelationAudit.Full.Sections11_17.icvscicIcRateEqMarcusCertificate | 0 | deliveredModule | retained-legacy-support |
| cert-D-S12-ss-s0Surface-eq-kernel | definitionalIdentity | PhotoLean.RelationAudit.Full.Sections11_17.ssS0SurfaceEqKernelCertificate | 0 | deliveredModule | retained-legacy-support |
| cert-D-S12-ss-s1Surface-eq-kernel | definitionalIdentity | PhotoLean.RelationAudit.Full.Sections11_17.ssS1SurfaceEqKernelCertificate | 0 | deliveredModule | retained-legacy-support |
| cert-D-S15-egBarrier-zero-iff-emEnergy-zero | equivalence | PhotoLean.RelationAudit.Full.Sections11_17.egBarrierZeroIffEmEnergyZeroCertificate | 1 | deliveredModule | retained-legacy-support |
| cert-D-S15-icvscic-icRate-eq-eg-nrRate | definitionalIdentity | PhotoLean.RelationAudit.Full.Sections11_17.icvscicIcRateEqEnergyGapLawCertificate | 0 | deliveredModule | retained-legacy-support |
| cert-a1 | equivalence | PhotoLean.RelationAudit.betaHalfEquivalence | 1 | deliveredModule | retained-legacy-support |
| cert-composition | conditionalComposition | PhotoLean.RelationAudit.marcusNormalNoInvertedComposition | 2 | deliveredModule | reused-active-certificate |
| cert-composition-parent-epbounds | equivalence | PhotoLean.RelationAudit.epBoundsNoInvertedEquivalence | 1 | deliveredModule | retained-legacy-support |
| cert-composition-parent-normal | implication | PhotoLean.RelationAudit.marcusNormalImplication | 2 | deliveredModule | retained-legacy-support |
| cert-hammond-bep | implication | PhotoLean.RelationAudit.epBoundsImplication | 1 | deliveredModule | reused-active-certificate |
| cert-kasha-marcus | equivalence | PhotoLean.RelationAudit.kashaMarcusEquivalence | 9 | deliveredModule | retained-legacy-support |
| cert-kernel-pin | definitionalIdentity | PhotoLean.RelationAudit.kernelBarrierPin | 0 | deliveredModule | retained-legacy-support |
| cert-kv-refutation | refutation | PhotoLean.RelationAudit.kvRefutation | 2 | deliveredModule | retained-legacy-support |
| cert-sv-decoder | distinguishability | PhotoLean.RelationAudit.svJointIdentifiability | 3 | deliveredModule | retained-legacy-support |
| cert-sv-intensity | observationalEquivalence | PhotoLean.RelationAudit.svIntensityObservationalEquivalence | 2 | deliveredModule | retained-legacy-support |
| cert-sv-joint | distinguishability | PhotoLean.RelationAudit.svJointDistinguishability | 2 | deliveredModule | retained-legacy-support |

# Appendix B. Statement objects

These are the 92 normalized active statement endpoints. A statement ID is versioned by theory, declaration, and the explicit version field. Supporting theorems cited by evidence are not silently inserted as active endpoints.

**Table S11.** Active statement objects.

| Statement ID | Theory | Declaration | Version |
|---|---|---|---|
| bep.epBounds | BEP | PhotoLean.BEP.EPBounds | 1.0.0 |
| bep.noInvertedDirection | BEP | PhotoLean.BEP.epBounds_iff_no_inverted_direction | 1.0.0 |
| goldschmidt.conformance | Goldschmidt | PhotoLean.Goldschmidt.GoldschmidtConforms | 1.0.0 |
| hammond.reactionRegion | Hammond | PhotoLean.Hammond.ReactionRegion | 1.0.0 |
| kasha.kashaRule | Kasha | PhotoLean.Kasha.KashaRule | 1.0.0 |
| kasha.kashaWithin | Kasha | PhotoLean.Kasha.KashaWithin | 1.0.0 |
| marcus.barrier | Marcus | PhotoLean.Marcus.barrier | 1.0.0 |
| marcus.normalRegion | Marcus | PhotoLean.Marcus.NormalRegion | 1.0.0 |
| sternvolmer.intensityDyn | SternVolmer | PhotoLean.SternVolmer.svRatioDyn | 1.0.0 |
| stmt-BEP-PhotoLean-BEP-EPConformsOnWindow | BEP | PhotoLean.BEP.EPConformsOnWindow | 1.0.0 |
| stmt-BEP-PhotoLean-BEP-bepLine | BEP | PhotoLean.BEP.bepLine | 1.0.0 |
| stmt-BEP-PhotoLean-BEP-bepRadius | BEP | PhotoLean.BEP.bepRadius | 1.0.0 |
| stmt-BEP-PhotoLean-BEP-eact | BEP | PhotoLean.BEP.eact | 1.0.0 |
| stmt-BEP-PhotoLean-BEP-rate_eq_exp_neg_eact | BEP | PhotoLean.BEP.rate_eq_exp_neg_eact | 1.0.0 |
| stmt-BEP-PhotoLean-BEP-secSlope | BEP | PhotoLean.BEP.secSlope | 1.0.0 |
| stmt-BEP-PhotoLean-BEP-transfer | BEP | PhotoLean.BEP.transfer | 1.0.0 |
| stmt-Einstein-PhotoLean-Einstein-aOfB | Einstein | PhotoLean.Einstein.aOfB | 1.0.0 |
| stmt-Einstein-PhotoLean-Einstein-radFactor | Einstein | PhotoLean.Einstein.radFactor | 1.0.0 |
| stmt-Einstein-PhotoLean-Einstein-tauR | Einstein | PhotoLean.Einstein.tauR | 1.0.0 |
| stmt-Einstein-PhotoLean-Einstein-yield_radiative | Einstein | PhotoLean.Einstein.yield_radiative | 1.0.0 |
| stmt-EnergyGapLaw-PhotoLean-EnergyGapLaw-InvertedGap | EnergyGapLaw | PhotoLean.EnergyGapLaw.InvertedGap | 1.0.0 |
| stmt-EnergyGapLaw-PhotoLean-EnergyGapLaw-egZoneQ | EnergyGapLaw | PhotoLean.EnergyGapLaw.egZoneQ | 1.0.0 |
| stmt-EnergyGapLaw-PhotoLean-EnergyGapLaw-nrBarrier | EnergyGapLaw | PhotoLean.EnergyGapLaw.nrBarrier | 1.0.0 |
| stmt-EnergyGapLaw-PhotoLean-EnergyGapLaw-nrRate | EnergyGapLaw | PhotoLean.EnergyGapLaw.nrRate | 1.0.0 |
| stmt-FluorPhos-PhotoLean-FluorPhos-fpZoneQ | FluorPhos | PhotoLean.FluorPhos.fpZoneQ | 1.0.0 |
| stmt-FluorPhos-PhotoLean-FluorPhos-iscBranch | FluorPhos | PhotoLean.FluorPhos.iscBranch | 1.0.0 |
| stmt-FluorPhos-PhotoLean-FluorPhos-phiF | FluorPhos | PhotoLean.FluorPhos.phiF | 1.0.0 |
| stmt-FluorPhos-PhotoLean-FluorPhos-phiP | FluorPhos | PhotoLean.FluorPhos.phiP | 1.0.0 |
| stmt-FluorPhos-PhotoLean-FluorPhos-s1Decay | FluorPhos | PhotoLean.FluorPhos.s1Decay | 1.0.0 |
| stmt-Forster-PhotoLean-Forster-fretEff6 | Forster | PhotoLean.Forster.fretEff6 | 1.0.0 |
| stmt-Forster-PhotoLean-Forster-fretEff_self | Forster | PhotoLean.Forster.fretEff_self | 1.0.0 |
| stmt-Forster-PhotoLean-Forster-fretRate | Forster | PhotoLean.Forster.fretRate | 1.0.0 |
| stmt-Forster-PhotoLean-Forster-kappaSq | Forster | PhotoLean.Forster.kappaSq | 1.0.0 |
| stmt-Forster-PhotoLean-Forster-r0six | Forster | PhotoLean.Forster.r0six | 1.0.0 |
| stmt-Goldschmidt-PhotoLean-Goldschmidt-InBand | Goldschmidt | PhotoLean.Goldschmidt.InBand | 1.0.0 |
| stmt-Goldschmidt-PhotoLean-Goldschmidt-contact_iff_tolFac_one | Goldschmidt | PhotoLean.Goldschmidt.contact_iff_tolFac_one | 1.0.0 |
| stmt-Goldschmidt-PhotoLean-Goldschmidt-gapA | Goldschmidt | PhotoLean.Goldschmidt.gapA | 1.0.0 |
| stmt-Goldschmidt-PhotoLean-Goldschmidt-goldschmidtZone | Goldschmidt | PhotoLean.Goldschmidt.goldschmidtZone | 1.0.0 |
| stmt-Goldschmidt-PhotoLean-Goldschmidt-latticeOf | Goldschmidt | PhotoLean.Goldschmidt.latticeOf | 1.0.0 |
| stmt-Goldschmidt-PhotoLean-Goldschmidt-tauGoldschmidt | Goldschmidt | PhotoLean.Goldschmidt.tauGoldschmidt | 1.0.0 |
| stmt-Goldschmidt-PhotoLean-Goldschmidt-tolFac | Goldschmidt | PhotoLean.Goldschmidt.tolFac | 1.0.0 |
| stmt-Hammond-PhotoLean-Hammond-HammondDescriptor | Hammond | PhotoLean.Hammond.HammondDescriptor | 1.0.0 |
| stmt-Hammond-PhotoLean-Hammond-Rat-hammondZoneQ | Hammond | PhotoLean.Hammond.Rat.hammondZoneQ | 1.0.0 |
| stmt-Hammond-PhotoLean-Hammond-gapReactant | Hammond | PhotoLean.Hammond.gapReactant | 1.0.0 |
| stmt-Hammond-PhotoLean-Hammond-lefflerSecant | Hammond | PhotoLean.Hammond.lefflerSecant | 1.0.0 |
| stmt-Hammond-PhotoLean-Hammond-productSurface | Hammond | PhotoLean.Hammond.productSurface | 1.0.0 |
| stmt-Hammond-PhotoLean-Hammond-tsCoord | Hammond | PhotoLean.Hammond.tsCoord | 1.0.0 |
| stmt-ICvsISC-PhotoLean-ICvsISC-fcBarrier | ICvsISC | PhotoLean.ICvsISC.fcBarrier | 1.0.0 |
| stmt-ICvsISC-PhotoLean-ICvsISC-icRate | ICvsISC | PhotoLean.ICvsISC.icRate | 1.0.0 |
| stmt-ICvsISC-PhotoLean-ICvsISC-iscRate | ICvsISC | PhotoLean.ICvsISC.iscRate | 1.0.0 |
| stmt-ICvsISC-PhotoLean-ICvsISC-isc_dominates_iff | ICvsISC | PhotoLean.ICvsISC.isc_dominates_iff | 1.0.0 |
| stmt-ICvsISC-PhotoLean-ICvsISC-rate_ratio_eq | ICvsISC | PhotoLean.ICvsISC.rate_ratio_eq | 1.0.0 |
| stmt-ICvsISC-PhotoLean-ICvsISC-spin_discount | ICvsISC | PhotoLean.ICvsISC.spin_discount | 1.0.0 |
| stmt-Kasha-PhotoLean-Kasha-emitYield | Kasha | PhotoLean.Kasha.emitYield | 1.0.0 |
| stmt-Kasha-PhotoLean-Kasha-fluoYield | Kasha | PhotoLean.Kasha.fluoYield | 1.0.0 |
| stmt-Kasha-PhotoLean-Kasha-funnelRatio | Kasha | PhotoLean.Kasha.funnelRatio | 1.0.0 |
| stmt-Kasha-PhotoLean-Kasha-icBranch | Kasha | PhotoLean.Kasha.icBranch | 1.0.0 |
| stmt-Kasha-PhotoLean-Kasha-marcusIC | Kasha | PhotoLean.Kasha.marcusIC | 1.0.0 |
| stmt-Kasha-PhotoLean-Kasha-radBranch | Kasha | PhotoLean.Kasha.radBranch | 1.0.0 |

| Statement ID | Theory | Declaration | Version |
|---|---|---|---|
| stmt-KashaVavilov-PhotoLean-KashaVavilov-SpecSame | KashaVavilov | PhotoLean.KashaVavilov.SpecSame | 1.0.0 |
| stmt-KashaVavilov-PhotoLean-KashaVavilov-antiKasha_observable_iff | KashaVavilov | PhotoLean.KashaVavilov.antiKasha_observable_iff | 1.0.0 |
| stmt-KashaVavilov-PhotoLean-KashaVavilov-d2_verdict | KashaVavilov | PhotoLean.KashaVavilov.d2_verdict | 1.0.0 |
| stmt-KashaVavilov-PhotoLean-KashaVavilov-fluoYield_eq_emitYield_zero_add_upperYield | KashaVavilov | PhotoLean.KashaVavilov.fluoYield_eq_emitYield_zero_add_upperYield | 1.0.0 |
| stmt-KashaVavilov-PhotoLean-KashaVavilov-lossless_separates_kasha_vavilov | KashaVavilov | PhotoLean.KashaVavilov.lossless_separates_kasha_vavilov | 1.0.0 |
| stmt-Marcus-PhotoLean-Marcus-InvertedRegion | Marcus | PhotoLean.Marcus.InvertedRegion | 1.0.0 |
| stmt-Marcus-PhotoLean-Marcus-Rat-zoneQ | Marcus | PhotoLean.Marcus.Rat.zoneQ | 1.0.0 |
| stmt-Marcus-PhotoLean-Marcus-descriptor_sharp | Marcus | PhotoLean.Marcus.descriptor_sharp | 1.0.0 |
| stmt-Marcus-PhotoLean-Marcus-rate | Marcus | PhotoLean.Marcus.rate | 1.0.0 |
| stmt-QuantumYield-PhotoLean-QuantumYield-yieldOf | QuantumYield | PhotoLean.QuantumYield.yieldOf | 1.0.0 |
| stmt-QuantumYield-PhotoLean-QuantumYield-yieldOf_div_yieldOf | QuantumYield | PhotoLean.QuantumYield.yieldOf_div_yieldOf | 1.0.0 |
| stmt-RACI-PhotoLean-RACI-accessGap | RACI | PhotoLean.RACI.accessGap | 1.0.0 |
| stmt-RACI-PhotoLean-RACI-barrierRate | RACI | PhotoLean.RACI.barrierRate | 1.0.0 |
| stmt-RACI-PhotoLean-RACI-competitionRatio | RACI | PhotoLean.RACI.competitionRatio | 1.0.0 |
| stmt-RACI-PhotoLean-RACI-quantumYield | RACI | PhotoLean.RACI.quantumYield | 1.0.0 |
| stmt-RACI-PhotoLean-RACI-torsionH | RACI | PhotoLean.RACI.torsionH | 1.0.0 |
| stmt-RACI-PhotoLean-RACI-torsionH_conicalSet | RACI | PhotoLean.RACI.torsionH_conicalSet | 1.0.0 |
| stmt-Sabatier-PhotoLean-Sabatier-NearOptimal | Sabatier | PhotoLean.Sabatier.NearOptimal | 1.0.0 |
| stmt-Sabatier-PhotoLean-Sabatier-VolcanoDescriptor | Sabatier | PhotoLean.Sabatier.VolcanoDescriptor | 1.0.0 |
| stmt-Sabatier-PhotoLean-Sabatier-activity | Sabatier | PhotoLean.Sabatier.activity | 1.0.0 |
| stmt-Sabatier-PhotoLean-Sabatier-apexPar | Sabatier | PhotoLean.Sabatier.apexPar | 1.0.0 |
| stmt-Sabatier-PhotoLean-Sabatier-parabolicBarrier | Sabatier | PhotoLean.Sabatier.parabolicBarrier | 1.0.0 |
| stmt-Sabatier-PhotoLean-Sabatier-volcanoBarrier | Sabatier | PhotoLean.Sabatier.volcanoBarrier | 1.0.0 |
| stmt-SternVolmer-PhotoLean-SternVolmer-KSV | SternVolmer | PhotoLean.SternVolmer.KSV | 1.0.0 |
| stmt-SternVolmer-PhotoLean-SternVolmer-dynDecay | SternVolmer | PhotoLean.SternVolmer.dynDecay | 1.0.0 |
| stmt-SternVolmer-PhotoLean-SternVolmer-svZoneQ | SternVolmer | PhotoLean.SternVolmer.svZoneQ | 1.0.0 |
| stmt-StokesShift-PhotoLean-StokesShift-absEnergy | StokesShift | PhotoLean.StokesShift.absEnergy | 1.0.0 |
| stmt-StokesShift-PhotoLean-StokesShift-emEnergy | StokesShift | PhotoLean.StokesShift.emEnergy | 1.0.0 |
| stmt-StokesShift-PhotoLean-StokesShift-s0Surface | StokesShift | PhotoLean.StokesShift.s0Surface | 1.0.0 |
| stmt-StokesShift-PhotoLean-StokesShift-stokesShift | StokesShift | PhotoLean.StokesShift.stokesShift | 1.0.0 |
| stmt-SymmetryFactor-PhotoLean-SymmetryFactor-asymReactantSurface | SymmetryFactor | PhotoLean.SymmetryFactor.asymReactantSurface | 1.0.0 |
| stmt-SymmetryFactor-PhotoLean-SymmetryFactor-tsCoordZero | SymmetryFactor | PhotoLean.SymmetryFactor.tsCoordZero | 1.0.0 |
| symmetry.betaHalfReading | SymmetryFactor | PhotoLean.SymmetryFactor.BetaHalfReading | 1.0.0 |

# Appendix C. Active audit records

This appendix lists all 152 active records. For certified rows the kind column is the relation kind; for unresolved rows it is the expected candidate kind; otherwise it is a dash. `pairId` is read directly from the record. Scope and snapshot are printed per row, preserving all query-level snapshots. No row is converted into a binary edge/non-edge claim.

**Table S12.** Complete active audit-record index.

| Record ID | Status | Kind/candidate | Query ID | Pair ID | Scope | Snapshot |
|---|---|---|---|---|---|---|
| rec-bounded-Q-CI-BP-01 | unresolved | observationalEquivalence | Q-CI-BP-01 | P-BEP-RACI | scope-CI-candidate-bridge | snap-bounded-audit-2026-11-04 |
| rec-bounded-Q-CI-EB-01 | certified | observationalEquivalence | Q-CI-EB-01 | P-Einstein-RACI | scope-CI-radiative-anchor | snap-bounded-audit-2026-11-04 |
| rec-bounded-Q-CI-EG-01 | certified | observationalEquivalence | Q-CI-EG-01 | P-EnergyGapLaw-RACI | scope-CI-rate-law | snap-bounded-audit-2026-11-04 |
| rec-bounded-Q-CI-FO-01 | unresolved | conditionalComposition | Q-CI-FO-01 | P-Forster-RACI | scope-CI-candidate-bridge | snap-bounded-audit-2026-11-04 |
| rec-bounded-Q-CI-FP-01 | certified | observationalEquivalence | Q-CI-FP-01 | P-FluorPhos-RACI | scope-CI-fluor-phos-s1 | snap-bounded-audit-2026-11-04 |
| rec-bounded-Q-CI-GS-01 | outOfScope | – | Q-CI-GS-01 | P-Goldschmidt-RACI | scope-CI-out-of-jurisdiction | snap-bounded-audit-2026-11-04 |
| rec-bounded-Q-CI-HM-01 | unresolved | implication | Q-CI-HM-01 | P-Hammond-RACI | scope-CI-candidate-bridge | snap-bounded-audit-2026-11-04 |
| rec-bounded-Q-CI-IC-01 | certified | observationalEquivalence | Q-CI-IC-01 | P-ICvsISC-RACI | scope-CI-rate-law | snap-bounded-audit-2026-11-04 |
| rec-bounded-Q-CI-KS-01 | certified | definitionalIdentity | Q-CI-KS-01 | P-Kasha-RACI | scope-CI-ladder-yield | snap-bounded-audit-2026-11-04 |
| rec-bounded-Q-CI-KV-01 | notEstablished | – | Q-CI-KV-01 | P-KashaVavilov-RACI | scope-CI-kv-scoped-search | snap-bounded-audit-2026-11-04 |
| rec-bounded-Q-CI-MR-01 | certified | observationalEquivalence | Q-CI-MR-01 | P-Marcus-RACI | scope-CI-rate-law | snap-bounded-audit-2026-11-04 |
| rec-bounded-Q-CI-QY-01 | certified | observationalEquivalence | Q-CI-QY-01 | P-QuantumYield-RACI | scope-CI-yield-two-channel | snap-bounded-audit-2026-11-04 |
| rec-bounded-Q-CI-SA-01 | unresolved | implication | Q-CI-SA-01 | P-Sabatier-RACI | scope-CI-candidate-bridge | snap-bounded-audit-2026-11-04 |
| rec-bounded-Q-CI-SF-01 | unresolved | equivalence | Q-CI-SF-01 | P-SymmetryFactor-RACI | scope-CI-candidate-bridge | snap-bounded-audit-2026-11-04 |
| rec-bounded-Q-CI-SS-01 | unresolved | equivalence | Q-CI-SS-01 | P-StokesShift-RACI | scope-CI-candidate-bridge | snap-bounded-audit-2026-11-04 |
| rec-bounded-Q-CI-SV-01 | certified | observationalEquivalence | Q-CI-SV-01 | P-SternVolmer-RACI | scope-CI-quench-dilution | snap-bounded-audit-2026-11-04 |
| rec-bounded-Q-E2-BE-01 | outOfScope | – | Q-E2-BE-01 | P-BEP-Einstein | scope-E2-jurisdiction | snap-bounded-audit-2026-11-04-r4e2 |
| rec-bounded-Q-E2-BF-01 | outOfScope | – | Q-E2-BF-01 | P-BEP-Forster | scope-E2-jurisdiction | snap-bounded-audit-2026-11-04-r4e2 |
| rec-bounded-Q-E2-BI-01 | certified | definitionalIdentity | Q-E2-BI-01 | P-BEP-ICvsISC | scope-E2-bep-ic-barrier | snap-bounded-audit-2026-11-04-r4e2 |
| rec-bounded-Q-E2-BS-01 | certified | observationalEquivalence | Q-E2-BS-01 | P-BEP-StokesShift | scope-E2-bep-stokes-thermoneutral | snap-bounded-audit-2026-11-04-r4e2 |
| rec-bounded-Q-E2-KS-01 | unresolved | implication | Q-E2-KS-01 | P-Kasha-SymmetryFactor | scope-E2-candidate-bridge | snap-bounded-audit-2026-11-04-r4e2 |
| rec-bounded-Q-E2-SE-01 | certified | observationalEquivalence | Q-E2-SE-01 | P-SymmetryFactor-EnergyGapLaw | scope-E2-sc-egl-thermoneutral | snap-bounded-audit-2026-11-04-r4e2 |
| rec-bounded-Q-E2-SEIn-01 | outOfScope | – | Q-E2-SEIn-01 | P-SymmetryFactor-Einstein | scope-E2-jurisdiction | snap-bounded-audit-2026-11-04-r4e2 |
| rec-bounded-Q-E2-SF-01 | unresolved | implication | Q-E2-SF-01 | P-SymmetryFactor-FluorPhos | scope-E2-candidate-bridge | snap-bounded-audit-2026-11-04-r4e2 |
| rec-bounded-Q-E2-SFo-01 | outOfScope | – | Q-E2-SFo-01 | P-SymmetryFactor-Forster | scope-E2-jurisdiction | snap-bounded-audit-2026-11-04-r4e2 |
| rec-bounded-Q-E2-SI-01 | certified | observationalEquivalence | Q-E2-SI-01 | P-SymmetryFactor-ICvsISC | scope-E2-sc-ic-thermoneutral | snap-bounded-audit-2026-11-04-r4e2 |
| rec-bounded-Q-E2-SK-01 | outOfScope | – | Q-E2-SK-01 | P-SymmetryFactor-KashaVavilov | scope-E2-jurisdiction | snap-bounded-audit-2026-11-04-r4e2 |
| rec-bounded-Q-E2-SQ-01 | unresolved | implication | Q-E2-SQ-01 | P-SymmetryFactor-QuantumYield | scope-E2-candidate-bridge | snap-bounded-audit-2026-11-04-r4e2 |
| rec-bounded-Q-E2-SS-01 | certified | definitionalIdentity | Q-E2-SS-01 | P-SymmetryFactor-StokesShift | scope-E2-sc-stokes-surfaces | snap-bounded-audit-2026-11-04-r4e2 |
| rec-bounded-Q-E2-SV-01 | unresolved | implication | Q-E2-SV-01 | P-SymmetryFactor-SternVolmer | scope-E2-candidate-bridge | snap-bounded-audit-2026-11-04-r4e2 |
| rec-bounded-Q-R3-KF-01 | unresolved | implication | Q-R3-KF-01 | P-Kasha-FluorPhos | scope-R3-candidate-bridge | snap-bounded-audit-2026-11-04 |
| rec-bounded-Q-R3-KQ-01 | certified | observationalEquivalence | Q-R3-KQ-01 | P-Kasha-QuantumYield | scope-R3-kasha-yield | snap-bounded-audit-2026-11-04 |
| rec-bounded-Q-R3-KV-01 | unresolved | implication | Q-R3-KV-01 | P-Kasha-SternVolmer | scope-R3-candidate-bridge | snap-bounded-audit-2026-11-04 |
| rec-bounded-Q-R3-MF-01 | unresolved | implication | Q-R3-MF-01 | P-Marcus-FluorPhos | scope-R3-candidate-bridge | snap-bounded-audit-2026-11-04 |
| rec-bounded-Q-R3-MK-01 | certified | equivalence | Q-R3-MK-01 | P-Marcus-Kasha | scope-R3-marcus-kasha-window | snap-bounded-audit-2026-11-04 |
| rec-bounded-Q-R3-MQ-01 | unresolved | implication | Q-R3-MQ-01 | P-Marcus-QuantumYield | scope-R3-candidate-bridge | snap-bounded-audit-2026-11-04 |
| rec-bounded-Q-R3-MS-01 | certified | observationalEquivalence | Q-R3-MS-01 | P-Marcus-SymmetryFactor | scope-R3-surface-thermoneutral | snap-bounded-audit-2026-11-04 |

| Record ID | Status | Kind/candidate | Query ID | Pair ID | Scope | Snapshot |
|---|---|---|---|---|---|---|
| rec-bounded-Q-R3-MV-01 | unresolved | implication | Q-R3-MV-01 | P-Marcus-SternVolmer | scope-R3-candidate-bridge | snap-bounded-audit-2026-11-04 |
| rec-bounded-Q-R3-QF-01 | certified | observationalEquivalence | Q-R3-QF-01 | P-QuantumYield-FluorPhos | scope-R3-qy-fp-composition | snap-bounded-audit-2026-11-04 |
| rec-bounded-Q-R3-QF-02 | certified | observationalEquivalence | Q-R3-QF-02 | P-QuantumYield-FluorPhos | scope-R3-qy-fp-composition | snap-bounded-audit-2026-11-04 |
| rec-bounded-Q-R3-SF-01 | certified | observationalEquivalence | Q-R3-SF-01 | P-SternVolmer-FluorPhos | scope-R3-sv-fp-invariance | snap-bounded-audit-2026-11-04 |
| rec-bounded-Q-R3-SQ-01 | certified | observationalEquivalence | Q-R3-SQ-01 | P-SternVolmer-QuantumYield | scope-R3-sv-qy-dilution | snap-bounded-audit-2026-11-04 |
| rec-bounded-Q-R4-E1-BEG-01 | certified | equivalence | Q-R4-E1-BEG-01 | P-BEP-EnergyGapLaw | scope-R4-E1-barrierless | snap-bounded-audit-2026-11-04 |
| rec-bounded-Q-R4-E1-BFP-01 | unresolved | implication | Q-R4-E1-BFP-01 | P-BEP-FluorPhos | scope-R4-E1-candidate-bridge | snap-bounded-audit-2026-11-04 |
| rec-bounded-Q-R4-E1-BK-01 | certified | equivalence | Q-R4-E1-BK-01 | P-BEP-Kasha | scope-R4-E1-rate-geometry | snap-bounded-audit-2026-11-04 |
| rec-bounded-Q-R4-E1-BKV-01 | unresolved | conditionalComposition | Q-R4-E1-BKV-01 | P-BEP-KashaVavilov | scope-R4-E1-candidate-bridge | snap-bounded-audit-2026-11-04 |
| rec-bounded-Q-R4-E1-BQY-01 | unresolved | implication | Q-R4-E1-BQY-01 | P-BEP-QuantumYield | scope-R4-E1-candidate-bridge | snap-bounded-audit-2026-11-04 |
| rec-bounded-Q-R4-E1-BSV-01 | unresolved | implication | Q-R4-E1-BSV-01 | P-BEP-SternVolmer | scope-R4-E1-candidate-bridge | snap-bounded-audit-2026-11-04 |
| rec-bounded-Q-R4-E1-HEG-01 | certified | equivalence | Q-R4-E1-HEG-01 | P-Hammond-EnergyGapLaw | scope-R4-E1-barrierless | snap-bounded-audit-2026-11-04 |
| rec-bounded-Q-R4-E1-HEIN-01 | unresolved | implication | Q-R4-E1-HEIN-01 | P-Hammond-Einstein | scope-R4-E1-candidate-bridge | snap-bounded-audit-2026-11-04 |
| rec-bounded-Q-R4-E1-HFO-01 | outOfScope | – | Q-R4-E1-HFO-01 | P-Hammond-Forster | scope-R4-E1-jurisdiction | snap-bounded-audit-2026-11-04 |
| rec-bounded-Q-R4-E1-HFP-01 | unresolved | implication | Q-R4-E1-HFP-01 | P-Hammond-FluorPhos | scope-R4-E1-candidate-bridge | snap-bounded-audit-2026-11-04 |
| rec-bounded-Q-R4-E1-HIC-01 | certified | definitionalIdentity | Q-R4-E1-HIC-01 | P-Hammond-ICvsISC | scope-R4-E1-barrier-pin | snap-bounded-audit-2026-11-04 |
| rec-bounded-Q-R4-E1-HK-01 | certified | equivalence | Q-R4-E1-HK-01 | P-Hammond-Kasha | scope-R4-E1-rate-geometry | snap-bounded-audit-2026-11-04 |
| rec-bounded-Q-R4-E1-HKV-01 | unresolved | conditionalComposition | Q-R4-E1-HKV-01 | P-Hammond-KashaVavilov | scope-R4-E1-candidate-bridge | snap-bounded-audit-2026-11-04 |
| rec-bounded-Q-R4-E1-HQY-01 | unresolved | implication | Q-R4-E1-HQY-01 | P-Hammond-QuantumYield | scope-R4-E1-candidate-bridge | snap-bounded-audit-2026-11-04 |
| rec-bounded-Q-R4-E1-HSS-01 | certified | observationalEquivalence | Q-R4-E1-HSS-01 | P-Hammond-StokesShift | scope-R4-E1-surface-observable | snap-bounded-audit-2026-11-04 |
| rec-bounded-Q-R4-E1-HSV-01 | unresolved | implication | Q-R4-E1-HSV-01 | P-Hammond-SternVolmer | scope-R4-E1-candidate-bridge | snap-bounded-audit-2026-11-04 |
| rec-bounded-Q-RAD1-KEB-01 | certified | observationalEquivalence | Q-RAD1-KEB-01 | P-Kasha-Einstein | scope-RAD1-kasha-einstein | snap-bounded-audit-2026-11-04-r5rad1 |
| rec-bounded-Q-RAD1-KEG-01 | certified | definitionalIdentity | Q-RAD1-KEG-01 | P-Kasha-EnergyGapLaw | scope-RAD1-kasha-egl-rate | snap-bounded-audit-2026-11-04-r5rad1 |
| rec-bounded-Q-RAD1-KFO-01 | outOfScope | – | Q-RAD1-KFO-01 | P-Kasha-Forster | scope-RAD1-jurisdiction | snap-bounded-audit-2026-11-04-r5rad1 |
| rec-bounded-Q-RAD1-KIC-01 | unresolved | conditionalComposition | Q-RAD1-KIC-01 | P-Kasha-ICvsISC | scope-RAD1-candidate-bridge | snap-bounded-audit-2026-11-04-r5rad1 |
| rec-bounded-Q-RAD1-KKV-01 | certified | implication | Q-RAD1-KKV-01 | P-Kasha-KashaVavilov | scope-RAD1-kasha-vav-spec | snap-bounded-audit-2026-11-04-r5rad1 |
| rec-bounded-Q-RAD1-KSS-01 | outOfScope | – | Q-RAD1-KSS-01 | P-Kasha-StokesShift | scope-RAD1-jurisdiction | snap-bounded-audit-2026-11-04-r5rad1 |
| rec-bounded-Q-RAD1-MEB-01 | outOfScope | – | Q-RAD1-MEB-01 | P-Marcus-Einstein | scope-RAD1-jurisdiction | snap-bounded-audit-2026-11-04-r5rad1 |
| rec-bounded-Q-RAD1-MEG-01 | certified | definitionalIdentity | Q-RAD1-MEG-01 | P-Marcus-EnergyGapLaw | scope-RAD1-egl-rate | snap-bounded-audit-2026-11-04-r5rad1 |
| rec-bounded-Q-RAD1-MEG-02 | certified | equivalence | Q-RAD1-MEG-02 | P-Marcus-EnergyGapLaw | scope-RAD1-egl-zone | snap-bounded-audit-2026-11-04-r5rad1 |
| rec-bounded-Q-RAD1-MFO-01 | outOfScope | – | Q-RAD1-MFO-01 | P-Marcus-Forster | scope-RAD1-jurisdiction | snap-bounded-audit-2026-11-04-r5rad1 |
| rec-bounded-Q-RAD1-MIC-01 | certified | definitionalIdentity | Q-RAD1-MIC-01 | P-Marcus-ICvsISC | scope-RAD1-ic-rate | snap-bounded-audit-2026-11-04-r5rad1 |
| rec-bounded-Q-RAD1-MKVV-01 | certified | implication | Q-RAD1-MKVV-01 | P-Marcus-KashaVavilov | scope-RAD1-marcus-kv-window | snap-bounded-audit-2026-11-04-r5rad1 |
| rec-bounded-Q-RAD1-MSS-01 | certified | equivalence | Q-RAD1-MSS-01 | P-Marcus-StokesShift | scope-RAD1-ss-window | snap-bounded-audit-2026-11-04-r5rad1 |
| rec-bounded-Q-RAD1-VEG-01 | unresolved | conditionalComposition | Q-RAD1-VEG-01 | P-KashaVavilov-EnergyGapLaw | scope-RAD1-candidate-bridge | snap-bounded-audit-2026-11-04-r5rad1 |
| rec-bounded-Q-RAD1-VFP-01 | certified | observationalEquivalence | Q-RAD1-VFP-01 | P-KashaVavilov-FluorPhos | scope-RAD1-kv-fp-lump | snap-bounded-audit-2026-11-04-r5rad1 |
| rec-bounded-Q-RAD1-VQY-01 | certified | observationalEquivalence | Q-RAD1-VQY-01 | P-KashaVavilov-QuantumYield | scope-RAD1-kv-qy-decomp | snap-bounded-audit-2026-11-04-r5rad1 |
| rec-bounded-Q-RAD1-VSV-01 | certified | observationalEquivalence | Q-RAD1-VSV-01 | P-KashaVavilov-SternVolmer | scope-RAD1-kv-sv-quench | snap-bounded-audit-2026-11-04-r5rad1 |
| rec-bounded-Q-RAD2-FPE-01 | unresolved | conditionalComposition | Q-RAD2-FPE-01 | P-FluorPhos-EnergyGapLaw | scope-RAD2-candidate-bridge | snap-bounded-audit-2026-11-04-r5rad2 |
| rec-bounded-Q-RAD2-FPS-01 | outOfScope | – | Q-RAD2-FPS-01 | P-FluorPhos-StokesShift | scope-RAD2-jurisdiction | snap-bounded-audit-2026-11-04-r5rad2 |
| rec-bounded-Q-RAD2-KVE-01 | unresolved | conditionalComposition | Q-RAD2-KVE-01 | P-KashaVavilov-Einstein | scope-RAD2-candidate-bridge | snap-bounded-audit-2026-11-04-r5rad2 |

| Record ID | Status | Kind/candidate | Query ID | Pair ID | Scope | Snapshot |
|---|---|---|---|---|---|---|
| rec-bounded-Q-RAD2-KVF-01 | outOfScope | – | Q-RAD2-KVF-01 | P-KashaVavilov-Forster | scope-RAD2-jurisdiction | snap-bounded-audit-2026-11-04-r5rad2 |
| rec-bounded-Q-RAD2-KVI-01 | unresolved | conditionalComposition | Q-RAD2-KVI-01 | P-KashaVavilov-ICvsISC | scope-RAD2-candidate-bridge | snap-bounded-audit-2026-11-04-r5rad2 |
| rec-bounded-Q-RAD2-KVS-01 | outOfScope | – | Q-RAD2-KVS-01 | P-KashaVavilov-StokesShift | scope-RAD2-jurisdiction | snap-bounded-audit-2026-11-04-r5rad2 |
| rec-bounded-Q-RAD2-QYE-01 | certified | observationalEquivalence | Q-RAD2-QYE-01 | P-QuantumYield-Energy-GapLaw | scope-RAD2-qy-eg-branch | snap-bounded-audit-2026-11-04-r5rad2 |
| rec-bounded-Q-RAD2-QYEB-01 | certified | observationalEquivalence | Q-RAD2-QYEB-01 | P-QuantumYield-Einstein | scope-RAD2-qy-eb-radiative | snap-bounded-audit-2026-11-04-r5rad2 |
| rec-bounded-Q-RAD2-QYF-01 | certified | observationalEquivalence | Q-RAD2-QYF-01 | P-QuantumYield-Forster | scope-RAD2-qy-fo-efficiency | snap-bounded-audit-2026-11-04-r5rad2 |
| rec-bounded-Q-RAD2-QYI-01 | certified | observationalEquivalence | Q-RAD2-QYI-01 | P-QuantumYield-ICvsISC | scope-RAD2-qy-ic-branch | snap-bounded-audit-2026-11-04-r5rad2 |
| rec-bounded-Q-RAD2-QYS-01 | outOfScope | – | Q-RAD2-QYS-01 | P-QuantumYield-StokesShift | scope-RAD2-jurisdiction | snap-bounded-audit-2026-11-04-r5rad2 |
| rec-bounded-Q-RAD2-SVE-01 | unresolved | conditionalComposition | Q-RAD2-SVE-01 | P-SternVolmer-Energy-GapLaw | scope-RAD2-candidate-bridge | snap-bounded-audit-2026-11-04-r5rad2 |
| rec-bounded-Q-RAD2-SVEb-01 | unresolved | conditionalComposition | Q-RAD2-SVEb-01 | P-SternVolmer-Einstein | scope-RAD2-candidate-bridge | snap-bounded-audit-2026-11-04-r5rad2 |
| rec-bounded-Q-RAD2-SVF-01 | outOfScope | – | Q-RAD2-SVF-01 | P-SternVolmer-Forster | scope-RAD2-jurisdiction | snap-bounded-audit-2026-11-04-r5rad2 |
| rec-bounded-Q-RAD2-SVI-01 | unresolved | conditionalComposition | Q-RAD2-SVI-01 | P-SternVolmer-ICvsISC | scope-RAD2-candidate-bridge | snap-bounded-audit-2026-11-04-r5rad2 |
| rec-bounded-Q-RAD2-SVS-01 | outOfScope | – | Q-RAD2-SVS-01 | P-SternVolmer-StokesShift | scope-RAD2-jurisdiction | snap-bounded-audit-2026-11-04-r5rad2 |
| rec-bounded-Q-RAD3-EE-01 | outOfScope | – | Q-RAD3-EE-01 | P-EnergyGapLaw-Einstein | scope-RAD3-jurisdiction | snap-bounded-audit-2026-11-04 |
| rec-bounded-Q-RAD3-EF-01 | outOfScope | – | Q-RAD3-EF-01 | P-EnergyGapLaw-Forster | scope-RAD3-jurisdiction | snap-bounded-audit-2026-11-04 |
| rec-bounded-Q-RAD3-EI-01 | certified | definitionalIdentity | Q-RAD3-EI-01 | P-EnergyGapLaw-ICvsISC | scope-RAD3-rate-identity | snap-bounded-audit-2026-11-04 |
| rec-bounded-Q-RAD3-EI-02 | certified | conditionalComposition | Q-RAD3-EI-02 | P-EnergyGapLaw-ICvsISC | scope-RAD3-rate-identity | snap-bounded-audit-2026-11-04 |
| rec-bounded-Q-RAD3-ES-01 | certified | equivalence | Q-RAD3-ES-01 | P-EnergyGapLaw-StokesShift | scope-RAD3-gap-window | snap-bounded-audit-2026-11-04 |
| rec-bounded-Q-RAD3-ES-02 | certified | observationalEquivalence | Q-RAD3-ES-02 | P-EnergyGapLaw-StokesShift | scope-RAD3-gap-window | snap-bounded-audit-2026-11-04 |
| rec-bounded-Q-RAD3-FE-01 | certified | observationalEquivalence | Q-RAD3-FE-01 | P-FluorPhos-Einstein | scope-RAD3-radiative-yield | snap-bounded-audit-2026-11-04 |
| rec-bounded-Q-RAD3-FF-01 | certified | observationalEquivalence | Q-RAD3-FF-01 | P-FluorPhos-Forster | scope-RAD3-transfer-efficiency | snap-bounded-audit-2026-11-04 |
| rec-bounded-Q-RAD3-FI-01 | certified | observationalEquivalence | Q-RAD3-FI-01 | P-FluorPhos-ICvsISC | scope-RAD3-branching | snap-bounded-audit-2026-11-04 |
| rec-bounded-Q-RAD3-FI-02 | certified | equivalence | Q-RAD3-FI-02 | P-FluorPhos-ICvsISC | scope-RAD3-el-sayed | snap-bounded-audit-2026-11-04 |
| rec-bounded-Q-RAD3-FoE-01 | certified | observationalEquivalence | Q-RAD3-FoE-01 | P-Forster-Einstein | scope-RAD3-rate-identity | snap-bounded-audit-2026-11-04 |
| rec-bounded-Q-RAD3-IE-01 | outOfScope | – | Q-RAD3-IE-01 | P-ICvsISC-Einstein | scope-RAD3-jurisdiction | snap-bounded-audit-2026-11-04 |
| rec-bounded-Q-RAD3-IF-01 | unresolved | conditionalComposition | Q-RAD3-IF-01 | P-ICvsISC-Forster | scope-RAD3-scoped-search | snap-bounded-audit-2026-11-04 |
| rec-bounded-Q-RAD3-SE-01 | outOfScope | – | Q-RAD3-SE-01 | P-StokesShift-Einstein | scope-RAD3-jurisdiction | snap-bounded-audit-2026-11-04 |
| rec-bounded-Q-RAD3-SF-01 | notEstablished | – | Q-RAD3-SF-01 | P-StokesShift-Forster | scope-RAD3-scoped-search | snap-bounded-audit-2026-11-04 |
| rec-bounded-Q-RAD3-SI-01 | certified | observationalEquivalence | Q-RAD3-SI-01 | P-StokesShift-ICvsISC | scope-RAD3-surface-height | snap-bounded-audit-2026-11-04 |
| rec-bounded-Q-WPS01-BF-01 | certified | observationalEquivalence | Q-WPS01-BF-01 | P-BEP-SymmetryFactor | scope-WPS01-equal-curvature | snap-bounded-audit-2026-11-04 |
| rec-bounded-Q-WPS01-BS-01 | certified | observationalEquivalence | Q-WPS01-BS-01 | P-BEP-Sabatier | scope-WPS01-descriptor-axis | snap-bounded-audit-2026-11-04 |
| rec-bounded-Q-WPS01-HB-01 | certified | implication | Q-WPS01-HB-01 | P-Hammond-BEP | scope-WPS01-two-parabola | snap-bounded-audit-2026-11-04 |
| rec-bounded-Q-WPS01-HB-02 | certified | observationalEquivalence | Q-WPS01-HB-02 | P-Hammond-BEP | scope-WPS01-transfer | snap-bounded-audit-2026-11-04 |
| rec-bounded-Q-WPS01-HB-03 | certified | observationalEquivalence | Q-WPS01-HB-03 | P-Hammond-BEP | scope-WPS01-slope | snap-bounded-audit-2026-11-04 |
| rec-bounded-Q-WPS01-HB-04 | certified | implication | Q-WPS01-HB-04 | P-Hammond-BEP | scope-WPS01-window | snap-bounded-audit-2026-11-04 |
| rec-bounded-Q-WPS01-HF-01 | certified | observationalEquivalence | Q-WPS01-HF-01 | P-Hammond-SymmetryFactor | scope-WPS01-equal-curvature | snap-bounded-audit-2026-11-04 |
| rec-bounded-Q-WPS01-HS-01 | unresolved | implication | Q-WPS01-HS-01 | P-Hammond-Sabatier | scope-WPS01-candidate-bridge | snap-bounded-audit-2026-11-04 |
| rec-bounded-Q-WPS01-KS-01 | unresolved | equivalence | Q-WPS01-KS-01 | P-Kasha-Sabatier | scope-WPS01-candidate-bridge | snap-bounded-audit-2026-11-04 |
| rec-bounded-Q-WPS01-MB-01 | certified | conditionalComposition | Q-WPS01-MB-01 | P-Marcus-BEP | scope-WPS01-two-parabola | snap-bounded-audit-2026-11-04 |
| rec-bounded-Q-WPS01-MB-02 | certified | definitionalIdentity | Q-WPS01-MB-02 | P-Marcus-BEP | scope-WPS01-two-parabola | snap-bounded-audit-2026-11-04 |
| rec-bounded-Q-WPS01-MB-03 | certified | definitionalIdentity | Q-WPS01-MB-03 | P-Marcus-BEP | scope-WPS01-rate | snap-bounded-audit-2026-11-04 |

| Record ID | Status | Kind/candidate | Query ID | Pair ID | Scope | Snapshot |
|---|---|---|---|---|---|---|
| rec-bounded-Q-WPS01-MH-01 | certified | definitionalIdentity | Q-WPS01-MH-01 | P-Marcus-Hammond | scope-WPS01-two-parabola | snap-bounded-audit-2026-11-04 |
| rec-bounded-Q-WPS01-MH-02 | certified | equivalence | Q-WPS01-MH-02 | P-Marcus-Hammond | scope-WPS01-two-parabola | snap-bounded-audit-2026-11-04 |
| rec-bounded-Q-WPS01-MH-03 | certified | equivalence | Q-WPS01-MH-03 | P-Marcus-Hammond | scope-WPS01-secant | snap-bounded-audit-2026-11-04 |
| rec-bounded-Q-WPS01-MH-04 | certified | equivalence | Q-WPS01-MH-04 | P-Marcus-Hammond | scope-WPS01-rational | snap-bounded-audit-2026-11-04 |
| rec-bounded-Q-WPS01-MH-05 | certified | equivalence | Q-WPS01-MH-05 | P-Marcus-Hammond | scope-WPS01-sharp | snap-bounded-audit-2026-11-04 |
| rec-bounded-Q-WPS01-MS-01 | certified | definitionalIdentity | Q-WPS01-MS-01 | P-Marcus-Sabatier | scope-WPS01-rate | snap-bounded-audit-2026-11-04 |
| rec-bounded-Q-WPS01-SF-01 | unresolved | equivalence | Q-WPS01-SF-01 | P-Sabatier-SymmetryFactor | scope-WPS01-candidate-bridge | snap-bounded-audit-2026-11-04 |
| rec-bounded-Q-XD1-BG-01 | outOfScope | – | Q-XD1-BG-01 | P-BEP-Goldschmidt | scope-XD1-goldschmidt-jurisdiction | snap-bounded-audit-2026-11-04 |
| rec-bounded-Q-XD1-GSF-01 | outOfScope | – | Q-XD1-GSF-01 | P-Goldschmidt-SymmetryFactor | scope-XD1-goldschmidt-jurisdiction | snap-bounded-audit-2026-11-04 |
| rec-bounded-Q-XD1-HG-01 | outOfScope | – | Q-XD1-HG-01 | P-Hammond-Goldschmidt | scope-XD1-goldschmidt-jurisdiction | snap-bounded-audit-2026-11-04 |
| rec-bounded-Q-XD1-KG-01 | outOfScope | – | Q-XD1-KG-01 | P-Kasha-Goldschmidt | scope-XD1-goldschmidt-jurisdiction | snap-bounded-audit-2026-11-04 |
| rec-bounded-Q-XD1-MG-01 | outOfScope | – | Q-XD1-MG-01 | P-Marcus-Goldschmidt | scope-XD1-goldschmidt-jurisdiction | snap-bounded-audit-2026-11-04 |
| rec-bounded-Q-XD1-SE-01 | outOfScope | – | Q-XD1-SE-01 | P-Sabatier-Einstein | scope-XD1-sabatier-jurisdiction | snap-bounded-audit-2026-11-04 |
| rec-bounded-Q-XD1-SEG-01 | unresolved | implication | Q-XD1-SEG-01 | P-Sabatier-EnergyGapLaw | scope-XD1-sabatier-egl-candidate | snap-bounded-audit-2026-11-04 |
| rec-bounded-Q-XD1-SF-01 | outOfScope | – | Q-XD1-SF-01 | P-Sabatier-Forster | scope-XD1-sabatier-jurisdiction | snap-bounded-audit-2026-11-04 |
| rec-bounded-Q-XD1-SFP-01 | outOfScope | – | Q-XD1-SFP-01 | P-Sabatier-FluorPhos | scope-XD1-sabatier-jurisdiction | snap-bounded-audit-2026-11-04 |
| rec-bounded-Q-XD1-SG-01 | outOfScope | – | Q-XD1-SG-01 | P-Sabatier-Goldschmidt | scope-XD1-goldschmidt-jurisdiction | snap-bounded-audit-2026-11-04 |
| rec-bounded-Q-XD1-SIC-01 | unresolved | implication | Q-XD1-SIC-01 | P-Sabatier-ICvsISC | scope-XD1-sabatier-ic-candidate | snap-bounded-audit-2026-11-04 |
| rec-bounded-Q-XD1-SKV-01 | outOfScope | – | Q-XD1-SKV-01 | P-Sabatier-KashaVavilov | scope-XD1-sabatier-jurisdiction | snap-bounded-audit-2026-11-04 |
| rec-bounded-Q-XD1-SQY-01 | outOfScope | – | Q-XD1-SQY-01 | P-Sabatier-QuantumYield | scope-XD1-sabatier-jurisdiction | snap-bounded-audit-2026-11-04 |
| rec-bounded-Q-XD1-SSS-01 | unresolved | implication | Q-XD1-SSS-01 | P-Sabatier-StokesShift | scope-XD1-sabatier-stokes-candidate | snap-bounded-audit-2026-11-04 |
| rec-bounded-Q-XD1-SSV-01 | outOfScope | – | Q-XD1-SSV-01 | P-Sabatier-SternVolmer | scope-XD1-sabatier-jurisdiction | snap-bounded-audit-2026-11-04 |
| rec-bounded-Q-XD2-EG-01 | outOfScope | – | Q-XD2-EG-01 | P-Goldschmidt-EnergyGapLaw | scope-XD2-EG-jurisdiction | snap-bounded-audit-2026-11-04 |
| rec-bounded-Q-XD2-EG-02 | outOfScope | – | Q-XD2-EG-02 | P-Goldschmidt-EnergyGapLaw | scope-XD2-EG-jurisdiction | snap-bounded-audit-2026-11-04 |
| rec-bounded-Q-XD2-EI-01 | outOfScope | – | Q-XD2-EI-01 | P-Goldschmidt-Einstein | scope-XD2-EI-jurisdiction | snap-bounded-audit-2026-11-04 |
| rec-bounded-Q-XD2-FP-01 | outOfScope | – | Q-XD2-FP-01 | P-Goldschmidt-FluorPhos | scope-XD2-FP-jurisdiction | snap-bounded-audit-2026-11-04 |
| rec-bounded-Q-XD2-FR-01 | outOfScope | – | Q-XD2-FR-01 | P-Goldschmidt-Forster | scope-XD2-FR-jurisdiction | snap-bounded-audit-2026-11-04 |
| rec-bounded-Q-XD2-FR-02 | outOfScope | – | Q-XD2-FR-02 | P-Goldschmidt-Forster | scope-XD2-FR-jurisdiction | snap-bounded-audit-2026-11-04 |
| rec-bounded-Q-XD2-IC-01 | outOfScope | – | Q-XD2-IC-01 | P-Goldschmidt-ICvsISC | scope-XD2-IC-jurisdiction | snap-bounded-audit-2026-11-04 |
| rec-bounded-Q-XD2-KV-01 | outOfScope | – | Q-XD2-KV-01 | P-Goldschmidt-KashaVavilov | scope-XD2-KV-jurisdiction | snap-bounded-audit-2026-11-04 |
| rec-bounded-Q-XD2-QY-01 | outOfScope | – | Q-XD2-QY-01 | P-Goldschmidt-QuantumYield | scope-XD2-QY-jurisdiction | snap-bounded-audit-2026-11-04 |
| rec-bounded-Q-XD2-SS-01 | outOfScope | – | Q-XD2-SS-01 | P-Goldschmidt-StokesShift | scope-XD2-SS-jurisdiction | snap-bounded-audit-2026-11-04 |
| rec-bounded-Q-XD2-SV-01 | outOfScope | – | Q-XD2-SV-01 | P-Goldschmidt-SternVolmer | scope-XD2-SV-jurisdiction | snap-bounded-audit-2026-11-04 |

# Appendix D. Legacy dispositions

The additive ledger disposes every one of the 151 unchanged legacy records exactly once. An empty active-query list is intentional and is displayed as a dash. Neither a pair anchor nor an intra-theory record contributes active coverage; a reused certificate's active query is counted in Appendix C, not added again here.

**Table S13.** Complete legacy-disposition index.

| Legacy record | Legacy query | Status | Disposition | Pair | Active query IDs | Legacy scope |
|---|---|---|---|---|---|---|
| rec-kernel-pin | q-kernel-pin | certified | retained-intra-theory-evidence | – | – | global |
| rec-hammond-bep | q-hammond-bep | certified | reused-as-active-evidence | P-Hammond-BEP | Q-WPS01-HB-01 | global |
| rec-a1 | q-a1 | certified | retained-intra-theory-evidence | – | – | global |
| rec-kasha-marcus | q-kasha-marcus | certified | recertified-statement-level | P-Marcus-Kasha | Q-R3-MK-01 | global |
| rec-composition | q-composition | certified | reused-as-active-evidence | P-Marcus-BEP | Q-WPS01-MB-01 | global |

| Legacy record | Legacy query | Status | Disposition | Pair | Active query IDs | Legacy scope |
|---|---|---|---|---|---|---|
| rec-sv-intensity | q-sv-intensity | certified | retained-intra-theory-evidence | — | – | global |
| rec-sv-joint | q-sv-joint | certified | retained-intra-theory-evidence | — | – | global |
| rec-sv-decoder | q-sv-decoder | certified | retained-intra-theory-evidence | — | – | global |
| rec-kasha-vavilov | q-kasha-vavilov | certified | retained-refutation-evidence | P-Kasha-KashaVavilov | – | global |
| rec-kasha-sternvolmer-unresolved | q-kasha-sternvolmer-candidate | unresolved | retained-unresolved-candidate | P-Kasha-SternVolmer | Q-R3-KV-01 | global |
| rec-goldschmidt-marcus-noedge | q-goldschmidt-marcus | notEstablished | retained-scoped-absence | P-Marcus-Goldschmidt | – | goldschmidt-marcus |
| rec-D-S02-transfer-eq-tsCoord-bridge | q-D-S02-transfer-eq-tsCoord-bridge | certified | reused-as-active-evidence | P-Hammond-BEP | Q-WPS01-HB-02 | global |
| rec-D-S02-secSlope-eq-lefflerSecant | q-D-S02-secSlope-eq-lefflerSecant | certified | reused-as-active-evidence | P-Hammond-BEP | Q-WPS01-HB-03 | global |
| rec-D-S02-tsCoord-lt-zero-iff-inverted | q-D-S02-tsCoord-lt-zero-iff-inverted | certified | reused-as-active-evidence | P-Marcus-Hammond | Q-WPS01-MH-02 | global |
| rec-D-S02-lefflerSecant-neg-iff-inverted | q-D-S02-lefflerSecant-neg-iff-inverted | certified | reused-as-active-evidence | P-Marcus-Hammond | Q-WPS01-MH-03 | global |
| rec-D-S02-hammondZoneQ-beyondReactant-iff-inverted | q-D-S02-hammondZoneQ-beyondReactant-iff-inverted | certified | reused-as-active-evidence | P-Marcus-Hammond | Q-WPS01-MH-04 | global |
| rec-D-S03-hammondDescriptor-of-epConformsOnWindow | q-D-S03-hammondDescriptor-of-epConformsOnWindow | certified | reused-as-active-evidence | P-Hammond-BEP | Q-WPS01-HB-04 | global |
| rec-D-S05-bep-eact-eq-marcus-barrier | q-D-S05-bep-eact-eq-marcus-barrier | certified | reused-as-active-evidence | P-Marcus-BEP | Q-WPS01-MB-02 | global |
| rec-D-S05-bep-rate-eq-exp-neg-eact | q-D-S05-bep-rate-eq-exp-neg-eact | certified | repaired-and-reused | P-Marcus-BEP | Q-WPS01-MB-03 | global |
| rec-D-S05-hammond-barrier-eq-gapReactant | q-D-S05-hammond-barrier-eq-gapReactant | certified | reused-as-active-evidence | P-Marcus-Hammond | Q-WPS01-MH-01 | global |
| rec-D-S06-hammond-sharp-iff-marcus-sharp | q-D-S06-hammond-sharp-iff-marcus-sharp | certified | repaired-and-reused | P-Marcus-Hammond | Q-WPS01-MH-05 | global |
| rec-D-S12-eg-rate-eq-marcus | q-D-S12-eg-rate-eq-marcus | certified | recertified-statement-level | P-Marcus-EnergyGapLaw | Q-RAD1-MEG-01 | global |
| rec-D-S12-eg-invertedGap-iff-marcus | q-D-S12-eg-invertedGap-iff-marcus | certified | recertified-statement-level | P-Marcus-EnergyGapLaw | Q-RAD1-MEG-02 | global |
| rec-D-S12-icvscic-icRate-eq-marcus | q-D-S12-icvscic-icRate-eq-marcus | certified | recertified-statement-level | P-Marcus-ICvsISC | Q-RAD1-MIC-01 | global |
| rec-D-S15-egBarrier-zero-iff-emEnergy-zero | q-D-S15-egBarrier-zero-iff-emEnergy-zero | certified | repaired-and-restated | P-EnergyGapLaw-StokesShift | Q-RAD3-ES-01 | global |
| rec-D-S15-icvscic-icRate-eq-eg-nrRate | q-D-S15-icvscic-icRate-eq-eg-nrRate | certified | recertified-statement-level | P-EnergyGapLaw-ICvsISC | Q-RAD3-EI-01 | global |
| rec-notestablished-P-Marcus-Sabatier | q-pair-P-Marcus-Sabatier | notEstablished | reconciled-absence-with-active-evidence | P-Marcus-Sabatier | Q-WPS01-MS-01 | scope-P-Marcus-Sabatier |
| rec-notestablished-P-Marcus-SymmetryFactor | q-pair-P-Marcus-SymmetryFactor | notEstablished | reconciled-absence-with-active-evidence | P-Marcus-SymmetryFactor | Q-R3-MS-01 | scope-P-Marcus-SymmetryFactor |
| rec-notestablished-P-Marcus-KashaVavilov | q-pair-P-Marcus-KashaVavilov | notEstablished | reconciled-absence-with-active-evidence | P-Marcus-KashaVavilov | Q-RAD1-MKVV-01 | scope-P-Marcus-KashaVavilov |
| rec-notestablished-P-Marcus-SternVolmer | q-pair-P-Marcus-SternVolmer | notEstablished | retained-scoped-absence | P-Marcus-SternVolmer | Q-R3-MV-01 | scope-P-Marcus-SternVolmer |
| rec-notestablished-P-Marcus-QuantumYield | q-pair-P-Marcus-QuantumYield | notEstablished | retained-scoped-absence | P-Marcus-QuantumYield | Q-R3-MQ-01 | scope-P-Marcus-QuantumYield |
| rec-notestablished-P-Marcus-FluorPhos | q-pair-P-Marcus-FluorPhos | notEstablished | retained-scoped-absence | P-Marcus-FluorPhos | Q-R3-MF-01 | scope-P-Marcus-FluorPhos |
| rec-notestablished-P-Marcus-StokesShift | q-pair-P-Marcus-StokesShift | notEstablished | reconciled-absence-with-active-evidence | P-Marcus-StokesShift | Q-RAD1-MSS-01 | scope-P-Marcus-StokesShift |
| rec-notestablished-P-Marcus-Forster | q-pair-P-Marcus-Forster | notEstablished | retained-scoped-absence | P-Marcus-Forster | Q-RAD1-MFO-01 | scope-P-Marcus-Forster |
| rec-notestablished-P-Marcus-Einstein | q-pair-P-Marcus-Einstein | notEstablished | retained-scoped-absence | P-Marcus-Einstein | Q-RAD1-MEB-01 | scope-P-Marcus-Einstein |
| rec-notestablished-P-Marcus-RACI | q-pair-P-Marcus-RACI | notEstablished | reconciled-absence-with-active-evidence | P-Marcus-RACI | Q-CI-MR-01 | scope-P-Marcus-RACI |
| rec-notestablished-P-Hammond-Kasha | q-pair-P-Hammond-Kasha | notEstablished | reconciled-absence-with-active-evidence | P-Hammond-Kasha | Q-R4-E1-HK-01 | scope-P-Hammond-Kasha |
| rec-notestablished-P-Hammond-Sabatier | q-pair-P-Hammond-Sabatier | notEstablished | retained-scoped-absence | P-Hammond-Sabatier | Q-WPS01-HS-01 | scope-P-Hammond-Sabatier |
| rec-notestablished-P-Hammond-Goldschmidt | q-pair-P-Hammond-Goldschmidt | notEstablished | retained-scoped-absence | P-Hammond-Goldschmidt | Q-XD1-HG-01 | scope-P-Hammond-Goldschmidt |
| rec-notestablished-P-Hammond-SymmetryFactor | q-pair-P-Hammond-SymmetryFactor | notEstablished | reconciled-absence-with-active-evidence | P-Hammond-SymmetryFactor | Q-WPS01-HF-01 | scope-P-Hammond-SymmetryFactor |
| rec-notestablished-P-Hammond-KashaVavilov | q-pair-P-Hammond-KashaVavilov | notEstablished | retained-scoped-absence | P-Hammond-KashaVavilov | Q-R4-E1-HKV-01 | scope-P-Hammond-KashaVavilov |
| rec-notestablished-P-Hammond-SternVolmer | q-pair-P-Hammond-SternVolmer | notEstablished | retained-scoped-absence | P-Hammond-SternVolmer | Q-R4-E1-HSV-01 | scope-P-Hammond-SternVolmer |
| rec-notestablished-P-Hammond-QuantumYield | q-pair-P-Hammond-QuantumYield | notEstablished | retained-scoped-absence | P-Hammond-QuantumYield | Q-R4-E1-HQY-01 | scope-P-Hammond-QuantumYield |
| rec-notestablished-P-Hammond-FluorPhos | q-pair-P-Hammond-FluorPhos | notEstablished | retained-scoped-absence | P-Hammond-FluorPhos | Q-R4-E1-HFP-01 | scope-P-Hammond-FluorPhos |
| rec-notestablished-P-Hammond-EnergyGapLaw | q-pair-P-Hammond-EnergyGapLaw | notEstablished | reconciled-absence-with-active-evidence | P-Hammond-EnergyGapLaw | Q-R4-E1-HEG-01 | scope-P-Hammond-EnergyGapLaw |
| rec-notestablished-P-Hammond-StokesShift | q-pair-P-Hammond-StokesShift | notEstablished | reconciled-absence-with-active-evidence | P-Hammond-StokesShift | Q-R4-E1-HSS-01 | scope-P-Hammond-StokesShift |
| rec-notestablished-P-Hammond-ICvsISC | q-pair-P-Hammond-ICvsISC | notEstablished | reconciled-absence-with-active-evidence | P-Hammond-ICvsISC | Q-R4-E1-HIC-01 | scope-P-Hammond-ICvsISC |

| Legacy record | Legacy query | Status | Disposition | Pair | Active query IDs | Legacy scope |
|---|---|---|---|---|---|---|
| rec-notestablished-P-Hammond-Forster | q-pair-P-Hammond-Forster | notEstablished | retained-scoped-absence | P-Hammond-Forster | Q-R4-E1-HFO-01 | scope-P-Hammond-Forster |
| rec-notestablished-P-Hammond-Einstein | q-pair-P-Hammond-Einstein | notEstablished | retained-scoped-absence | P-Hammond-Einstein | Q-R4-E1-HEIN-01 | scope-P-Hammond-Einstein |
| rec-notestablished-P-Hammond-RACI | q-pair-P-Hammond-RACI | notEstablished | retained-scoped-absence | P-Hammond-RACI | Q-CI-HM-01 | scope-P-Hammond-RACI |
| rec-notestablished-P-BEP-Kasha | q-pair-P-BEP-Kasha | notEstablished | reconciled-absence-with-active-evidence | P-BEP-Kasha | Q-R4-E1-BK-01 | scope-P-BEP-Kasha |
| rec-notestablished-P-BEP-Sabatier | q-pair-P-BEP-Sabatier | notEstablished | reconciled-absence-with-active-evidence | P-BEP-Sabatier | Q-WPS01-BS-01 | scope-P-BEP-Sabatier |
| rec-notestablished-P-BEP-Goldschmidt | q-pair-P-BEP-Goldschmidt | notEstablished | retained-scoped-absence | P-BEP-Goldschmidt | Q-XD1-BG-01 | scope-P-BEP-Goldschmidt |
| rec-notestablished-P-BEP-SymmetryFactor | q-pair-P-BEP-SymmetryFactor | notEstablished | reconciled-absence-with-active-evidence | P-BEP-SymmetryFactor | Q-WPS01-BF-01 | scope-P-BEP-SymmetryFactor |
| rec-notestablished-P-BEP-KashaVavilov | q-pair-P-BEP-KashaVavilov | notEstablished | retained-scoped-absence | P-BEP-KashaVavilov | Q-R4-E1-BKV-01 | scope-P-BEP-KashaVavilov |
| rec-notestablished-P-BEP-SternVolmer | q-pair-P-BEP-SternVolmer | notEstablished | retained-scoped-absence | P-BEP-SternVolmer | Q-R4-E1-BSV-01 | scope-P-BEP-SternVolmer |
| rec-notestablished-P-BEP-QuantumYield | q-pair-P-BEP-QuantumYield | notEstablished | retained-scoped-absence | P-BEP-QuantumYield | Q-R4-E1-BQY-01 | scope-P-BEP-QuantumYield |
| rec-notestablished-P-BEP-FluorPhos | q-pair-P-BEP-FluorPhos | notEstablished | retained-scoped-absence | P-BEP-FluorPhos | Q-R4-E1-BFP-01 | scope-P-BEP-FluorPhos |
| rec-notestablished-P-BEP-EnergyGapLaw | q-pair-P-BEP-EnergyGapLaw | notEstablished | reconciled-absence-with-active-evidence | P-BEP-EnergyGapLaw | Q-R4-E1-BEG-01 | scope-P-BEP-EnergyGapLaw |
| rec-notestablished-P-BEP-StokesShift | q-pair-P-BEP-StokesShift | notEstablished | reconciled-absence-with-active-evidence | P-BEP-StokesShift | Q-E2-BS-01 | scope-P-BEP-StokesShift |
| rec-notestablished-P-BEP-ICvsISC | q-pair-P-BEP-ICvsISC | notEstablished | reconciled-absence-with-active-evidence | P-BEP-ICvsISC | Q-E2-BI-01 | scope-P-BEP-ICvsISC |
| rec-notestablished-P-BEP-Forster | q-pair-P-BEP-Forster | notEstablished | retained-scoped-absence | P-BEP-Forster | Q-E2-BF-01 | scope-P-BEP-Forster |
| rec-notestablished-P-BEP-Einstein | q-pair-P-BEP-Einstein | notEstablished | retained-scoped-absence | P-BEP-Einstein | Q-E2-BE-01 | scope-P-BEP-Einstein |
| rec-notestablished-P-BEP-RACI | q-pair-P-BEP-RACI | notEstablished | retained-scoped-absence | P-BEP-RACI | Q-CI-BP-01 | scope-P-BEP-RACI |
| rec-notestablished-P-Kasha-Sabatier | q-pair-P-Kasha-Sabatier | notEstablished | retained-scoped-absence | P-Kasha-Sabatier | Q-WPS01-KS-01 | scope-P-Kasha-Sabatier |
| rec-notestablished-P-Kasha-Goldschmidt | q-pair-P-Kasha-Goldschmidt | notEstablished | retained-scoped-absence | P-Kasha-Goldschmidt | Q-XD1-KG-01 | scope-P-Kasha-Goldschmidt |
| rec-notestablished-P-Kasha-SymmetryFactor | q-pair-P-Kasha-SymmetryFactor | notEstablished | retained-scoped-absence | P-Kasha-SymmetryFactor | Q-E2-KS-01 | scope-P-Kasha-SymmetryFactor |
| rec-notestablished-P-Kasha-QuantumYield | q-pair-P-Kasha-QuantumYield | notEstablished | reconciled-absence-with-active-evidence | P-Kasha-QuantumYield | Q-R3-KQ-01 | scope-P-Kasha-QuantumYield |
| rec-candidate-P-Kasha-FluorPhos | q-pair-P-Kasha-FluorPhos | unresolved | retained-unresolved-candidate | P-Kasha-FluorPhos | Q-R3-KF-01 | global |
| rec-notestablished-P-Kasha-EnergyGapLaw | q-pair-P-Kasha-EnergyGapLaw | notEstablished | reconciled-absence-with-active-evidence | P-Kasha-EnergyGapLaw | Q-RAD1-KEG-01 | scope-P-Kasha-EnergyGapLaw |
| rec-notestablished-P-Kasha-StokesShift | q-pair-P-Kasha-StokesShift | notEstablished | retained-scoped-absence | P-Kasha-StokesShift | Q-RAD1-KSS-01 | scope-P-Kasha-StokesShift |
| rec-notestablished-P-Kasha-ICvsISC | q-pair-P-Kasha-ICvsISC | notEstablished | retained-scoped-absence | P-Kasha-ICvsISC | Q-RAD1-KIC-01 | scope-P-Kasha-ICvsISC |
| rec-notestablished-P-Kasha-Forster | q-pair-P-Kasha-Forster | notEstablished | retained-scoped-absence | P-Kasha-Forster | Q-RAD1-KFO-01 | scope-P-Kasha-Forster |
| rec-notestablished-P-Kasha-Einstein | q-pair-P-Kasha-Einstein | notEstablished | reconciled-absence-with-active-evidence | P-Kasha-Einstein | Q-RAD1-KEB-01 | scope-P-Kasha-Einstein |
| rec-notestablished-P-Kasha-RACI | q-pair-P-Kasha-RACI | notEstablished | reconciled-absence-with-active-evidence | P-Kasha-RACI | Q-CI-KS-01 | scope-P-Kasha-RACI |
| rec-notestablished-P-Sabatier-Goldschmidt | q-pair-P-Sabatier-Goldschmidt | notEstablished | retained-scoped-absence | P-Sabatier-Goldschmidt | Q-XD1-SG-01 | scope-P-Sabatier-Goldschmidt |
| rec-notestablished-P-Sabatier-SymmetryFactor | q-pair-P-Sabatier-SymmetryFactor | notEstablished | retained-scoped-absence | P-Sabatier-SymmetryFactor | Q-WPS01-SF-01 | scope-P-Sabatier-SymmetryFactor |
| rec-notestablished-P-Sabatier-KashaVavilov | q-pair-P-Sabatier-KashaVavilov | notEstablished | retained-scoped-absence | P-Sabatier-KashaVavilov | Q-XD1-SKV-01 | scope-P-Sabatier-KashaVavilov |
| rec-notestablished-P-Sabatier-SternVolmer | q-pair-P-Sabatier-SternVolmer | notEstablished | retained-scoped-absence | P-Sabatier-SternVolmer | Q-XD1-SSV-01 | scope-P-Sabatier-SternVolmer |
| rec-notestablished-P-Sabatier-QuantumYield | q-pair-P-Sabatier-QuantumYield | notEstablished | retained-scoped-absence | P-Sabatier-QuantumYield | Q-XD1-SQY-01 | scope-P-Sabatier-QuantumYield |
| rec-notestablished-P-Sabatier-FluorPhos | q-pair-P-Sabatier-FluorPhos | notEstablished | retained-scoped-absence | P-Sabatier-FluorPhos | Q-XD1-SFP-01 | scope-P-Sabatier-FluorPhos |
| rec-notestablished-P-Sabatier-EnergyGapLaw | q-pair-P-Sabatier-EnergyGapLaw | notEstablished | retained-scoped-absence | P-Sabatier-EnergyGapLaw | Q-XD1-SEG-01 | scope-P-Sabatier-EnergyGapLaw |
| rec-notestablished-P-Sabatier-StokesShift | q-pair-P-Sabatier-StokesShift | notEstablished | retained-scoped-absence | P-Sabatier-StokesShift | Q-XD1-SSS-01 | scope-P-Sabatier-StokesShift |
| rec-notestablished-P-Sabatier-ICvsISC | q-pair-P-Sabatier-ICvsISC | notEstablished | retained-scoped-absence | P-Sabatier-ICvsISC | Q-XD1-SIC-01 | scope-P-Sabatier-ICvsISC |
| rec-notestablished-P-Sabatier-Forster | q-pair-P-Sabatier-Forster | notEstablished | retained-scoped-absence | P-Sabatier-Forster | Q-XD1-SF-01 | scope-P-Sabatier-Forster |
| rec-notestablished-P-Sabatier-Einstein | q-pair-P-Sabatier-Einstein | notEstablished | retained-scoped-absence | P-Sabatier-Einstein | Q-XD1-SE-01 | scope-P-Sabatier-Einstein |
| rec-notestablished-P-Sabatier-RACI | q-pair-P-Sabatier-RACI | notEstablished | retained-scoped-absence | P-Sabatier-RACI | Q-CI-SA-01 | scope-P-Sabatier-RACI |
| rec-notestablished-P-Goldschmidt-SymmetryFactor | q-pair-P-Goldschmidt-SymmetryFactor | notEstablished | retained-scoped-absence | P-Goldschmidt-SymmetryFactor | Q-XD1-GSF-01 | scope-P-Goldschmidt-SymmetryFactor |

| Legacy record | Legacy query | Status | Disposition | Pair | Active query IDs | Legacy scope |
|---|---|---|---|---|---|---|
| rec-notestablished-P-Goldschmidt-KashaVavilov | q-pair-P-Goldschmidt-KashaVavilov | notEstablished | retained-scoped-absence | P-Goldschmidt-KashaVavilov | Q-XD2-KV-01 | scope-P-Goldschmidt-KashaVavilov |
| rec-notestablished-P-Goldschmidt-SternVolmer | q-pair-P-Goldschmidt-SternVolmer | notEstablished | retained-scoped-absence | P-Goldschmidt-SternVolmer | Q-XD2-SV-01 | scope-P-Goldschmidt-SternVolmer |
| rec-notestablished-P-Goldschmidt-QuantumYield | q-pair-P-Goldschmidt-QuantumYield | notEstablished | retained-scoped-absence | P-Goldschmidt-QuantumYield | Q-XD2-QY-01 | scope-P-Goldschmidt-QuantumYield |
| rec-notestablished-P-Goldschmidt-FluorPhos | q-pair-P-Goldschmidt-FluorPhos | notEstablished | retained-scoped-absence | P-Goldschmidt-FluorPhos | Q-XD2-FP-01 | scope-P-Goldschmidt-FluorPhos |
| rec-notestablished-P-Goldschmidt-EnergyGapLaw | q-pair-P-Goldschmidt-EnergyGapLaw | notEstablished | retained-scoped-absence | P-Goldschmidt-EnergyGapLaw | Q-XD2-EG-01, Q-XD2-EG-02 | scope-P-Goldschmidt-EnergyGapLaw |
| rec-notestablished-P-Goldschmidt-StokesShift | q-pair-P-Goldschmidt-StokesShift | notEstablished | retained-scoped-absence | P-Goldschmidt-StokesShift | Q-XD2-SS-01 | scope-P-Goldschmidt-StokesShift |
| rec-notestablished-P-Goldschmidt-ICvsISC | q-pair-P-Goldschmidt-ICvsISC | notEstablished | retained-scoped-absence | P-Goldschmidt-ICvsISC | Q-XD2-IC-01 | scope-P-Goldschmidt-ICvsISC |
| rec-notestablished-P-Goldschmidt-Forster | q-pair-P-Goldschmidt-Forster | notEstablished | retained-scoped-absence | P-Goldschmidt-Forster | Q-XD2-FR-01, Q-XD2-FR-02 | scope-P-Goldschmidt-Forster |
| rec-notestablished-P-Goldschmidt-Einstein | q-pair-P-Goldschmidt-Einstein | notEstablished | retained-scoped-absence | P-Goldschmidt-Einstein | Q-XD2-EI-01 | scope-P-Goldschmidt-Einstein |
| rec-notestablished-P-Goldschmidt-RACI | q-pair-P-Goldschmidt-RACI | notEstablished | retained-scoped-absence | P-Goldschmidt-RACI | Q-CI-GS-01 | scope-P-Goldschmidt-RACI |
| rec-notestablished-P-SymmetryFactor-KashaVavilov | q-pair-P-SymmetryFactor-KashaVavilov | notEstablished | retained-scoped-absence | P-SymmetryFactor-KashaVavilov | Q-E2-SK-01 | scope-P-SymmetryFactor-KashaVavilov |
| rec-notestablished-P-SymmetryFactor-SternVolmer | q-pair-P-SymmetryFactor-SternVolmer | notEstablished | retained-scoped-absence | P-SymmetryFactor-SternVolmer | Q-E2-SV-01 | scope-P-SymmetryFactor-SternVolmer |
| rec-notestablished-P-SymmetryFactor-QuantumYield | q-pair-P-SymmetryFactor-QuantumYield | notEstablished | retained-scoped-absence | P-SymmetryFactor-QuantumYield | Q-E2-SQ-01 | scope-P-SymmetryFactor-QuantumYield |
| rec-notestablished-P-SymmetryFactor-FluorPhos | q-pair-P-SymmetryFactor-FluorPhos | notEstablished | retained-scoped-absence | P-SymmetryFactor-FluorPhos | Q-E2-SF-01 | scope-P-SymmetryFactor-FluorPhos |
| rec-notestablished-P-SymmetryFactor-EnergyGapLaw | q-pair-P-SymmetryFactor-EnergyGapLaw | notEstablished | reconciled-absence-with-active-evidence | P-SymmetryFactor-EnergyGapLaw | Q-E2-SE-01 | scope-P-SymmetryFactor-EnergyGapLaw |
| rec-notestablished-P-SymmetryFactor-StokesShift | q-pair-P-SymmetryFactor-StokesShift | notEstablished | reconciled-absence-with-active-evidence | P-SymmetryFactor-StokesShift | Q-E2-SS-01 | scope-P-SymmetryFactor-StokesShift |
| rec-notestablished-P-SymmetryFactor-ICvsISC | q-pair-P-SymmetryFactor-ICvsISC | notEstablished | reconciled-absence-with-active-evidence | P-SymmetryFactor-ICvsISC | Q-E2-SI-01 | scope-P-SymmetryFactor-ICvsISC |
| rec-notestablished-P-SymmetryFactor-Forster | q-pair-P-SymmetryFactor-Forster | notEstablished | retained-scoped-absence | P-SymmetryFactor-Forster | Q-E2-SFo-01 | scope-P-SymmetryFactor-Forster |
| rec-notestablished-P-SymmetryFactor-Einstein | q-pair-P-SymmetryFactor-Einstein | notEstablished | retained-scoped-absence | P-SymmetryFactor-Einstein | Q-E2-SEIn-01 | scope-P-SymmetryFactor-Einstein |
| rec-notestablished-P-SymmetryFactor-RACI | q-pair-P-SymmetryFactor-RACI | notEstablished | retained-scoped-absence | P-SymmetryFactor-RACI | Q-CI-SF-01 | scope-P-SymmetryFactor-RACI |
| rec-candidate-P-KashaVavilov-SternVolmer | q-pair-P-KashaVavilov-SternVolmer | unresolved | reconciled-candidate-with-active-evidence | P-KashaVavilov-SternVolmer | Q-RAD1-VSV-01 | global |
| rec-candidate-P-KashaVavilov-QuantumYield | q-pair-P-KashaVavilov-QuantumYield | unresolved | reconciled-candidate-with-active-evidence | P-KashaVavilov-QuantumYield | Q-RAD1-VQY-01 | global |
| rec-candidate-P-KashaVavilov-FluorPhos | q-pair-P-KashaVavilov-FluorPhos | unresolved | reconciled-candidate-with-active-evidence | P-KashaVavilov-FluorPhos | Q-RAD1-VFP-01 | global |
| rec-candidate-P-KashaVavilov-EnergyGapLaw | q-pair-P-KashaVavilov-EnergyGapLaw | unresolved | retained-unresolved-candidate | P-KashaVavilov-EnergyGapLaw | Q-RAD1-VEG-01 | global |
| rec-notestablished-P-KashaVavilov-StokesShift | q-pair-P-KashaVavilov-StokesShift | notEstablished | retained-scoped-absence | P-KashaVavilov-StokesShift | Q-RAD2-KVS-01 | scope-P-KashaVavilov-StokesShift |
| rec-notestablished-P-KashaVavilov-ICvsISC | q-pair-P-KashaVavilov-ICvsISC | notEstablished | retained-scoped-absence | P-KashaVavilov-ICvsISC | Q-RAD2-KVI-01 | scope-P-KashaVavilov-ICvsISC |
| rec-notestablished-P-KashaVavilov-Forster | q-pair-P-KashaVavilov-Forster | notEstablished | retained-scoped-absence | P-KashaVavilov-Forster | Q-RAD2-KVF-01 | scope-P-KashaVavilov-Forster |
| rec-notestablished-P-KashaVavilov-Einstein | q-pair-P-KashaVavilov-Einstein | notEstablished | retained-scoped-absence | P-KashaVavilov-Einstein | Q-RAD2-KVE-01 | scope-P-KashaVavilov-Einstein |
| rec-notestablished-P-KashaVavilov-RACI | q-pair-P-KashaVavilov-RACI | notEstablished | retained-scoped-absence | P-KashaVavilov-RACI | Q-CI-KV-01 | scope-P-KashaVavilov-RACI |
| rec-notestablished-P-SternVolmer-QuantumYield | q-pair-P-SternVolmer-QuantumYield | notEstablished | reconciled-absence-with-active-evidence | P-SternVolmer-QuantumYield | Q-R3-SQ-01 | scope-P-SternVolmer-QuantumYield |
| rec-notestablished-P-SternVolmer-FluorPhos | q-pair-P-SternVolmer-FluorPhos | notEstablished | reconciled-absence-with-active-evidence | P-SternVolmer-FluorPhos | Q-R3-SF-01 | scope-P-SternVolmer-FluorPhos |
| rec-notestablished-P-SternVolmer-EnergyGapLaw | q-pair-P-SternVolmer-EnergyGapLaw | notEstablished | retained-scoped-absence | P-SternVolmer-EnergyGapLaw | Q-RAD2-SVE-01 | scope-P-SternVolmer-EnergyGapLaw |
| rec-notestablished-P-SternVolmer-StokesShift | q-pair-P-SternVolmer-StokesShift | notEstablished | retained-scoped-absence | P-SternVolmer-StokesShift | Q-RAD2-SVS-01 | scope-P-SternVolmer-StokesShift |
| rec-notestablished-P-SternVolmer-ICvsISC | q-pair-P-SternVolmer-ICvsISC | notEstablished | retained-scoped-absence | P-SternVolmer-ICvsISC | Q-RAD2-SVI-01 | scope-P-SternVolmer-ICvsISC |
| rec-notestablished-P-SternVolmer-Forster | q-pair-P-SternVolmer-Forster | notEstablished | retained-scoped-absence | P-SternVolmer-Forster | Q-RAD2-SVF-01 | scope-P-SternVolmer-Forster |
| rec-notestablished-P-SternVolmer-Einstein | q-pair-P-SternVolmer-Einstein | notEstablished | retained-scoped-absence | P-SternVolmer-Einstein | Q-RAD2-SVEb-01 | scope-P-SternVolmer-Einstein |
| rec-notestablished-P-SternVolmer-RACI | q-pair-P-SternVolmer-RACI | notEstablished | reconciled-absence-with-active-evidence | P-SternVolmer-RACI | Q-CI-SV-01 | scope-P-SternVolmer-RACI |
| rec-notestablished-P-QuantumYield-FluorPhos | q-pair-P-QuantumYield-FluorPhos | notEstablished | reconciled-absence-with-active-evidence | P-QuantumYield-FluorPhos | Q-R3-QF-01, Q-R3-QF-02 | scope-P-QuantumYield-FluorPhos |

| Legacy record | Legacy query | Status | Disposition | Pair | Active query IDs | Legacy scope |
|---|---|---|---|---|---|---|
| rec-notestablished-P-QuantumYield-EnergyGapLaw | q-pair-P-QuantumYield-EnergyGapLaw | notEstablished | reconciled-absence-with-active-evidence | P-QuantumYield-EnergyGapLaw | Q-RAD2-QYE-01 | scope-P-QuantumYield-EnergyGapLaw |
| rec-notestablished-P-QuantumYield-StokesShift | q-pair-P-QuantumYield-StokesShift | notEstablished | retained-scoped-absence | P-QuantumYield-StokesShift | Q-RAD2-QYS-01 | scope-P-QuantumYield-StokesShift |
| rec-notestablished-P-QuantumYield-ICvsISC | q-pair-P-QuantumYield-ICvsISC | notEstablished | reconciled-absence-with-active-evidence | P-QuantumYield-ICvsISC | Q-RAD2-QYI-01 | scope-P-QuantumYield-ICvsISC |
| rec-notestablished-P-QuantumYield-Forster | q-pair-P-QuantumYield-Forster | notEstablished | reconciled-absence-with-active-evidence | P-QuantumYield-Forster | Q-RAD2-QYF-01 | scope-P-QuantumYield-Forster |
| rec-notestablished-P-QuantumYield-Einstein | q-pair-P-QuantumYield-Einstein | notEstablished | reconciled-absence-with-active-evidence | P-QuantumYield-Einstein | Q-RAD2-QYEB-01 | scope-P-QuantumYield-Einstein |
| rec-notestablished-P-QuantumYield-RACI | q-pair-P-QuantumYield-RACI | notEstablished | reconciled-absence-with-active-evidence | P-QuantumYield-RACI | Q-CI-QY-01 | scope-P-QuantumYield-RACI |
| rec-notestablished-P-FluorPhos-EnergyGapLaw | q-pair-P-FluorPhos-EnergyGapLaw | notEstablished | retained-scoped-absence | P-FluorPhos-EnergyGapLaw | Q-RAD2-FPE-01 | scope-P-FluorPhos-EnergyGapLaw |
| rec-notestablished-P-FluorPhos-StokesShift | q-pair-P-FluorPhos-StokesShift | notEstablished | retained-scoped-absence | P-FluorPhos-StokesShift | Q-RAD2-FPS-01 | scope-P-FluorPhos-StokesShift |
| rec-notestablished-P-FluorPhos-ICvsISC | q-pair-P-FluorPhos-ICvsISC | notEstablished | reconciled-absence-with-active-evidence | P-FluorPhos-ICvsISC | Q-RAD3-FI-01, Q-RAD3-FI-02 | scope-P-FluorPhos-ICvsISC |
| rec-notestablished-P-FluorPhos-Forster | q-pair-P-FluorPhos-Forster | notEstablished | reconciled-absence-with-active-evidence | P-FluorPhos-Forster | Q-RAD3-FF-01 | scope-P-FluorPhos-Forster |
| rec-notestablished-P-FluorPhos-Einstein | q-pair-P-FluorPhos-Einstein | notEstablished | reconciled-absence-with-active-evidence | P-FluorPhos-Einstein | Q-RAD3-FE-01 | scope-P-FluorPhos-Einstein |
| rec-notestablished-P-FluorPhos-RACI | q-pair-P-FluorPhos-RACI | notEstablished | reconciled-absence-with-active-evidence | P-FluorPhos-RACI | Q-CI-FP-01 | scope-P-FluorPhos-RACI |
| rec-notestablished-P-EnergyGapLaw-Forster | q-pair-P-EnergyGapLaw-Forster | notEstablished | retained-scoped-absence | P-EnergyGapLaw-Forster | Q-RAD3-EF-01 | scope-P-EnergyGapLaw-Forster |
| rec-notestablished-P-EnergyGapLaw-Einstein | q-pair-P-EnergyGapLaw-Einstein | notEstablished | retained-scoped-absence | P-EnergyGapLaw-Einstein | Q-RAD3-EE-01 | scope-P-EnergyGapLaw-Einstein |
| rec-notestablished-P-EnergyGapLaw-RACI | q-pair-P-EnergyGapLaw-RACI | notEstablished | reconciled-absence-with-active-evidence | P-EnergyGapLaw-RACI | Q-CI-EG-01 | scope-P-EnergyGapLaw-RACI |
| rec-notestablished-P-StokesShift-ICvsISC | q-pair-P-StokesShift-ICvsISC | notEstablished | reconciled-absence-with-active-evidence | P-StokesShift-ICvsISC | Q-RAD3-SI-01 | scope-P-StokesShift-ICvsISC |
| rec-notestablished-P-StokesShift-Forster | q-pair-P-StokesShift-Forster | notEstablished | retained-scoped-absence | P-StokesShift-Forster | Q-RAD3-SF-01 | scope-P-StokesShift-Forster |
| rec-notestablished-P-StokesShift-Einstein | q-pair-P-StokesShift-Einstein | notEstablished | retained-scoped-absence | P-StokesShift-Einstein | Q-RAD3-SE-01 | scope-P-StokesShift-Einstein |
| rec-notestablished-P-StokesShift-RACI | q-pair-P-StokesShift-RACI | notEstablished | retained-scoped-absence | P-StokesShift-RACI | Q-CI-SS-01 | scope-P-StokesShift-RACI |
| rec-notestablished-P-ICvsISC-Forster | q-pair-P-ICvsISC-Forster | notEstablished | retained-scoped-absence | P-ICvsISC-Forster | Q-RAD3-IF-01 | scope-P-ICvsISC-Forster |
| rec-notestablished-P-ICvsISC-Einstein | q-pair-P-ICvsISC-Einstein | notEstablished | retained-scoped-absence | P-ICvsISC-Einstein | Q-RAD3-IE-01 | scope-P-ICvsISC-Einstein |
| rec-notestablished-P-ICvsISC-RACI | q-pair-P-ICvsISC-RACI | notEstablished | reconciled-absence-with-active-evidence | P-ICvsISC-RACI | Q-CI-IC-01 | scope-P-ICvsISC-RACI |
| rec-notestablished-P-Forster-Einstein | q-pair-P-Forster-Einstein | notEstablished | reconciled-absence-with-active-evidence | P-Forster-Einstein | Q-RAD3-FoE-01 | scope-P-Forster-Einstein |
| rec-notestablished-P-Forster-RACI | q-pair-P-Forster-RACI | notEstablished | retained-scoped-absence | P-Forster-RACI | Q-CI-FO-01 | scope-P-Forster-RACI |
| rec-notestablished-P-Einstein-RACI | q-pair-P-Einstein-RACI | notEstablished | reconciled-absence-with-active-evidence | P-Einstein-RACI | Q-CI-EB-01 | scope-P-Einstein-RACI |